# Modelling the Impact of Quantum Circuit Imperfections on Networks and Computer Applications

*Savo Glisic, Senior Member, IEEE,*
*Department of Physics, Worcester Polytechnic Institute,*
*Worcester, Massachusetts, sglisic@wpi.edu*

*Abstract* - Post-Quantum (*q*-) and Quantum Cryptography schemes ($s^{ch}$*'s*) are feasible q-computer applications for 7G networks. These $s^{ch}$ 's could possibly replace RSA and ECC $s^{ch}$ 's. These algorithms have been compromised by advances in quantum search algorithms run on quantum computers like Shor's algorithm. Shor's algorithm is a quantum algorithm for finding the prime factors of an integer which is the basis of RSA algorithm. This has become an available quantum computer application putting the use of RSA algorithm at risk.

Our recent paper provides a detailed survey of the work on post-quantum (q-) and q-cryptography (*Cry*) algorithms ($a^{lg}$*'s*) with focus on their applicability in 7G networks ($n^{et}$*'s*).

Since the paper focuses on the cryptography algorithms ($a^{lg}$*'s*) as a follow up, in this paper, we provide a new frame-work for quantum network optimization and survey in detail the work on enabling technologies (q-hardware) for the practical implementation of these $a^{lg}$*'s* including the most important segments of quantum hardware ($h^{rdw}$) in 7G.

As always in engineering practice practical solutions are a compromise between the performance and complexity/cost of the implementation. For this reason, as the main contribution, the paper presents a network and computer applications optimization framework that includes implementation imperfections. The tools should be useful in optimizing future generation practical computer system design. After that a comprehensive survey of the existing work on quantum hardware is presented pointing out the sources of these imperfections. This enables us to make a fair assessment of how much investment into quantum hardware improvements contributes to the performance enhancement of the overall system. In this way a decision can be made on proper partitioning between the investment in hardware and system level complexity.

The paper is organized as follows: In Section II we start by presenting a q-network optimization framework that includes implementation imperfections. In order to present the sources of these imperfections in Section III we survey work on elementary component of the system, q-bit physics. Work on q- $h^{rdw}$ building blocks, q-computing gate libraries, are surveyed in Section IV including q-memories surveyed in Section V. Section VI surveys several implementation examples of q-key distribution (QKD). Finally, Section VII offers concluding remarks.

*Index Terms*- q- *Network and computer applications; modelling q-hardware imperfections in network and computer applications Q memories; Qubit Physics; Quantum Circuit Libraries; implementation Examples of cv QKD -*$h^{rdw}$

*NOTE: As it is common for a survey type of paper, here we do not propose specific solutions for different problems that 6/7G* $n^{et}$*'s will face, but rather a performance/complexity comparison of a variety of technology enablers to choose from when building up a specific solution*

*NOTE on the writing style: In this paper we use specific notation where some characteristic terms (*$t^{rm}$*'s), often repeated in the text, are replaced with corresponding acronyms representing the original* $t^{rm}$ *and its derivatives (conjugations). This approach (compressed language) enables more precise characterization of the system (*$s^{yst}$*) processes (*$p^{ro}$*'s) and operations (*$o^{per}$ *'s) and a specific* $t^{rm}$ *sounds more like a* $s^{yst}$ *parameter that can be used more efficiently throughout the text. While this opens new options for the* $s^{yst}$ *presentation the writing occasionally sounds like an AI synthesized text. We hope the readers will easily get used to this style. In anticipation of what is coming in the field of ML and AI, this approach of integration of classical language and language of acronyms, might be further studied to increase the efficiency of Human-AI communication, maybe in the long run resulting in H-AI language. Light acronymization used in this paper, only for illustration purposes, may be further intensified. The depth of acronymization would depend on specific application.*

## *Acronyms*

$a^{ct}$- *act, action(s)*
$\mathcal{A}^{nn}$-*annihilation, annihilate*
$\mathcal{A}^{tt}$-*attack*
$a^{lg}$- *algorithm*
$a^{loc}$- *allocation*
$a^{mpl}$- *amplitude*
$\mathcal{A}$-*ancilla*
$a^{prox}$- *approximation(s), approximate(ly)(d)*
$a^{pp}$- *application(s), apply*
$a^{sis}$- *assisted*
$a^{symp}$- *asymptotic*
$b^{nd}$-*band*
$b^{\wedge}$-*upper bound*
$b_{\wedge}$ -*lower bound*
$b^{arr}$- *barrier*
$b^{as}$- *basis*
$\mathcal{B}$- *bosonic*

$B^{ell}$- Bell
$b^{eam}$- beam
$B^{ool}$ -Boolean
$b^{split}$- beam splitter(s)
BS-balanced $b^{split}$
$c^{city}$- capacity
$\bar{C}$ – CNOT
$c^{hrg}$- charge
$C^{rri}$- carrier
$sC^{rri}$- sub-carrier
$mC^{rri}$- multiple-carrier
$c^{irc}$- circuit(s)
$\mathcal{C}$-communications, communicate
$c^{crc}$- concurrence
$\mathcal{C}^{omm}$ -commutation, commutative, commuting, commute
$c^{omp}$- composable, composing, composability
$c^{oll}$- collective
$C^{trol}$- control(led)
$\mathcal{C}^{ha}$- channel
c- classical, conventional
$c^{omb}$- combination
$\mathcal{C}^{ompl}$-complex(ity)
$c^{hrnt}$- coherent
$c^{or}$- correction, correct
$c^{ord}$- coordinate(s)
$c^{pld}$- coupled(ing)
$c^{resp}$- correspond(ing),
$c^{plex}$- complex
C,$C^{omp}$ -comput(ing)(e), computation(aly), computer
$\mathcal{C}^{re}$ – creation
Cry- cryptography
$d^{ta}$- data
$d^{comp}$ -decompositions, decompose(ing)
$d^{vic}$- device
$d^{pth}$- depth
$d^{gre}$- degree(s)
$d^{ir}$- direct
$d^{t}$- dot(s)
$d^{str}$- distribution, distributed
$d^{tect}$- detection(s), detector, detect
$d^{yna}$- dynamic(s)
$\mathcal{D}^{x}$-x-dimensional
$\mathcal{E}$-entanglement, entangled, entangling
e- excess
$e^{ffi}$ -efficiency, efficient(ly)
$e^{drop}$- eavesdropper, eavesdropping

$e^{lm}$-electromagnetic
$e^{ctr}$- electric(al)(ly)
$e^{lec}$-electron
$e^{cod}$- encod(ing), (ed), (er)
$e^{nrgy}$- energy
$e^{vir}$- environment
eq- equation
$e^{vlt}$- evolution
$e^{ig}$ -eigen
$e^{xp}$-experimental(y), experiment, experimented
exp- exponential(y)
$\varepsilon$-error
$f^{iel}$- field
F-function, functional
$f^{rct}$- fraction
$F^{red}$- FREDKIN
$f^{req}$- frequency
$\mathcal{G}$-Gaussian
G-$g^{at}$-gate(s)
$\mathcal{G}r$- group
G-general
$g^{ph}$-graph
$g^{ner}$ – generate(d), generating, generalized
$\mathcal{H}$- Hamiltonian
$h^{rdw}$- hardware
$\mathcal{H}^{eis}$-Heisenberg
$\mathcal{H}^{mard}$-Hadamard
$h^{mon}$-harmonic
$h^{et}$ -heterodyne
$h^{om}$-homodyne
$\mathcal{I}$ information
$i^{d}$- identity
$\mathcal{I}^{mpl}$- implementations, implementing, implement(ed)
$i^{np}$- input
$i^{nit}$- initial, initialization
$i^{nv}$- inverse
$I^{dep}$- independent(ly)
$i^{ntac}$- interact(ion)
$k^{\mathcal{R}}$- key rate
$\mathcal{K}$-key
$l^{gic}$ -logic
lG-logic gate(s)
l$\mathcal{G}$-set of logic gate
$l^{ine}$- linear
$l^{ibr}$- library(ies)
$\mathcal{L}^{ink}$-link
$\mathcal{L}^{os}$-loss
$\underline{b}$-lower bound
$\mathcal{M}$-measure(ments), measuring, measured
$\mathcal{M}^{\leftrightarrows}$-map(ping)(ped)
$m^{emo}$ -memory(ies)

$m^{gn}$- magnetic
ML- machine learning
$mC^{rri}$-multicarrier
$m^{ix}$-mixers
$m^{xin}$- mixing, mixed
$m^{tri}$-matrix, matrices
$m^{ax}$- maximal(ly), maximum, maximiz(ation)
$m^{ech}$- mechanical, mechanics
$m^{ini}$- minimization, minimize(d), minimizing, minimum
$\mathcal{M}^{om}$-momentum
$m^{od}$- mode(s)
$m^{del}$-model(s)
$m^{odu}$- modulation, modulated
$m^{lti}$- multi(ple), ply
$n^{et}$- network(ing)
$\mathcal{N}^{ois}$- noise, noisy
$n^{od}$- node
$n^{ber}$- number
npqg-non-permutative q-$g^{at}$ 's
$\mathcal{O}$-optimization, optimal(y), optimize
$o^{rde}$- order(ing)
$o^{rtg}$- orthogonal
$o^{per}$- operator, operations, operational
$o^{pti}$- optical, optics
$o^{sc}$- oscillator
$o^{ut}$- output
$p^{erm}$- permutation
$p^{met}$- parameter
$p^{erf}$- performance, perform
$p^{rob}$-probability(ies)
$\mathcal{P}^{os}$-position
$\mathcal{P}^{th}$-path
$P^{au}$- Pauli
$p^{has}$-phase
$p^{hys}$- physics
$\mathcal{P}^{rol}$-protocol
$p^{duc}$- product
$p^{ten}$- potential
$p^{ton}$- photon
$p^{rep}$- prepare, preparation
$p^{ro}$-process(ing), process(or)
$p^{blem}$-problem
$p^{lar}$- polarization, polariz(ed)(ing)
$p^{ls}$- pulse
$p^{ur}$- pure
$p^{wer}$-power
Q, q-quantum
$\mathcal{Q}$-quadrature

qb-qubit(s)
QKD-quantum key distribution
$q^{za}$ -quantization
$\mathcal{R}$-rate
$\mathcal{R}^{conc}$-reconciliation
r-repeater
$r^{lay}$- relay
$r^{ceiv}$ -receiv(er), received
$r^{nd}$-random
$r^{eg}$- register
$\mathcal{R}e$- real
$r^{eso}$- resource
$r^{ot}$- rotate(ion)
$r^{vers}$- reversible, reverse
s-secure, secret
$s^{equ}$- sequence
$s^{ch}$- scheme
$\mathcal{S}^{gn}$- signal
$\mathcal{S}^{sec}$-security
$S^{epa}$- separability, separable, separated separate, separator, separation
$s^{ynth}$- synthesis, synthesize
$s^{tate}$ , $s^{ta}$ – state(s)
$s^{tr}$- stretch(ing), stretchable
$S^{po}$- superposition
$s^{pin}$- spin
$S_{\wedge}$- subset
$s^{con}$- superconducting
$s^{tr}$- stor(ed)
$s^{uff}$- sufficient
$s^{yst}$- system(s)
$s^{ymb}$- symbol
$s^{wa}$- swap, swapping
$t^{rgt}$- target
$t^{chn}$- technique
$t^{rm}$- term
$t^{hrm}$- thermal
$T^{off}$- TOFFOLI
$t^{rans}$- transformations, transform, transformed
$t^{port}$ -teleportation, teleport
$t^{miss}$ -transmission,
$t^{mit}$-transmitted, transmit(er), transfer
U- unitary
u-universal, universality
$u^{nwn}$- unknown
$\mathcal{V}$-vector
$v^{ria}$- variable
$v^{ar}$-variance
$w^{ve}$- wave_______________

## *I INTRODUCTION*

*Motivation*

While QKD enables the unconditional solution to the security ($\mathcal{S}^{sec}$-) problem ($p^{blem}$) , due to a number ($n^{ber}$) of open $p^{blem}$ ‘s , its optimum $s^{ch}$ is still hard to implement ($\mathcal{J}^{mpl}$). On the other side, QKD protocols ($\mathcal{P}^{col}$*’s*) , which are fully- device ($d^{vic}$) independent ($I^{dep}$) [1,2] , offering the maximum ($m^{ax}$) quantum (*q*-) $\mathcal{S}^{sec}$ , are difficult to $\mathcal{J}^{mpl}$ and have rather low secret (*s*-) key rate’s ($\mathcal{R}$) ($k^{\mathcal{R}}$*’s*). More feasible QKD $\mathcal{P}^{col}$ ‘s require using trusted $d^{vic}$ ‘s, enabling them to get better $\mathcal{R}$’s, but makes them more vulnerable to possible *side-channel* ($\mathcal{C}^{ha}$) attacks ($\mathcal{A}^{tt}$). Besides a need to compromise between $\mathcal{S}^{sec}$ and $\mathcal{R}$, they also need to compromise between $\mathcal{R}$ and distance, restricting any $\mathcal{J}^{mpl}$ of QKD. As a solution, we need to use *q*- repeaters (*r’s*) [ 3,4] and *q*- $n^{et}$ ‘s [5]. This improves distance and further enhance the $\mathcal{R}$’s by deploying more complex ($c^{plex}$) routing $\mathcal{P}^{col}$ ‘s. The analysis of *q*- *r* ‘s and secure (*s*-) QKD $n^{et}$ ‘s attracts a lot of attention today.

To optimize ($\mathcal{O}$) *q*- cryptography (*Cry*) design in $c^{plex}$ -$n^{et}$ ‘s, we need to be aware of mutual dependency and limitations of possible choices of the solutions in different segments of $c^{plex}$ -$n^{et}$ ‘s. Having discussed *Encryption/ Decryption* $\mathcal{P}^{col}$ *‘s for q- Cry* in [6] here we survey the relevant work on the related $p^{blem}$ ‘s of technology enablers for QKD focusing on QC -$h^{rdw}$ and its imperfections and present a network optimization framework for such scenario.

This framework, discussed in Section II, includes A. *Optimum Design of LEO* $s^{atel}$ *-*$n^{et}$ *for QKD, B. Network Cost Optimization framework, and C. Optimum Resource Allocation in q-Networks* Then in Section III covering qubit (*qb*) physics ($p^{hys}$), we discuss imperfections in superconducting ($s^{con}$) *qb* ‘s, *qb* -gates (*G’s*) based on spin ($s^{pin}$) states ($s^{ta}$) of coupled ($c^{pld}$) single-electron ($e^{lec}$) *q*- dots ($d^{t}$*’s*), *q*- logic ($l^{gic}$) by polarizing ($p^{lar}$) beam ($b^{eam}$) splitters, *q*- *G* -$\mathcal{J}^{mpl}$ ‘ed by Trapped ions, *q*- *G* and *q*- $l^{gic}$ -$n^{et}$ ‘s
In Section IV, we discuss *q*- computing ($C^{omp}$) *G* libraries ($l^{ibr}$*’s*) including *q*- *G* -$l^{ibr}$, depth ($d^{pth}$) -$\mathcal{O}$- *q*- circuits ($c^{irc}$*’s*), exact minimization ($m^{ini}$) of *q*- $c^{irc}$ *s* and decomposing ($d^{comp}$*’ing*) CV operations ($o^{per}$*’ions*) into a universal (*u*-) $g^{at}$ -$l^{ibr}$. In Section V on q- memories ($m^{emo}$*’s)*), we cover topics on integrated local unitaries (U’s), factorization U and basis ($b^{as}$) partitioning U. Finally in Section VI we present several $\mathcal{J}^{mpl}$ examples of CV QKD including discussion on imperfect channel included in effective $\mathcal{C}^{ha}$ model ($m^{del}$), $m^{del}$ ‘ling transceiver component, $\mathcal{P}^{col}$ ‘s and noise ($\mathcal{N}^{ois}$), QKD $\mathcal{J}^{mpl}$ at Terahertz bands ($b^{nd}$*’s*) and *q*- receivers ($r^{ceiv}$*’s*).

*Contribution*

As already indicated, this paper provides a survey of the work on QC -$h^{rdw}$ for implementing *Cry* algorithms for 7G $n^{et}$ ‘s emphasizing the sources of imperfections having impact on overall system performance.

The paper will provide to the $n^{et}$ designers data ($d^{ta}$) for:

a) Generating $s^{con}$ -*qb* ‘s,
b) Designing *qb* -$g^{at}$ based on $s^{pin}$ -$s^{ta}$ of $c^{pld}$ single- $e^{lec}$ -*q*- $d^{t}$ ‘s, *q*- $l^{gic}$ by $p^{lar}$ beam splitters ($b^{split}$), *q*- $g^{at}$ -$\mathcal{J}^{mpl}$‘ed by Trapped Ions, *q*- $g^{at}$ and *q*- $l^{gic}$ -$n^{et}$ ‘s
c) Using *q*- $C^{omp}$-$g^{at}$ for designing $d^{pth}$ -$\mathcal{O}$- *q*-$c^{irc}$ ‘s
d) Exact $m^{ini}$ of *q*- $c^{irc}$ ‘s and $d^{comp}$ ‘ing CV $o^{per}$ ‘ions into a *u*- $g^{at}$ -$l^{ibr}$.
e) Information ($\mathcal{J}$) on constraints imposed by q- $m^{emo}$ ‘ies
f) Generating integrated local U’s, factorization U and $b^{as}$ partitioning U.
g) $\mathcal{J}^{mpl}$ examples of CV QKD including discussion on effective $\mathcal{C}^{ha}$ -$m^{del}$, $m^{del}$ ‘ling transceiver component, $\mathcal{P}^{col}$ ‘s and $\mathcal{N}^{ois}$, QKD $\mathcal{J}^{mpl}$ at Terahertz $b^{nd}$ ‘s and *q*- $r^{ceiv}$ ‘s.
*h) Network optimization framework including* $h^{rdw}$ *imperfections.*

## *II q-NETWORK OPTIMIZATION IN THE PRESENCE OF QUANTUM* $h^{rdw}$ *IMPERFECTIONS*

### *A. Optimum Design of LEO* $s^{atel}$ *-*$n^{et}$ *for QKD*

*Here we present an option, for QKD,* $b^{se}$ *‘d only on LEO* $s^{atel}$ *‘s since reaching the LEO orbit requires only about 1% of the* $t^{rns}$ *-*$p^{wer}$ *required for reaching the GEO orbit. That is the major motivation for presenting a methodology for optimum design of such constellations. On the other hand, LEO -*$n^{et}$ *is dynamic and more demanding when it comes to its* $\mathcal{O}^{ptmz}$*. The objective is to design the LEO constellation with certain requirements on mutual visibility between* $s^{atel}$ *‘s because due to limited* $p^{erf}$ *of the q- memories any waiting in the* $n^{et}$ *nodes should be avoided. To formally formulate the* $p^{rblm}$ *let us first introduce some* $d^{fin}$ *‘itions. Definitions:*

*Orbit i: Approximately* $d^{fin}$ *‘ed by three angles, diameter, velocity and the time offset with* $r^{spct}$ *to the* $r^{ef}$ *time/*$\mathcal{P}^{os}$ *of the starting point on the orbit* $\mathcal{O}_i=(\alpha_i,\beta_i,\gamma_i, d_i, v_i,\Delta t)$

*The constellations:*$\mathcal{C}=\{\mathcal{O}_i\}=\mathcal{C}'\cup\mathcal{C}''$ *overall constellation,* $\mathcal{C}'=\{\mathcal{O}_i\}$ *existing constellation* $\mathcal{C}''=\{\mathcal{O}_i\}$ *augmented Constellation*

*Inter-satellite visibility:* $v_{i,j}=T(i,j)/T_o$ *, T(i,j)-overall time in an orbit cycle when there is visibility between satellite i and j,* $T_o$*-orbiting time*

*Network: S=S’+S’’- overall* $n^{umb}$ *of satellites (orbits) (size of the constellation), S’-* $n^{umb}$ *of satellites in existing constellation , S’’-* $n^{umb}$ of satellites in augmented constellation, $\mathcal{N}$-set of satellites in the $n^{et}$, $\mathcal{K}$, K -set and $n^{umb}$ of nodes in the $n^{et}$,

$V_{k,m}$ -visibility between the nodes

$\mathcal{O}^{ptmz}$ ‘ation: At this point different $\mathcal{O}^{ptmz}$ -$p^{rblm}$ ‘s can be formulated. Below are only some possible examples:

**#1** maximize the sum of visibilities between the satellites

$$\mathcal{P}_1 = max_{\mathcal{C}} \sum_{i,j} v_{i,j} \ ; \forall i, j \in \mathcal{C}$$

**#2** minimize the $n^{umb}$ of satellites that can make all nodes mutually visible all the time ($m^{lti}$ -hop zero wfv latency routs)

$$\mathcal{P}_2 = min_{\mathcal{C}} S \ ; \ s.t: V_{k,m} = 1; \ \forall k, m \in \mathcal{N}$$

**#3** minimize the $n^{umb}$ of satellites in augmented $n^{et}$ that can make all nodes mutually visibly all the time ($m^{lti}$ -hop zero wfv latency routs)

$$\mathcal{P}_3 = min_{\mathcal{C}} S" \ ; \ s.t: V_{k,m} = 1; \ \forall k, m \in \mathcal{N}$$

**#4** minimize the $n^{umb}$ of visible satellite’s neighbors to minimize the longest zero wfv latency route in hops

$\mathcal{N}_i^n = s_{i,j}^n$ -set of satellite’s neighboring satellites i ,

$\mathcal{N}_i^{n,v} = s_{i,j}^{n,v}$ -set of zero wfv latency satellites neighboring satellite i. Portion of the satellite’s neighbors with no interrupted visibility

$v_i = \left|\mathcal{N}_i^{n,v}\right| / \left|\mathcal{N}_i^n\right|$ , $\mathcal{V} = v_i$ , $\mathbb{R} = r_{k,m}$ -set of routs between the nodes in the $n^{et}$, $r_{k,m}$ - $m^{lti}$ -hop zero wfv latency rout,

$\mathbb{R}^v = r_{k,m}^v$ -set of routs between the nodes in the $n^{et}$

$$\mathcal{P}_4 = \min_{\mathcal{V}} max \, r_{k,m}^v \, ; \forall k, m$$

The $\mathcal{O}^{ptmz}$ -$p^{rblm}$ ‘s $d^{fin}$ ‘ed above are combinatorial in nature and NP hard. Here is where q- computing again comes into the picture. We can use either help from q- search $\mathcal{A}^{lgrt}$ ‘s, like Grover’s $\mathcal{A}^{lgrt}$ or q- $a^{prx}$ -$\mathcal{O}^{ptmz}$ -$\mathcal{A}^{lgrt}$ ‘s designed for $\mathcal{O}^{ptmz}$ of combinatorial $p^{rblm}$ ‘s [189].

In addition to the significant speed up in the computation ($c^{ompt}$) due to the parallelism in the $o^{prt}$ (Google has developed a Q-$c^{ompt}$ that can $p^{rfm}$ -$c^{ompt}$ $10^8$ times faster than the c- one) q- $\mathcal{I}$ theory offers additional advantages:

1. QSA $\mathcal{A}^{lgrt}$ like Grover’s $\mathcal{A}^{lgrt}$ can find the $m^{ax}$ -$v^{lu}$ of the component in the set of N entries in $\sim N^{1/2}$ iterations while the c- approach with exhaustive search would require ~N iterations. So, if for example $N=10^6$ , Grover’s $\mathcal{A}^{lgrt}$ would find the $m^{ax}$ (optimum $v^{lu}$) $10^3$ time faster than the c- approach.
2. In G, QAOA $\mathcal{A}^{lgrt}$ ‘s can find the $m^{ax}$ of the combinatorial $\mathcal{O}^{ptmz}$ -$p^{rblm}$ in $p^{oly}$ -times so turning the NP hard $p^{rblm}$ ‘s with exponential times into faster $p^{rblm}$ ‘s with the price that the optimum $v^{lu}$ is an $a^{prx}$ ‘ion. The compromise between the accuracy and speed up in the execution of the $\mathcal{O}^{ptmz}$ -$\mathcal{A}^{lgrt}$ is the design $p^{mtr}$.

B. Network Cost Optimization framework

Network specification:
Parameters relevant for the network optimization are

$\tau$- processing cycle

$\tau_{a,c}$-processing time for the algorithm (a=1,2,...,A), assuming there are A different options, mainly compromising between serial and parallel processing. This includes known and new algorithms that might be discovered in the future and evaluated using this optimization framework.
At the moment, in addition to the significant speed up in the computation due to the parallelism in the operation (Google has announced a Q-computer that can perform computing $10^8$ times faster than the classical one) quantum information theory offers additional advantages:
QSA algorithm like Grover’s algorithm can find the maximum/minimum value of the component in the set of N entries in $\sim N^{1/2}$ iterations while the classical approach with exhaustive search would require ~N iterations. So, if for example $N=10^6$ , Grover’s algorithm would find the maximum (optimum value) $10^3$ time faster than the classical approach.
In general, quantum approximative optimization algorithms (QAOA) can find the maximum of the combinatorial optimization problem in polynomial times so turning the NP hard problems with exponential times into faster problems with the price that the optimum value is an approximation. The compromise between the accuracy and speed up in the execution of the optimization algorithm is the design parameter.

c-computer type (c=1,2,..,C), assuming C different computers type are available ranging from simple processor laptop/ desktop computer to near future and full size quantum computer.

$\tau_p$-propagation time over network diameter (p=1,2,..,P). P different options are considered. This time might be negligible when locally optimizing NN design and relevant when optimizing distributed network like routing in large size network or federated ML.

$\tau_e$- size of the encoding slots (e=1,2,..,E). The larger size of the slot enables encoding the amplitudes in SNN with high precision.

$\tau_{e,a}$- aggregate size of the encoding slots ($e_a=1,2,..,E_a$).

$\tau_s$-sync acquisition time (s=1,2,..,S)

$\mathcal{I}^{A\times P\times Ea\times S\times C}$={a, p, e, s, c}-set of indices, includes all combination of the indices a, p, e, s, c

i=( a, p, e, s) given combination of indices a, p, e, s
$p_{c,p}$-network coherency probability for a given p. The larger p increases the probability that something might go wrong in the network reducing the probability of network coherency. For the faster computer the processing time is reduced so reducing the probability of incoherency.

By using above notation, the optimization processing cycle can be expressed as

$$\tau(i)=(\tau_a+\tau_p+\tau_{e,a})\, p_{c,p}+(\tau_a+\tau_p+\tau_{e,a}+\tau_s)(1- p_{c,p})$$

*Processing cost*
*$N_{a,c}$ -number of operations for algorithm a on a computer c*
*$C_c$ -cost per operation on computer type c*
*$\$^c$=$N_{a,c}$ $C_c$ - Processing cost for computer c*
*Optimization problem*

$\mathcal{P}_1 = \min_{i\in\mathcal{J}} \$^c(i) \times \tau(i)$

$\mathcal{P}_2 = \min_{i\in\mathcal{J}} \$^c(i)$ ; w.c. $\tau_{max} \geq \tau$

$\mathcal{P}_3 = \min_{i\in\mathcal{J}} \$^c(i)$ ; w.c. $\tau_{max} \geq \tau$ and $\tau_e \geq \tau_{emin}$

*C. Optimum Resource Allocation in q-Networks*

*By considering capacity and secrete key rates as a major network communication resource, here, besides the contributions listed in the introduction, we provide a specific contribution by designing an optimization framework for sharing these resources among $m^{lti}$ 'ple users in the network. Besides managing quantum resources of the network, the framework is also based on using quantum computing, and quantum optimization tools for these purposes.*

*Definitions/notations:*

*p - protocol $\mathcal{P}^{rol}$/algorithm $a^{lg}$ , p∈$\mathcal{P}$.*

*a- allocation option a∈$\mathcal{A}$*

*r- resources either capacity or secrete key rate*

*l- link (channel) in the network l∈$\mathcal{L}$*

*i- user i∈$\mathcal{N}$*

$r_{i,l}^a$-$r_i^a$, *resources allocated to user i, at link l (same for all l), at allocation a*

$h_{i,l}^a$- *indicator if user i is using link l at allocation a (=1 if yes, =0 otherwise)*

$R_i^a = \sum_{l=1}^{\mathcal{L}} r_{i,l}^a h_{i,l}^a$ *-overall resources allocated to user i, for the complete route in the network, at allocation a*

$h_i^a$- $\sum_{l=1}^{\mathcal{L}} h_{i,l}^a$ *number of links (hopes) used by user i at allocation a*

$r_l^a = \sum_{i=1}^{\mathcal{N}} r_{i,l}^a h_{i,l}^a \leq R_l$- *overall resources of link l used by all users at allocation a*

$R_l$ *- resources limits on link l*

$\mathcal{P}_1$: $\mathcal{U}_{max}$=$max_{a,p} \sum_{i=1}^{\mathcal{N}} r_i^a$; $\mathcal{N}$=constant

$\mathcal{P}_2$: $\mathcal{U}_{\mathcal{N}}$= $max_{a,p}$ $\mathcal{N}$; $r_i^a$=constant

$\mathcal{P}_3$: $\mathcal{U}^{eliability}$= $max_{a,p} \sum_{i=1}^{\mathcal{N}} \frac{r_i^a}{h_i^a}$

$\mathcal{P}_4$: $\mathcal{U}^{ransparency}$= $max_{a,p}\, min_l\, (R_l - r_l^a)$

*When it comes to the quantum optimization tools see comments from Section VIA.*

## III QUBIT PHYSICS

Steady progress in the *q*- systems (*$s^{yst}$'s*) theory and technology have advanced the *q*- computation (*$C^{omp}$*), toward the creation (*$\mathcal{C}^{re}$*) and manipulation of multi (*$m^{lti}$*) - *qb* - *$p^{ro}$*'s [5,6]. The technology shifted, from theory to the development of practical solutions in *design, control ($C^{trol}$), and readout* of *$m^{lti}$* -*qb* -*q*- *$s^{yst}$* 's creating a new discipline referred to as *q*-engineering.

One way for building a *$m^{lti}$* - *qb*- *q*- *$p^{ro}$* 'or is based on using *$s^{con}$* -*qb* 's, where $\mathcal{J}$ is encoded in the *q*- *$d^{gre}$* 's of freedom of, harmonic (*$h^{mon}$*) oscillators (*$o^{sc}$'s*) built from *$s^{con}$* -*$c^{irc}$* elements [1-4,10-15]. Alternative ways, e.g. *$e^{lec}$* -*$s^{pin}$* 's in silicon [16-21] and *q*- *$d^{t}$'s* [22 − 25], trapped ions [26-29], ultracold atoms [ 30 − 33], nitrogen - vacancies in diamonds [34,35], and *$p^{lar}$* photons (*$p^{ton}$'s*) [36-39] , have been also developed. Here we will review main characteristics of these groups ($\mathcal{G}r's$) of technologies.

*A. Superconducting qb 's*

Here, we will discuss how *q*- *$s^{yst}$* 's based on *$s^{con}$* -*$c^{irc}$* 's can be constructed and how to construct the interactions (*$i^{ntac}$'s*) (interfaces) between distinct *q*- *$s^{yst}$* 's, llike *qb* - *qb* and *qb* -resonator couplings.

A *q*- mechanical (*$m^{ech}$*) *$s^{yst}$* is described by the Schrödinger (*$S^{chrö}$*) equation (*eq*) [7],

$$\hat{H}|\psi(t)\rangle = i\hbar\frac{\partial}{\partial t}|\psi(t)\rangle,$$

Here $|\psi(t)\rangle$ represents the *$s^{ta}$* of the *q*- *$s^{yst}$* , ħ is the Planck's constant, and $\hat{H}$ is the *Hamiltonian* ($\mathcal{H}$) that represents the total energy of the *$s^{yst}$*. The "hat" indicates that $H$ is a *q*- *$o^{per}$* 'or. Since the *$S^{chrö}$* -*eq* is a first-order (*$o^{rde}$*) linear (*$l^{ine}$*) differential *eq*, its formal solution is,

$$|\psi(t)\rangle = e^{-i\hat{H}t/\hbar}|\psi(0)\rangle.$$

The time- *$I^{dep}$* -$\mathcal{H}$- $\hat{H}$ defines the evolution (*$e^{vlt}$*) in time of the *$s^{yst}$* by the *$o^{per}$*'or $e^{-i\hat{H}t/\hbar}$. Like in classical (*c*-) *$s^{yst}$* 's, getting the $\mathcal{H}$ of a *$s^{yst}$* , either the *c*-$\mathcal{H}$- $H$ or *q*- $\hat{H}$ , is needed to derive its *$d^{yna}$* 'al behavior.

To describe the *$d^{yna}$* of a *$s^{con}$* – *qb*- *$c^{irc}$*, we can use the *c*- *$l^{ine}$* LC -*$c^{irc}$* [Fig.1(a)]. Here, we have oscillation of the energy (*$e^{nrgy}$*) between electrical (*$e^{ctr}$'al*) *$e^{nrgy}$* in the capacitor $C$ and magnetic (*$m^{gn}$*) -*$e^{nrgy}$* in the inductor $L$. In the sequel, we associate the *$e^{ctr}$* -*$e^{nrgy}$* with the "kinetic *$e^{nrgy}$*" and the *$m^{gn}$* -*$e^{nrgy}$* with the "potential (*$p^{ten}$*) *$e^{nrgy}$*" of the *$o^{sc}$*. As a function of time, *$e^{nrgy}$* in each *$c^{irc}$* is function of its current $I(t')$ and voltage $V(t')$ ,

$$E(t) = \int_{-\infty}^{t} V(t')I(t')dt', \quad (1)$$

The Lagrange-Hamilton formulation is used when deriving the *c*- $\mathcal{H}$. Here, the *$c^{irc}$* elements are represented in *$t^{rm}$* 's of charge (*$c^{hrg}$*) or flux. We use flux, defined as

$$\Phi(t) = \int_{-\infty}^{t} V(t')dt'. \quad (2)$$

For more details see [40].

We can replace the use of kinetic *$e^{nrgy}$* (momentum ($\mathcal{M}^{om}$) coordinates- *$c^{ord}$*) and *$p^{ten}$* -*$e^{nrgy}$* (position ($\mathcal{P}^{os}$) *$c^{ord}$*), by the *$c^{hrg}$* variable (*$v^{ria}$*) $Q(t) = \int_{-\infty}^{t} I(t')dt'$ .

From (1) and (2), and $V = LdI/dt$ and $I = CdV/dt$, and using the partial integration formula, we get $e^{nrgy}$ -$t^{rm}$ 's for the C and L component in $t^{rm}$ 's of the $n^{od}$ flux,

$$\mathcal{T}_C = \frac{1}{2} C\dot{\Phi}^2, \quad (3)$$

$$\mathcal{U}_L = \frac{1}{2L} \Phi^2. \quad (4)$$

The Lagrangian is defined as

$$\mathcal{L} = \mathcal{T}_C - \mathcal{U}_L = \frac{1}{2} C\dot{\Phi}^2 - \frac{1}{2L} \Phi^2. \quad (5)$$

From (5), by using the Legendre transformation ($t^{rans}$), we get

$$Q = \frac{\partial \mathcal{L}}{\partial \dot{\Phi}} = C\dot{\Phi}. \quad (6)$$

The $\mathcal{H}$ of the $s^{yst}$ is

$$H = Q\dot{\Phi} - \mathcal{L} = \frac{Q^2}{2C} + \frac{\Phi^2}{2L} \equiv \frac{1}{2} CV^2 + \frac{1}{2} LI^2, \quad (7)$$

as expected for an $e^{ctr}$ -LC- $c^{irc}$. This $\mathcal{H}$ is equivalent to that of a $m^{ech}$ -$h^{mon}$ -$o^{sc}$, with mass $m = C$, and resonant frequency ($f^{req}$) $\omega = 1/\sqrt{LC}$, which expressed in $\mathcal{P}^{os}$, $x$, and

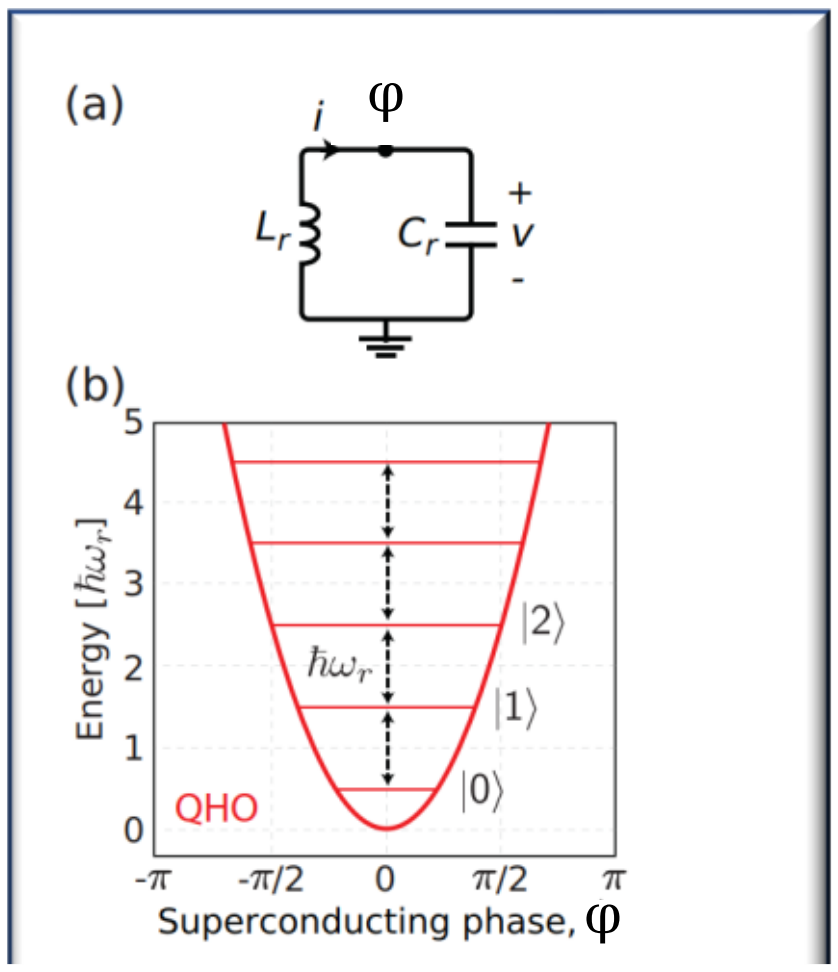

*Fig1. (a) Parallel LC oscillator (b) Energy potentials [125]*

$\mathcal{M}^{om}$, $p$, $c^{ord}$ 's becomes $H = p^2/2m + m\omega^2 x^2/2$[7]. The $\mathcal{H}$ derived above is $c$-.

For a $q$ -$m^{ech}$ representation of the $s^{yst}$, we use $c^{hrg}$ and flux $c^{ord}$ 's for $q$- $o^{per}$ 'ors. Whereas the $c$- $c^{ord}$ 's satisfy the Poisson bracket:

$$\{f, g\} = \frac{\delta f}{\delta \Phi}\frac{\delta g}{\delta Q} - \frac{\delta g}{\delta \Phi}\frac{\delta f}{\delta Q} \quad (8)$$

$$\rightarrow \{\Phi, Q\} = \frac{\delta \Phi}{\delta \Phi}\frac{\delta Q}{\delta Q} - \frac{\delta Q}{\delta \Phi}\frac{\delta \Phi}{\delta Q} = 1 - 0 = 1, \quad (9)$$

the $q$- $o^{per}$ 'ors commute ($\mathcal{C}^{omm}$):

$$[\hat{\Phi}, \hat{Q}] = \hat{\Phi}\hat{Q} - \hat{Q}\hat{\Phi} = i\hbar, \quad (10)$$

where hats indicate $o^{per}$ 'ors, which may be omitted for simplicity.

---

*EXAMPLE*: In Fig. 1(a), both L and C are $l^{ine}$ -$c^{irc}$'s. With the normalized flux $\varphi \equiv 2\pi\Phi/\Phi_0$ and the $c^{hrg}$- n $= Q/2e$, the $q$-$m^{ech}$ -$\mathcal{H}$ for the $c^{irc}$ becomes,

$$H = 4E_C n^2 + \frac{1}{2} EL\, \varphi^2, \quad (11)$$

$E_C = e^2/(2C)$ is the $e^{nrgy}$ needed to add each $e^{lec}$ of the Cooper-pair to the island and $E_L = (\Phi_0/2\pi)^2/L$ is the inductive $e^{nrgy}$, and $\Phi_0 = h/(2e)$ is the $s^{con}$ - $m^{gn}$ flux quantum. The $q$- $o^{per}$ 'or n is the $e$- $n^{ber}$ of Cooper-pairs on the island, and φ — the reduced flux —denotes the "gauge-invariant phase ($p^{has}$)" across the L $c^{irc}$. These two $o^{per}$ 'ors $\mathcal{C}^{comm}$ with $[\varphi, n] = i$. Factor 4 in (11) is because this $e^{nrgy}$ scale was first defined for single- $e^{lec}$ -$s^{yst}$ 's and then adopted to two- $e^{lec}$ Cooper-pair $s^{yst}$ 's.

The $\mathcal{H}$ in Eq. (11) is same as for a particle in a one-dimensional ($\mathcal{D}^{1}$)) quadratic $p^{ten}$, a $q$- $h^{mon}$ -$o^{sc}$. We can consider φ as the $g^{ner}$ 'ed $\mathcal{P}^{os}$ -$c^{ord}$, where the first $t^{rm}$ is the kinetic $e^{nrgy}$, and the second $t^{rm}$ is the $p^{ten}$-$e^{nrgy}$. The functional ($F$'$al$) form of the $p^{ten}$- $e^{nrgy}$ has an impact on the eigen ($e^{ig}$)-solutions. For instance, the fact that ($U_L \propto \varphi^2$) in Eq. (11) shapes the $p^{ten}$ in Fig. 1(b). The solution to this $e^{ig}$ -value $p^{blem}$ gives an infinite series of $e^{ig}$ -$s^{ta}$ $|k\rangle$, $(k = 0,1,2,\ldots)$ , with corresponding ($c^{resp}$'ing) $e^{ig}$ -energies $E_k$ obeying the relation $E_{k+1} - E_k = \hbar\omega_r$, where $\omega_r = \sqrt{8E_L E_C}/\hbar = 1/\sqrt{LC}$ is the resonant $f^{req}$ of the $s^{yst}$, see Fig. 1(b). This gives (second quantization ($q^{za}$)) for the $q$- $h^{mon}$ -$o^{sc}$ -$\mathcal{H}$ as

$$H = \hbar\omega_r(\, a^\dagger a{+}1/2) \quad (12)$$

with $a^\dagger(a)$ being the $\mathcal{C}^{re}$ (annihilation ($\mathcal{A}^{nn}$)) $o^{per}$ 'or of a single excitation of the resonator. The $\mathcal{H}$ in Eq. (12) is represented as an $e^{nrgy}$ and if divided by ħ has units of radian $f^{req}$. This is useful since in the sequel we will drive transitions at a particular $f^{req}$ at which two $s^{yst}$ 's mutually interact. So, in the sequel, ħ will be omitted.

The $c^{hrg}$- $n^{ber}$ and $p^{has}$ -$o^{per}$ 'ors are given as $n = n_{zpf} \times i(a - a^\dagger)$ and $\varphi = \varphi_{zpf} \times (a + a^\dagger)$ , where $n_{zpf} = [E_L/(32E_C)]^{1/4}$ and $\varphi_{zpf} = (2E_C/E_L)^{1/4}$ are the *zero-point fluctuations* of the $c^{hrg}$ and $p^{has}$ -$v^{ria}$ 's, respectively. Q- $m^{ech}$ 'ly, the $q$- $s^{ta}$ are written as wave ($w^{ve}$) -$F$ 's that are $G$ 'ly distributed ($d^{str}$'ed) over $n$ and φ and, as a consequence, the $w^{ve}$ -$F$ 's have non-zero standard deviations. Such $w^{ve}$ - $F$ -$d^{str}$ 'ions are "$q$-fluctuations," and are present, even in the ground $s^{ta}$ (zero-point fluctuations).

Before the $s^{yst}$ can be used as a $qb$, we need to define a $C^{omp}$ subspace including only two $e^{nrgy}$ -$s^{ta}$ (usually the two-lowest $e^{nrgy}$ -$e^{ig}$ -$s^{ta}$) in between which transitions can be driven without also exciting other levels in the $s^{yst}$. Since many $g^{at}$ -$o^{per}$ 'ions, like a single-$qb$ -$g^{at}$, depend on $f^{req}$ selectivity, the equidistant level-spacing of the q-harmonic oscillator QHO, illustrated in Fig. 1(b), poses a practical limitation.

---

To deal with the $p^{blem}$ of unwanted $d^{yna}$ involving non-$C^{omp}$ -$s^{ta}$, an'$h^{mon}$ 'ity (or nonlinearity) needs to be added into the $s^{yst}$. In short, we require the transition frequencies $\omega_q^{0\to1}$ and $\omega_q^{1\to2}$ be different enough to be individually adressable. In $G$, the larger the an'$h^{mon}$ 'ity the better. A limit on how short the $p^{ls}$ 's used to drive the *qb* can be, depends on the amount of an'$h^{mon}$ 'ity.

The nonlinearity required to modify the $h^{mon}$-$p^{ten}$, is introduced by using the Josephson junction —a non' $l^{ine}$, dissipationless $c^{irc}$ element that forms the backbone in $s^{con}$ -circ's [40,41]. By replacing the $l^{ine}$ inductor of the QHO with a Josephson junction, playing the role of a non' $l^{ine}$ inductor, we can modify the *F* 'al form of the $p^{ten}$- $e^{nrgy}$. The $p^{ten}$- $e^{nrgy}$ of the Josephson junction can be derived from Eq. (1) and the two Josephson relations

$$I = I_c \sin(\varphi), V = \frac{\hbar}{2e}\frac{d\varphi}{dt}, \quad (13)$$

resulting in a modified $\mathcal{H}$

$$H = 4E_C n^2 - E_J \cos(\varphi), \quad (14)$$

where $E_C = e^2/(2C_\Sigma)$, $C_\Sigma = C_s + C_J$, $C_s$ is shunt capacitance, $C_J$ is the self-capacitance of the junction, and $E_J = I_c\Phi_0/2\pi$ is the Josephson $e^{nrgy}$, with $I_c$ being the critical current of the junction (the maximum supercurrent that the junction can support before it switches to the resistive $s^{ta}$ with non-zero voltage). With introduction of the Josephson junction in the $c^{irc}$, the $p^{ten}$- $e^{nrgy}$ no longer has parabolic form (resulting in the $h^{mon}$ spectrum), but rather a cosinusoidal form, see Eq. (14), which makes the $e^{nrgy}$ spectrum non-degenerate. So, the Josephson junction is the key ingredient that makes the $o^{sc}$ an' $h^{mon}$ and thus enables us to identify a uniquely addressable *q*- two-level $s^{yst}$.

After the non' $l^{ine}$'ity has been added, the $s^{yst}$ -$d^{yna}$ depends on the dominant $e^{nrgy}$ in Eq. (14), reflected in the $E_J/E_C$ ratio. The $s^{con}$ -*qb* engineering has converged towards $c^{irc}$ constructions with $E_J \gg E_C$. If $E_J \leq E_C$, the *qb* is highly sensitive to $c^{hrg}$ -$\mathcal{N}^{ois}$, which is more difficult to mitigate than flux $\mathcal{N}^{ois}$, making it challenging to achieve high coherence. Another reason is that there is more flexibility in engineering the inductive (or $p^{ten}$) part of the $\mathcal{H}$. So, working in the $E_J \leq E_C$ limit, makes the $s^{yst}$ more sensitive to the change in the $p^{ten}$- $\mathcal{H}$. As a consequence, we will focus here on the $s^{ta}$ -of-the-art *qb* solutions with $E_J \gg E_C$.

A way to achieve $E_J \gg E_C$ is to make $E_C$ small by shunting the junction with a large capacitor, $C_s \gg C_J$, which makes the *qb* less sensitive to $c^{hrg}$ -$\mathcal{N}^{ois}$ — a $c^{irc}$ known as the *transmon qb* [40]. In this limit, the $s^{con}$ -$p^{has}$ -φ is a good *q*- $n^{ber}$, i.e. the spread (or *q*- f*luctuation*) of φ represented by the *q*- $w^{ve}$ -*F* is small. So low- $e^{nrgy}$ -$e^{ig}$ - $s^{ta}$ are, to a good approximation ($a^{prox}$), localized $s^{ta}$ in the $p^{ten}$- well. By expanding the $p^{ten}$- $t^{rm}$ of Eq.(14) into a power ($p^{wer}$) series (since φ is small), as

$$E_J \cos(\varphi) = \frac{1}{2} E_J \varphi^2 - \frac{1}{24} E_J \varphi^4 + \mathcal{O}(\varphi^6). \quad (15)$$

we may get better insight. The leading $\varphi^2$ =$t^{rm}$ in Eq. (15) alone will result in a q-harmonic oscillator QHO, see Eq. (11). The $\varphi^4$ -$t^{rm}$ modifies the $e^{ig}$ -solution and disrupts the otherwise $h^{mon}$ -$e^{nrgy}$ structure. Note that, the negative coefficient of the $\varphi^4$ -$t^{rm}$ indicates that the an' $h^{mon}$'ity $\alpha = \omega_q^{1\to2} - \omega_q^{0\to1}$ is negative. For transmon, $\alpha = -E_C$ is usually chosen to be 100—300 MHz, to keep a desirable *qb*- $f^{req}$ $\omega_q = \left(\sqrt{8E_J E_C} - E_C\right)/\hbar$ = 3-6 GHz, while keeping an $e^{nrgy}$ ratio $s^{uff}$ 'ly large $\left(E_J/E_C \geq 50\right)$ to suppress $c^{hrg}$ sensitivity [42]. The good news is that, the $c^{hrg}$ sensitivity is exponentially (*exp'ly*) suppressed for increased $E_J/E_C$, while the reduction in an' $h^{mon}$'ity shows a weak- $p^{wer}$ law, resulting in an acceptable $d^{vic}$.

Using $t^{rm}$ 's up to $\varphi^4$ and the QHO $e^{ig}$ - $b^{as}$, the $s^{yst}$ -$\mathcal{H}$ becomes similar to that of a Duffing $o^{sc}$.

$$H = \omega_q a a^\dagger + \frac{\alpha}{2} a^\dagger a^\dagger a a. \quad (16)$$

Since $|\alpha| \ll \omega_q$, the transmon *qb* is basically a weakly an' $h^{mon}$ -$o^{sc}$ (AHO). If excitation to higher non- $C^{omp}$ -$s^{ta}$ is suppressed over any $g^{at}$ -$o^{per}$ 'ions ( by a large $|\alpha|$ or by $C^{trol}$ techniques ($t^{chn}$*'s*) such as the DRAG (Derivative Removal by Adiabatic $g^{at}$) $p^{ls}$, discussed later) the AHO may be treated as a *q*- two-level $s^{yst}$, reducing the $\mathcal{H}$ to

$$H = \omega_q \sigma_z/2 \quad (17)$$

Here $\sigma_z$ is the Pauli ($P^{au}$)-z -$o^{per}$ 'or. Never the less the higher levels physically ($p^{hys}$*'ly*) exist [43] and their impact on $s^{yst}$ -$d^{yna}$ should be considered when constructing the $s^{yst}$ and its $C^{trol}$ -$p^{ro}$ 's. In many cases the higher levels are useful to $\mathcal{J}^{mpl}$ more efficient ($e^{ffi}$) $g^{at}$ -$o^{per}$ 'ions [44].

The large shunt capacitor is used to decrease the $c^{hrg}$ dispersion. When forming the 3D transmon [45] it was shown that increasing the distance between the capacitor plates prolongs the coherence time [46-51].

*Tunable qb split transmon*: To $\mathcal{J}^{mpl}$ fast $g^{at}$ -$o^{per}$ 'ions with high-fidelity, required for $\mathcal{J}^{mpl}$ -*q*- $l^{gic}$, many (though not all [52] ) of the *q*- $p^{ro}$ 'or $\mathcal{J}^{mpl}$ 'ed today use tunable *qb* frequencies [53-56]. As example, sometimes we need two *qb* 's in resonance ,to exchange (swap- $s^{wa}$) $e^{nrgy}$, while we want to keep them separated ($S^{epa}$) during idling periods, to minimize their $i^{ntac}$ 's. For this, a parameter ($p^{met}$) is needed allowing us to access one of the degrees ($d^{gre}$*'s*) of freedom of the $s^{yst}$ in a $C^{trol}$ 'lable way. A possible $t^{chn}$ is to replace the single Josephson junction with a loop interrupted by two identical junctions-creating a *dc* $s^{con}$ -*q*-*interference* $d^{vic}$ (dc-SQUID) [57].

To generate ($g^{ner}$) entanglement ($\mathcal{E}$) between individual *q*- $s^{yst}$ 's - we need to have an $i^{ntac}$ -$\mathcal{H}$ that connects $d^{gre}$ 's of freedom in those individual $s^{yst}$ 's. For more details see [58-69].

*Qubit Gates:* Here, we discuss how $s^{con}$-*qb* 's are manipulated to $\mathcal{J}^{mpl}$ -*q*- algorithms ($a^{lg}$*'s*).

By assuming that the reader is familiar with the $g^{at}$ used in $c$-$C^{omp}$ here we discuss the $g^{at}$ used in $q$- $C^{omp}$, and the notion of $u$'ity. Then we describe the $t^{chn}$ of designing single $qb$- $g^{at}$ by using a capacitive coupling of a micro- $w^{ve}$ line, $c^{pld}$ to the $qb$. We use the concept of "virtual Z $g^{at}$" and "DRAG" pulsing. In the sequel, we review some $\mathcal{J}^{mpl}$ 's of two $qb$ -$g^{at}$ in both tunable and fixed- $f^{req}$ transmon $qb$- s. The 1- $qb$ and 2- $qb$ -$o^{per}$ 'ions together are the $b^{as}$ of a number of $s^{con}$- $q$- $p^{ro}$ 'ors nowadays.

---

*EXAMPLE:* By using $P^{au}$ -$o^{per}$ 'ors,

$$\sigma_x = \begin{pmatrix} 0 & 1 \\ 1 & 0 \end{pmatrix}, \sigma_y = \begin{pmatrix} 0 & -i \\ i & 0 \end{pmatrix} \text{ and } \sigma_z = \begin{pmatrix} 1 & 0 \\ 0 & -1 \end{pmatrix}$$

we express everything in the $C^{omp}$ -$b^{as}$ $\{|0\rangle, |1\rangle\}$ where $|0\rangle$ is the $+1$ $e^{ig}$ - $s^{ta}$ of $\sigma_z$ and $|1\rangle$ is the $-1$ $e^{ig}$ - $s^{ta}$. With uppercase fonts we denote the rotation ($r^{ot}$) -$o^{per}$ 'or of a $qb$- $s^{ta}$, e.g. $r^{ot}$ 's around the $x$-axis by an angle $\theta$ is expressed as

$$X_\theta = R_X(\theta) = e^{-i\frac{\theta}{2}\sigma_x} = \cos(\theta/2)\mathbb{I} - i\sin(\theta/2)\sigma_x$$

and with X a $\pi$ -$r^{ot}$ around the $x$ axis (equivalently for $\mathrm{Y} := \mathrm{Y}_\pi$ and $\mathrm{Z} := \mathrm{Z}_\pi$).

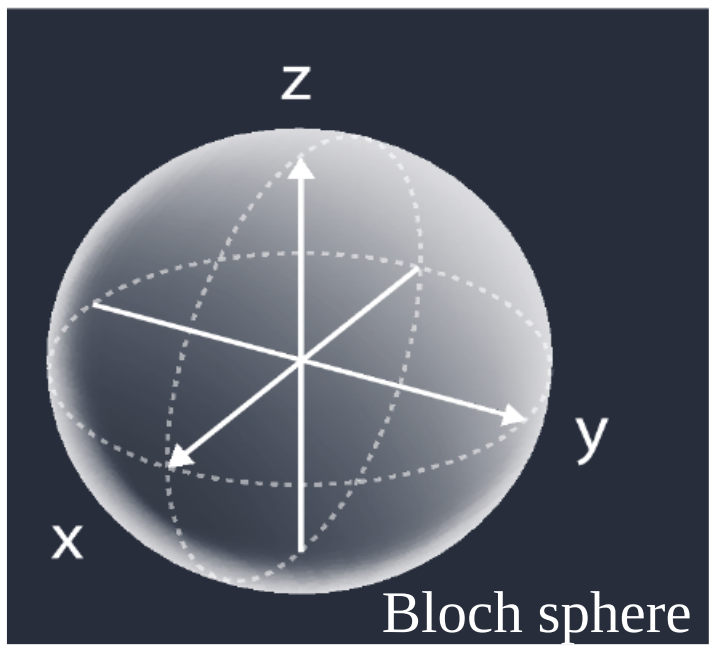

*Fig 2: Bloch sphere*

$Q$- $l^{gic}$ -$g^{at}$ 's (lG): Similar to $c$- $g^{at}$ 's, $q$- $l^{gic}$ can be $p^{erf}$ 'ed by a small set of 1- $qb$ and 2- $qb$ -$g^{at}$ 's. Qb's can be in $c$- $s^{ta}$ $|0\rangle$ and $|1\rangle$, at the upper pole and lower pole of the Bloch sphere (Fig.2), and also in arbitrary superpositions ($S^{po}$ 's) $\alpha|0\rangle + \beta|1\rangle$, $c^{resp}$ 'ing to any other $\mathcal{P}^{os}$ on the sphere.

---

Single- $qb$ -$o^{per}$ 'ions move an arbitrary $q$- $s^{ta}$ from one point on the Bloch sphere to another point by rotating the Bloch vector ($\mathcal{V}$) ($s^{pin}$) a certain angle about a particular axis.

---

| GATE | MATRIX REPRESENTATION |
|---|---|
| I Identity-gate: no rotation is performed. [I] | $I = \begin{pmatrix} 1 & 0 \\ 0 & 1 \end{pmatrix}$; Input → Output: $|0\rangle \to |0\rangle$, $|1\rangle \to |1\rangle$ |
| X gate: rotates the qubit state by π radians (180°) about the x-axis. [X] | $X = \begin{pmatrix} 0 & 1 \\ 1 & 0 \end{pmatrix}$; Input → Output: $|0\rangle \to |1\rangle$, $|1\rangle \to |0\rangle$ |
| Y gate: rotates the qubit state by π radians (180°) about the y-axis. [Y] | $Y = \begin{pmatrix} 0 & -i \\ i & 0 \end{pmatrix}$; Input → Output: $|0\rangle \to i\,|1\rangle$, $|1\rangle \to -i\,|0\rangle$ |
| Z gate: rotates the qubit state by π radians (180°) about the z-axis. [Z] | $Z = \begin{pmatrix} 1 & 0 \\ 0 & -1 \end{pmatrix}$; Input → Output: $|0\rangle \to |0\rangle$, $|1\rangle \to -|1\rangle$ |
| S gate: rotates the qubit state by $\frac{\pi}{2}$ radians (90°) about the z-axis. [S] | $S = \begin{pmatrix} 1 & 0 \\ 0 & e^{i\frac{\pi}{2}} \end{pmatrix}$; Input → Output: $|0\rangle \to |0\rangle$, $|1\rangle \to e^{i\frac{\pi}{2}}|1\rangle$ |
| T gate: rotates the qubit state by $\frac{\pi}{4}$ radians (45°) about the z-axis. [T] | $T = \begin{pmatrix} 1 & 0 \\ 0 & e^{i\frac{\pi}{4}} \end{pmatrix}$; Input → Output: $|0\rangle \to |0\rangle$, $|1\rangle \to e^{i\frac{\pi}{4}}|1\rangle$ |
| H gate: rotates the qubit state by π radians (180°) about an axis diagonal in the x-z plane. This is equivalent to an X-gate followed by a $\frac{\pi}{2}$ rotation about the y-axis. [H] | $H = \frac{1}{\sqrt{2}}\begin{pmatrix} 1 & 1 \\ 1 & -1 \end{pmatrix}$; Input → Output: $|0\rangle \to \frac{|0\rangle + |1\rangle}{\sqrt{2}}$, $|1\rangle \to \frac{|0\rangle - |1\rangle}{\sqrt{2}}$ |

Fig 3 Quantum single-qubit gates. Matrices are defined in the basis spanned by the state vectors $|0\rangle \equiv [10]^T$ and $|1\rangle \equiv [01]^T$. The numerical values in the truth table correspond to the quantum states $|0\rangle$ and $|1\rangle$. [88]

*EXAMPLE*: As indicated in Fig. 3, there are several single- $qb$ -$o^{per}$ 'ions, each defined by a matrix ($m^{tri}$) that describes the $q$-$o^{per}$ 'ion in the $C^{omp}$ -$b^{as}$ represented by the $e^{ig}$ -$\mathcal{V}$'s of the $\sigma_z$ -$o^{per}$ 'or, i.e. $|0\rangle \equiv [10]^T$ and $|1\rangle \equiv [01]^T$ .

For example, the *identity* ($i^{d}$) $g^{at}$ -$p^{erf}$ 's no $r^{ot}$ on the $s^{ta}$ of the *qb*. This is described by a 2×2- $i^{d}$ -$m^{tri}$. The X- $g^{at}$ -$p^{erf}$ 's a $\pi$ -$r^{ot}$ around the $x$ axis. In the same way, the Y- $g^{at}$ and Z- $g^{at}$ -$p^{erf}$ a $\pi$ -$r^{ot}$ around the $y$ and $z$ axis, respectively.

The S- $g^{at}$ -$p^{erf}$ 's a $\pi/2$ -$r^{ot}$ around the $z$ axis, and the T- $g^{at}$ -$p^{erf}$ 's a $r^{ot}$ of $\pi/4$ around the $z$ axis. The Hadamard ($\mathcal{H}^{mard}$) $g^{at}$ H is also a common single- *qb* -$g^{at}$ that $p^{erf}$ 's a $\pi$ -$r^{ot}$ around an axis diagonal in the *x*-*z* plane, see Fig. 3.

---

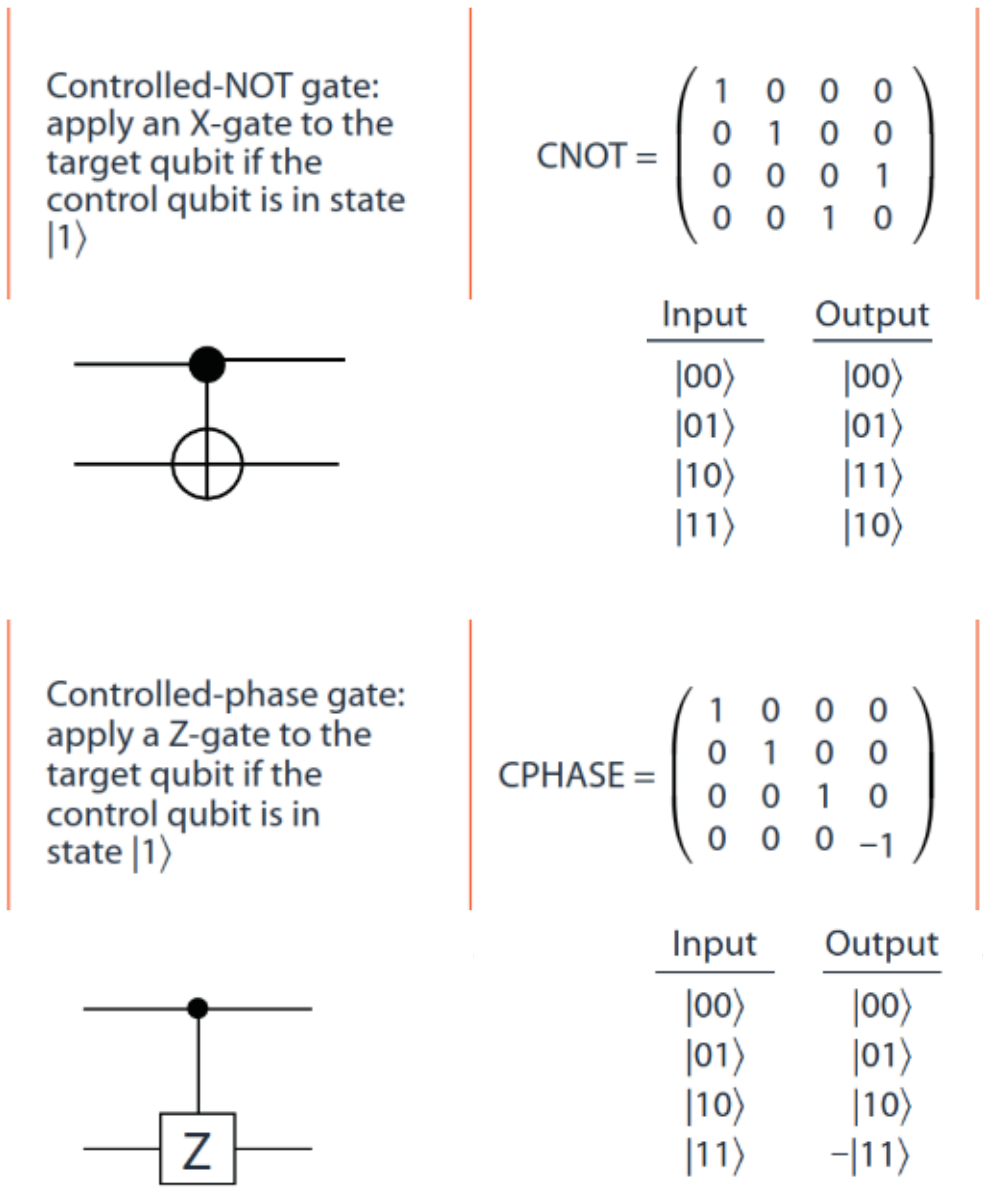

*Fig. 4 Q- 2- qb -$g^{at}$ 's*

Two- *qb* -*q*- *lG* are *G* 'ly *conditional* $g^{at}$ 's using two *qb* 's as inputs ($i^{np}$'s). Usually, the first *qb* is the $C^{trol}$ -*qb*, and the second is the *target* ($t^{rgt}$) -*qb*. A U- $o^{per}$ 'or is applied ($a^{pp}$) to the $t^{rgt}$ -*qb*, conditioned on the $s^{ta}$ of the $C^{trol}$ -*qb*.

---

*EXAMPLE:* The two examples shown in Fig. 4 are the $C^{trol}$ 'ed NOT ($\bar{C}$- $g^{at}$) and $C^{trol}$ 'ed $p^{has}$ (CZ or CPHASE $g^{at}$). The $\bar{C}$- $g^{at}$ flips the $s^{ta}$ of the $t^{rgt}$ -*qb* if the $C^{trol}$ *qb* is in $s^{ta}$ $|1\rangle$. The CPHASE- $g^{at}$ -$a^{pp}$ a Z $g^{at}$ to the $t^{rgt}$ -*qb*, if the $C^{trol}$ -*qb* is in $s^{ta}$ $|1\rangle$. As we will show later, the $is^{wa}$ -$g^{at}$ — another two- *qb*- $g^{at}$ -can be obtained from the $\bar{C}$- $g^{at}$ and single- *qb* -$g^{at}$ 's.

The U - $o^{per}$ 'or of the $\bar{C}$ -$g^{at}$ can be represented in a useful way, indicating that it $a^{pp}$ an X depending on the $s^{ta}$ of the $C^{trol}$ -*qb*.

$$U_{\bar{C}} = \begin{pmatrix} 1 & 0 & 0 & 0 \\ 0 & 1 & 0 & 0 \\ 0 & 0 & 0 & 1 \\ 0 & 0 & 1 & 0 \end{pmatrix} = |0\rangle\langle 0| \otimes \mathbb{I} + |1\rangle\langle 1| \otimes \mathrm{X}$$

and similarly, for the CPHASE $g^{at}$,

$$U_{\mathrm{CPHASE}} = \begin{pmatrix} 1 & 0 & 0 & 0 \\ 0 & 1 & 0 & 0 \\ 0 & 0 & 1 & 0 \\ 0 & 0 & 0 & -1 \end{pmatrix} = |0\rangle\langle 0| \otimes \mathbb{I} + |1\rangle\langle 1| \otimes \mathrm{Z}$$

---

Comparing the two U's, one can see that the two $g^{at}$ 's are related, a $\bar{C}$ can be $g^{ner}$ 'd from a CPHASE by $a^{pp}$ two $\mathcal{H}^{mard}$ -$g^{at}$ 's,

$$U_{\bar{C}} = (\mathbb{I} \otimes \mathrm{H}) U_{\mathrm{CPHASE}} (\mathbb{I} \otimes \mathrm{H})\ ,$$

since HZH = X. The CPHASE $g^{at}$ can be also denoted the CZ $g^{at}$, since it $a^{pp}$ a $C^{trol}$ -ed Z -$o^{per}$ 'or, like $\bar{C}$ (a $C^{trol}$ -ed $a^{pp}$ of X -$o^{per}$ 'or). From the picture of CPHASE in Fig. 4 one cannot say which *qb* is the $t^{rgt}$ and which is the $C^{trol}$, so, the $c^{irc}$ can be drawn in a symmetric way

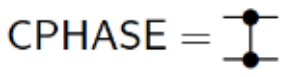

The $\bar{C}$ in $t^{rm}$ 's of CPHASE can then be implemented as

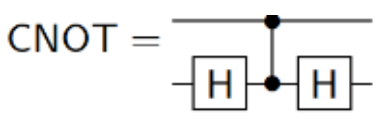

The $g^{at}$ 's -$\bar{C}$ and CPHASE are also called *$\mathcal{E}$'ing -$g^{at}$ 's*, since they can take product ($p^{duc}$) $s^{ta}$ as $i^{np}$ 's and output ($o^{ut}$) $\mathcal{E}$ - $s^{ta}$. So, they are important components of a *u*- $g^{at}$ set for *q*- $l^{gic}$. As an illustration, let us look at two *qb* 's $A$ and $B$ in the $s^{ta}$:

$$|\psi\rangle = \frac{1}{\sqrt{2}}(|0\rangle + |1\rangle)_A |0\rangle_B .$$

After $a^{pp}$ of a $\bar{C}$- $g^{at}$, $U_{\bar{C}}$, on this $s^{ta}$, with *qb* A the $C^{trol}$ -*qb*, and *qb* B the $t^{rgt}$ -*qb*, we get (see Fig. 4):

$$U_{\bar{C}}|\psi\rangle = \frac{1}{\sqrt{2}}(|0\rangle_A|0\rangle_B + |1\rangle_A|1\rangle_B) \neq (\,.\,.\,)_A\,(\,.\,.\,)_B,$$

which cannot be represented as a product of isolated *qb* -A and a *qb* -B. This is one of the two- *qb* -$\mathcal{E}$ *Bell* ($B^{ell}$) $s^{ta}$.

A *u*- set of 1- *qb* and 2- *qb* -$g^{at}$ 's is sufficient ($s^{uff}$) to $\mathcal{J}^{mpl}$ arbitrary *q*- $l^{gic}$. In other words, this $g^{at}$ set can generate *any* $s^{ta}$ in the $m^{lti}$ - *qb*- $s^{ta}$ -space. The $e^{ffi}$ of construction is related to the choice of *q*- $g^{at}$ 's that comprise the $g^{at}$ set. Recall that the 1- *qb* and 2- *qb*- $g^{at}$ 's are $r^{vers}$ *'ible*, that is, given the $o^{ut}$-$s^{ta}$, one knows the $i^{np}$ -$s^{ta}$. This distinction between *c*- and *q*- $g^{at}$ 's is due to the fact that *q*- $g^{at}$ 's are based on U -$o^{per}$ 'ions $U$. If a U -$o^{per}$ 'ion $U$ is a particular $g^{at}$ -$a^{pp}$ 'ed to a *qb*, then its hermitian conjugate $U^{\dagger}$can be $a^{pp}$ 'ed to recover the original $s^{ta}$, since $U^{\dagger}U = I$ resolves an $i^{d}$ - $o^{per}$ 'ion.

The $g^{at}$ -sequences ($s^{equ}$'s) used to build *q*- $a^{lg}$ 's resembles *c*- $C^{omp}$, with a few significant differences. If we, compare the *c*-NOT $g^{at}$, and the related *q*- $c^{irc}$ version we can see that while the c- bit-flip $g^{at}$ inverts any $i^{np}$ -$s^{ta}$, the *q*- bit-flip does not in *G* produce the antipodal $s^{ta}$ (on the Bloch sphere), but rather exchange the pre-factors of the $w^{ve}$ - *F* written in the $C^{omp}$ -$b^{as}$.

The X -$o^{per}$ ‘or is sometimes referred to as ′the *q*- NOT’ (or ‘*q*-bit-flip‘), although X only acts similar to the *c*- NOT $g^{at}$ in the case of *c*- $d^{ta}$ stored ($s^{tr}$’*ed*) in the *q*- bit, i.e. $X|g\rangle = |\overline{g}\rangle$ for $g \in \{0,1\}$.

Here we recall that, *all q- $g^{at}$* ‘s are *reversible ($r^{vers}$)*, due to the U nature of the $o^{per}$ ‘ors $\mathcal{J}^{mpl}$ ‘ing the $l^{gic}$ ‘al $o^{per}$ ‘ions. Certain other $p^{ro}$ ‘es used in *q*- $\mathcal{J}$- $p^{ro}$ ‘ing, however, are ir- $r^{vers}$ like measurements ($\mathcal{M}$) and $e^{nrgy}$ -$\mathcal{L}^{oss}$ to the environment ($e^{vir}$) (if the resulting $s^{ta}$ of the $e^{vir}$ is unknown). For modeling such $p^{ro}$ ‘s see [71] . Here we will only consider U- $C^{trol}$ -$o^{per}$ ‘ions. We also note that *q*- $c^{irc}$ ‘s are aranged left-to-right (in $o^{rde}$ of $a^{pp}$), while the calculation of the result of a $g^{at}$ - $s^{equ}$ ‘s, e.g the $c^{irc}$

$$|\psi_{\text{in}}\rangle \;\rangle\; -\boxed{U_0}-\boxed{U_1}- \cdots -\boxed{U_n}- \;|\,|\psi_{\text{out}}\rangle$$

is $p^{erf}$ ‘ed right-to-left, i.e.

$$|\psi_{\text{out}}\rangle = U_n \cdots U_1 U_0 |\psi_{\text{in}}\rangle.$$

Earlier we saw that both, the NOR and NAND $g^{at}$ ‘s are *u*- $g^{at}$ ‘s for *c*- $C^{omp}$. Since they have no direct ($d^{ir}$) *q*- analogue (because they are not $r^{vers}$), the question is which $g^{at}$ ‘s *are* needed to build a *u*- *q*- $C^{omp}$. It can be shown that the $r^{ot}$ around arbitrary axes on the Bloch-sphere (i.e. a complete 1- *qb*- $g^{at}$ set), together with any $\mathcal{E}$-ing 2- *qb* -$o^{per}$ ‘ion suffices for *u*’ity [70-72]. By using the (Krauss-Cirac $d^{comp}$, any 2- *qb*- $g^{at}$ can be $d^{comp}$ ‘ed into a series of $\bar{C}$ -$o^{per}$ ′tions [71 − 73].

*Gate $l^{ibr}$’s and $g^{at}$ synthesis ($s^{ynth}$):* A common *u*- *q*- $g^{at}$ set, that we refer to as a $l^{ibr}$, is

$$\mathcal{G}_0 = \{X_\theta, Y_\theta, Z_\theta, \text{Ph}_\theta, \bar{C}\}$$

where $\text{Ph}_\theta = e^{i\theta}\mathbb{I}$ app’s a $p^{has}$ $-\theta$ to a single *qb*. Here we mention another *u*- $g^{at}$ -$l^{ibr}$ (set),

$$\mathcal{G}_1 = \{\text{H, S, T}, \bar{C}\},$$

Any other 1- *qb*- $g^{at}$ can be $a^{prox}$ ‘ed to an error $\varepsilon$ using only $\mathcal{O}(\log^c(1/\varepsilon))$ (where $c > 0$) single- *qb*- $g^{at}$ ‘s from $\mathcal{G}_1$ [74,75].The $g^{at}$ -set $\mathcal{G}_1$ is commonly called the ‘Clifford +*T*’ set, where H, S and $\bar{C}$ are all Clifford $g^{at}$ ‘s.

For capacitive coupling of micro- $w^{ve}$ ‘s to a $s^{con}$- $c^{irc}$ used to drive single- *qb* -$g^{at}$ ‘s see [76-80].
The 2- *qb*- $g^{at}$ ‘s in the transmon like $s^{con}$- *qb* architecture can be split into two families: one $\mathcal{G}r$ using local $m^{gn}$ -$f^{iel}$ ‘s to tune the transition $f^{req}$ of *qb* -s and one $\mathcal{G}r$ consisting of all-micro- $w^{ve}$ $C^{trol}$. Hybrid $s^{ch}$ ‘s are also used in modern $s^{con}$- *qb* -$p^{ro}$ ‘ors [52,56,64,81-89].

*B. Qubit $g^{at}$ ‘s using the $s^{pin}$ -$s^{ta}$ of $c^{pld}$ - 1- $e^{lec}$ -q- $d^{t}$ ‘s*

*Q- Dots:* In this concept, *q*- $d^{t}$ ‘s are important so we make some *G* comments about these $s^{yst}$ ‘s here. A basic $s^{ch}$ is shown in Fig. 5.

---

*EXAMPE*: In Fig.5 (a) the *qb*’s are $e^{cod}$ ‘ed in the 1- $e^{lec}$ -$s^{pin}$ -$s^{ta}$ of *q*- $d^{t}$ ‘s (QDs), (presented by doted circles) with the barrier ($b^{arr}$) between QDs is $C^{trol}$ ‘lable via $e^{ctr}$ -$g^{at}$ ‘s. When the $b^{arr}$ is lowered, the $e^{lec}$ -$w^{ve}$ -*F* ‘s overlap and the $s^{pin}$ ‘s $S_L$ and $S_R$ interact in accordance with the Heisenberg ($\mathcal{H}^{eis}$) exchange coupling $J(t)S_L{\cdot}S_R$, where $J(t)$ is a *F* of $g^{at}$ voltage and $c^{resp}$ ‘s to the $e^{nrgy}$ splitting of the $s^{pin}$ singlet and triplet $s^{ta}$.

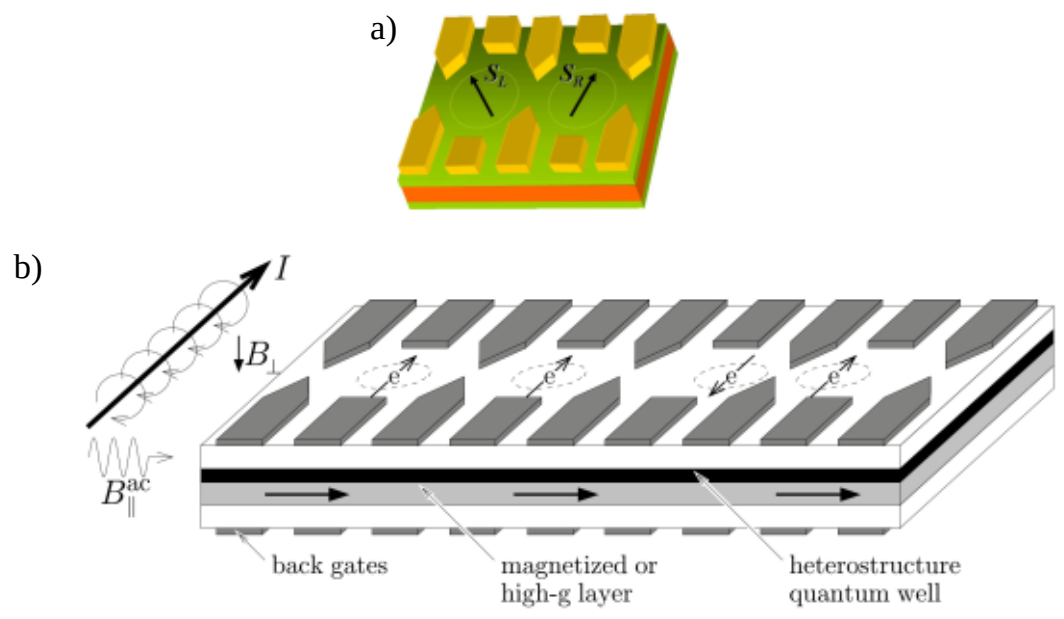

*Fig. 5 Basic $s^{ch}$ for the $p^{hys}$ -$\mathcal{J}^{mpl}$ of an all- $e^{ctr}$ ‘ly $C^{trol}$ ‘ed q- $d^{t}$ array [147].*

This enables $e^{ctr}$ ‘ly $C^{trol}$ ‘ed two- *qb*- $g^{at}$ ‘s with fast $o^{per}$ ‘ion times. In Fig 5(b) a QD array is based on panel a); here only the *qb* -s in the two right $d^{t}$ ‘s are $c^{pld}$ and $r^{ot}$ ‘s of $s^{pin}$ ‘s is obtained by bringing the desired $e^{lec}$ into a region of high magnetization or high *g* factor via back $g^{at}$ ‘s, such that the Zeeman splitting and, hence, the resonance condition changes for this $e^{lec}$ in the presence of a static $m^{gn}$ field $B_\perp$.

A resonantly $a^{pp}$’ed oscillating $B_\perp$ -$p^{ls}$ -$B_\parallel^{ac}$ then $r^{ot}$ the specific *qb* ($e^{lec}$ -$s^{pin}$ resonance, ESR), while all others are unchanged. Leveraging $s^{pin}$ -orbit $i^{ntac}$, the $r^{ot}$ ‘s may also be controlled fully $e^{ctr}$ ‘ly via $e^{ctr}$ -dipole-induced $s^{pin}$ resonance (EDSR). Alternatively, fast single- *qb*- $g^{at}$ ‘s may be $\mathcal{J}^{mpl}$ ‘ed via exchange- $C^{trol}$ ‘ed $s^{pin}$ $r^{ot}$ ‘s [90]. The combination of 1- and 2- *qb*- $g^{at}$ ‘s results in a *u*- set of *q*- $g^{at}$ ‘s, so that the $s^{ch}$ ‘s enable fast and purely ($p^{ur}$’*ly*)-$e^{ctr}$‘ly -$C^{trol}$ ‘ed *q*- $C^{omp}$ with $e^{lec}$ -$s^{pin}$ ‘s in QDs [90].

---

2- *qb*- *q*- $g^{at}$ -$o^{per}$ ‘es by a $p^{ur}$ ‘ly -$e^{ctr}$ gating of the tunneling $b^{arr}$ between adjacent *q*- $d^{t}$ ‘s, instead of spectroscopic choices as in the $m^{del}$ ‘s from the previous section. $C^{trol}$ ‘ed gating of the tunneling $b^{arr}$ between adjacent 1- $e^{lec}$ -*q*- $d^{t}$ ‘s in patterned $\mathcal{D}^2$ -$e^{lec}$ -gas structures have been implemented in practice using a split- $g^{at}$ -$t^{chn}$ [91]. If the $b^{arr}$ -$p^{ten}$ is “high”, tunneling is disabled between $d^{t}$ ‘s, and the *qb*- $s^{ta}$ does not $e^{vlt}$ in time (*t*). If the $b^{arr}$ is $p^{ls}$ ‘ed “low”, the $p^{hys}$ of the Hubbard $m^{del}$ [92] says that the $s^{pin}$ ‘s will be subject to a transient $\mathcal{H}^{eis}$ coupling,

$$H_s(t) = J(t)\vec{S}_1 \cdot \vec{S}_2 \quad (18)$$

Here $J(t) = 4t_0^2(t)/u$ is exchange time dependent constant, resulted from switching on and off the tunneling $m^{tri}$ element $t_0(t)$ . We can also construct a super-exchange $s^{ch}$ to get a $\mathcal{H}^{eis}$ - $i^{ntac}$ by using three aligned $q$- $d^t$ ‘s with the middle one having a higher $e^{nrgy}$ level (by the amount $\varepsilon$) such that the $e^{lec}$ -$s^{pin}$ ‘s of the outer two $d^t$ ‘s are also $\mathcal{H}^{eis}$ -$c^{pld}$, with the exchange coupling $J = 4t_0^4(1/\varepsilon^2 u + 1/2\varepsilon^3)$. Here $u$ is the charging $e^{nrgy}$ of a single $d^t$, and $\vec{S}_i$ is the $s^{pin}$ -1/2 $o^{per}$ ‘or [9] for $d^t$- $i$.

Eq. (18) describes the $q$- $d^t$- $s^{yst}$ well if: 1) Higher-lying 1-particle $s^{ta}$ of the $d^t$ ‘s can be neglected, which is met if $\Delta E \gg kT$, with $\Delta E$ being the level spacing and $T$ is the temperature (see Section II A). 2) The intervals $\tau_s$ for pulsing the $g^{at}$ -$p^{ten}$ “low” should be longer than $\hbar/\Delta E$, in $o^{rde}$ to avoid moving to higher orbital levels. 3) $u > t_0(t)$ for all $t$; in order for the $\mathcal{H}^{eis}$ -exchange $a^{prox}$ to be valid. 4) The decoherence time $\Gamma^{-1}$ should be much longer than the switching time $\tau_s$. *Parameter $\Gamma^{-1}$, is included in the objective function for the network level performance optimization discussed in Section I.*

For this reason here we further analyze the effect of a decohering $e^{vir}$. The $s^{pin}$ -1/2 -$d^{gre}$ ‘s of freedom in $q$- $d^t$*’s* are expected to have longer $\Gamma^{-1}$ than $c^{hrg}$- $d^{gre}$ ‘s of freedom due to insensitivity to any $e^{vir}$ ‘al variations of the $e^{ctr}$ -$p^{ten}$. The $c^{hrg}$ transport in such $c^{pld}$ -$q$- $d^t$ ‘s has $r^{ceiv}$ ‘ed much attention [91,93] as well as work on their non-equilibrium $s^{pin}$ -$d^{yna}$ [90]. The effect of $m^{gn}$ coupling to the $e^{vir}$ should be carefully considered.

Long $\Gamma^{-1}$ enables the ideal $q$- $C^{omp}$, wherein the effect of the $\mathcal{H}$ is to $a^{pp}$ an U time $e^{vlt}$ -$o^{per}$ $U_s(t) = T\exp\left\{-i\int_0^t H_s(t')dt'\right\}$ to the initial ($i^{nit}$) $s^{ta}$ of the two $s^{pin}$ ‘s: $|\Psi(t)\rangle = U_s|\Psi(0)\rangle$. The $p^{ls}$ ‘ed $\mathcal{H}^{eis}$ -$c^{pld}$ ‘ing gives a special form for $U_s$: For a given time $\tau_s$ of the $s^{pin}$ - $s^{pin}$ -$c^{pld}$ ‘ing where $\int d\,tJ(t) = J_0\tau_s = \pi(\mathrm{mod}\ 2\pi)$ [94], $U_s(J_0\tau_s = \pi) = U_{sw}$ is the “$s^{wa}$” -$o^{per}$ ‘or: if $|ij\rangle$ labels the $b^{as}$ -$s^{ta}$ of two $s^{pin}$ ‘s in the $S_z$- $b^{as}$ with $i, j = 0,1$, then $U_{sw}|ij\rangle = |ji\rangle$. Here, for simplicity, we assume that the shape of the $a^{pp}$-ed pulse ($p^{ls}$) is approximately ($a^{prox}$) rectangular with $J_0\tau_s$ constant. Since it conserves the total angular $\mathcal{M}^{om}$ of the $s^{yst}$, $U_{sw}$ is not by itself $s^{uff}$ to $p^{erf}$ useful $q$- $C^{omp}$, but if the $i^{ntac}$ is $p^{ls}$ ‘ed on for just half the duration, the resulting “square root of $s^{wa}$” is used as a basic $q$- $g^{at}$: for example, a $q$- XOR- $g^{at}$ is created by a simple $s^{equ}$ of $o^{per}$ ‘ions.

$$U_{XOR} = e^{i\frac{\pi}{2}S_1^z}e^{-i\frac{\pi}{2}S_2^z}U_{sw}^{\frac{1}{2}}e^{i\pi S_1^z}U_{sw}^{\frac{1}{2}}, \quad (19)$$

where $e^{i\pi S_1^z}$ etc. are 1- $qb$ -$o^{per}$ ‘ions only, which can be obtained e.g. by $a^{pp}$ local $m^{gn}$ -$f^{iel}$ ‘s. We note that explicitly $U_{XOR} = \frac{1}{2} + S_1^z + S_2^z - 2S_1^zS_2^z$, with the $c^{resp}$ ‘ing XOR $\mathcal{H}$, $\int_0^t d\,t' H_{XOR} = \pi[1 - 2S_1^z - 2S_2^z + 4S_1^zS_2^z]/4$. Another way to get the XOR-$o^{per}$ ‘ion is given by $U_{XOR} = e^{i\pi S_1^z}U_{sw}^{-\frac{1}{2}}e^{-i\frac{\pi}{2}S_1^z}U_{sw}e^{i\frac{\pi}{2}S_1^z}U_{sw}^{\frac{1}{2}}$. Here the 1-$qb$ -$o^{per}$ ‘ions involve only $s^{pin}$ 1. XOR with 1- $qb$ -$o^{per}$ ‘ions may be combined to do any $q$- $C^{omp}$ [95]. The XOR of Eq. (19) is presented in the $b^{as}$ where it has the form of a conditional $p^{has}$ -shift $o^{per}$ ‘ion; the standard XOR is obtained by a simple $b^{as}$ change for $qb$ 2. For analytical $m^{del}$ ‘s see [90,96-101]

*C. Q- $l^{gic}$ by $p^{lar}$ -$b^{split}$*

Several elementary $q$- $l^{gic}$ constructions, like $q$- parity check and a $q$- $e^{cod}$ ‘er, including their combination to $\mathcal{J}^{mpl}$ a $C^{trol}$ ‘ed-NOT ($\bar{C}$) $g^{at}$ can be built using $p^{lar}$-$b^{split}$ ‘s that completely transmit ($t^{mit}$) one $s^{ta}$ of $p^{lar}$ and totally reflect the orthogonal ($o^{rtg}$) $s^{ta}$ of $p^{lar}$, which allows a simple explanation of each $o^{per}$ ‘ion.

*$\bar{C}$ Using Four- $p^{ton}$ -$\mathcal{E}$- $s^{ta}$:* A $\bar{C}$ -$o^{per}$ ‘ion could be $p^{erf}$ ‘ed using a modified form of $q$- teleportation [102]. Although the required $B^{ell}$ - $s^{ta}$ -$\mathcal{M}$ are inherently non’ $l^{ine}$ [103-105], a partial or incomplete set of $B^{ell}$ -$s^{ta}$ -$\mathcal{M}$ can be $\mathcal{J}^{mpl}$ ‘ed using $l^{ine}$ -$o^{pti}$ ‘al elements and post- selection [106-108].

This combination was supposed to enable the $\mathcal{J}^{mpl}$ of a probabilistic $\bar{C}$, motivating the work in [109-111]. Here we discuss this approach using $p^{lar}$ -$b^{split}$ ‘s and a $q$- erasure $t^{chn}$ [111]. For this an ancilla ($\mathcal{A}$) is needed in the form of a four- $p^{ton}$ -$\mathcal{E}$ - $s^{ta}$ $|\chi\rangle$ [102,109,110]. The $\mathcal{J}^{mpl}$ given here is similar in spirit to those of [109,110], but the approach presented in [111] is more practical.

---

*EXAMPLE:*

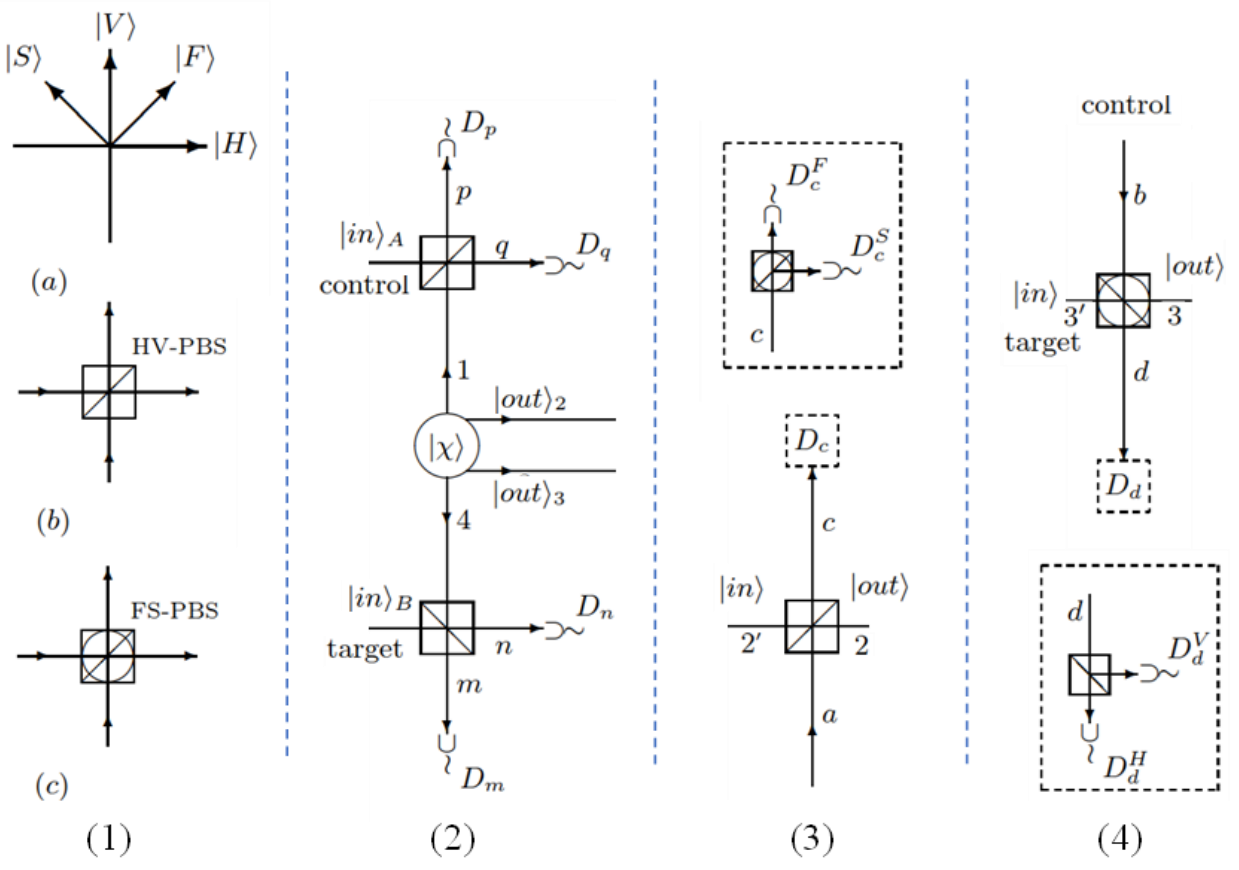

*__ Fig. 6 (1): (a) HV and FS $p^{lar}$ -$b^{as}$. (b) and (c) show the symbols ($s^{ymb}$’s) used to represent $p^{lar}$ -$b^{split}$ ‘s in the HV and FS $b^{as}$, respectively. (2) This $s^{ch}$ relies on a four- $p^{ton}$ -$\mathcal{E}$’ed $s^{ta}$, $|\chi\rangle$, to $p^{erf}$ a probabilistic $\bar{C}$ -$o^{per}$ ‘ion on the two $i^{np}$ - $p^{ton}$ ‘ic qb ‘s in $m^{od}$ ‘s A and B. Four $p^{lar}$ -sensitive $d^{tect}$ ‘s in the FS $b^{as}$ are labelled $D_m$, $D_n$, $D_p$, and $D_q$. (3) $\mathcal{J}^{mpl}$ of a probabilistic q- parity check of the qb ‘s in $m^{od}$ -2′ and $m^{od}$ -a using a $p^{lar}$ -$b^{split}$ in the HV $b^{as}$ and a $p^{lar}$ -sensitive $d^{tect}$ -$D_c$ in the FS $b^{as}$. The dashed-box insert shows the details of $D_c$, which consists of a $p^{lar}$ -$b^{split}$ in the FS $b^{as}$ followed by two ordinary single- $p^{ton}$ $d^{tect}$ ‘s. (4) $\mathcal{J}^{mpl}$ of a destructive-$\bar{C}$ that $p^{erf}$ ‘s a $s^{ta}$ -flip on the $p^{ton}$ in $m^{od}$ -3′ that is $C^{trol}$ ‘ed by the $p^{lar}$ of a single $p^{ton}$ in $m^{od}$ -b. Its successful $o^{per}$ ‘ion requires the destruction of the $C^{trol}$ -$p^{ton}$. The dashed box inset shows the details of the $p^{lar}$ sensitive $d^{tect}$- $D_d$. [37]*

*The $p^{lar}$ terminology used in the section is illustrated in Fig. 6 (1). The $p^{ton}$ 'ic $qb$ 's $|0\rangle$ and $|1\rangle$ will be represented by horizontal $|H\rangle$ and vertical $|V\rangle$ $p^{lar}$ 's, respectively, but $\mathcal{M}$ will also be carried on in the $|F\rangle$ and $|S\rangle$ $b^{as}$ shown in the figure Fig. 6(1c). $p^{lar}$ -$b^{split}$ 's (PBS) oriented in the HV $b^{as}$ will always $t^{mit}$- H- $p^{lar}$ -$p^{ton}$ 's and reflect V- $p^{lar}$ -$p^{ton}$ 's, while $p^{lar}$ -$b^{split}$ 's oriented in the FS $b^{as}$ will $t^{mit}$ F -$p^{lar}$ -$p^{ton}$ 's and reflect S- $p^{lar}$ -$p^{ton}$ 's. Here the $b^{split}$ 's and detectors ($d^{tect}$'s) are assumed to be ideal and all the $p^{ton}$ 's in a given $o^{pti}$ 'al path ($\mathcal{P}^{th}$) to be in the same spatial mode ($m^{od}$) [112-115]. Fig. 6(2) shows an $\mathcal{I}^{mpl}$ of a $\bar{C}$- $g^{at}$ using the Gas Chromatography (GC) $\mathcal{P}^{rol}$ [102] with $p^{lar}$ -$e^{cod}$ 'ed -$p^{ton}$ 'ic -$qb$ 's and partial $B^{ell}$ - $s^{ta}$ -$\mathcal{M}$. A source emits the 4- $p^{ton}$ -$\mathcal{E}$- $s^{ta}$ $|\chi\rangle$ where the $s^{ta}$ $|\chi\rangle$ is created in such a way that one $p^{ton}$ is emitted into each of the four $m^{od}$ 's labelled 1 through 4.*

---

The $i^{np}$ - $p^{ton}$ in $m^{od}$ -A is mixed ($m^{xin}$*'ed*) with the $p^{ton}$ in $m^{od}$ -1 by a $p^{lar}$ -$b^{split}$ in the HV $b^{as}$ with ideal 1- $p^{ton}$ -$d^{tect}$ 's, $D_p$ and $D_q$, in its $o^{ut}$-ports. Similarly, The $p^{ton}$ 's in $m^{od}$ 's B and 4 are $m^{xin}$*'ed* in another $p^{lar}$ -$b^{split}$. The other two $p^{ton}$ 's of $|\chi\rangle$ in $m^{od}$'s -2 and 3 will serve as the $o^{ut}$- $qb$ 's.

In a $\bar{C}$- $g^{at}$ the $t^{rgt}$ -$qb$ is to be $r^{vers}$ 'ed ($0 \leftrightarrow 1$) if the $C^{trol}$ -$qb$ has the value 1. If the $i^{np}$- $p^{ton}$ 's in $m^{od}$ 's A and B are the "$C^{trol}$" and "$t^{rgt}$" $p^{ton}$ 's, respectively, then the desired $\bar{C}$- $g^{at}$ -$o^{per}$ 'ion $c^{resp}$ 's to the following $s^{ta}$ -$t^{rans}$:

$$\alpha_1|H_A\rangle|H_B\rangle + \alpha_2|H_A\rangle|V_B\rangle + \alpha_3|V_A\rangle|H_B\rangle + \alpha_4|V_A\rangle|V_B\rangle$$

$$\rightarrow \alpha_1|H_2\rangle|H_3\rangle + \alpha_2|H_2\rangle|V_3\rangle + \alpha_3|V_2\rangle|V_3\rangle + \alpha_4|V_2\rangle|H_3\rangle \quad (20)$$

where the $\alpha_i$ are arbitrary coefficients $\left(\sum_{i=1}^{4} |\alpha_i|^2 = 1\right)$, and, for instance, $|H_A\rangle$ denotes a single horizontally $p^{lar}$ -$p^{ton}$ in $m^{od}$ -A. For simplicity in the sequel, the kets will be dropped from the notation.

In Fig. 6 (2), the partial $B^{ell}$ -$\mathcal{M}$ at the upper and lower $b^{split}$ 's consist of only accepting the $o^{ut}$'s in $m^{od}$ 's -2 and 3 iff one $p^{ton}$ is $d^{tect}$ 'ed in each of the four $d^{tect}$ 's. This condition, together with the $p^{lar}$ -dependent reflections and transmissions ($t^{miss}$*'s*) at the $b^{split}$ 's, enables the needed properties ($p^{pert}$) of the $s^{ta}$ $\chi$ to be read off from the $\bar{C}$- $s^{ta}$ -$t^{rans}$ in *eq* (20).

For instance, for the $i^{np}$- $a^{mpl}$ -$\alpha_1 H_A H_B$, the partial $B^{ell}$ -$\mathcal{M}$ will be correct if each of the $p^{ton}$ 's in $m^{od}$ 's -1 and 4 are also H- $p^{lar}$ (if one of them were V- $p^{lar}$ would result in two $p^{ton}$ 's at one $d^{tect}$ and zero at another). The $\bar{C}$ -$t^{rans}$ for this $i^{np}$- $a^{mpl}$ requires the $o^{ut}$- $p^{ton}$ 's in $m^{od}$ 's -2 and 3 to be H $p^{lar}$, so that the $s^{ta}$ $\chi$ must contain an $a^{mpl}$ of the form $H_1H_4H_2H_3$.

The other three $a^{mpl}$ 's can be read off in a similarly to reveal the required form of $\chi$:

$$\chi = \frac{1}{2}(H_1H_4H_2H_3 + H_1V_4H_2V_3 + V_1H_4V_2V_3 + V_1V_4V_2H_3) \quad (21)$$

From this one can see that the needed $p^{lar}$ will be $g^{ner}$ 'd in the $o^{ut}$- $m^{od}$ 's iff each of the $d^{tect}$ registers ($r^{eg}$*'s*) contains only one $p^{ton}$.

Although this defines the needed form of the $s^{ta}$ -$\chi$, the way in which the $o^{ut}$- $p^{ton}$ 's are $\mathcal{E}$'ed with the $p^{lar}$ of the $p^{ton}$ 's in the $\mathcal{P}^{th}$ 's $m, n, p$, and $q$ leading to the $d^{tect}$ 's has to be considered. The $\mathcal{E}$ would result in a more complex final $s^{ta}$ than the one given in Eq. (20). $\mathcal{E}$ of this kind would provide $\mathcal{J}$ regarding the $s^{ta}$ of the $i^{np}$- $qb$ 's, which would destroy the coherence of a $q$-$C^{omp}$ -$a^{lg}$. This case is discussed in [111], suggesting that this kind of $\mathcal{J}$ can be "erased" if all $d^{tect}$ 's -$\mathcal{M}$ the $p^{lar}$ of the $p^{ton}$ 's in the FS $b^{as}$ since an F- $p^{lar}$ -$p^{ton}$ is an equal $S^{po}$ of H and V- $p^{lar}$, for instance, so that such a $\mathcal{M}$ provides no $\mathcal{J}$ regarding the original values of the $qb$ 's. So here, the use of a $q$- erasure $t^{chn}$ is equivalent to a $|\varphi^+\rangle$ $B^{ell}$ - $s^{ta}$ -$\mathcal{M}$ [106].

Ref. [111] shows that we will get the specified $o^{ut}$ from the $d^{tect}$ 's 25% of the time, so that this $\bar{C}$- $g^{at}$ succeeds with a probability ($p^{rob}$) of 25%, given that the $s^{ta}$ -$\chi$ has been created with certainty. Details of designing of Q- Parity Check and destructive- $\bar{C}$- $g^{at}$ 's, shown in Figs 6(3) and 6(4) respectively, are presented in [111].

### *D. Q- Gates $\mathcal{I}^{mpl}$ 'ed by Trapped Ions*

*Ion Trapping:* Due to the $c^{hrg}$ of atomic ions, they can be confined by specific forms of electro- $m^{gn}$ ($e^{lm}$)-$f^{iel}$ 's. *Penning trap* [116] uses an arrangement of static $e^{ctr}$ and $m^{gn}$ -$f^{iel}$ 's and *Paul trap* [117-119] controls ions by oscillating $e^{ctr}$ -$f^{iel}$ 's. The $o^{per}$ 'ion of a number of ion traps is presented in [120].

---

*EXAMPLE:* Here we discuss only the $l^{ine}$ *Paul trap* (Fig.7) and refer the reader to [119-121] for the analytical modeling.

For typical $o^{per}$ 'ing $p^{met}$ 's used in $e^{xp}$ 's see [122-128]. It has been shown in the previous sections, that any U $o^{per}$ 'ion can be composed ($c^{omp}$) of single- $qb$ -$r^{ot}$ 's and two- $qb$ $C^{trol}$ 'ed-NOT $g^{at}$ 's [129]. For discussion on how these, and some more $c^{plex}$ -$q$- $g^{at}$ 's can be $\mathcal{I}^{mpl}$ 'ed on cold trapped ions the reader is referred to [130-133]. For speed of $q$- $g^{at}$ 's see [134-136].

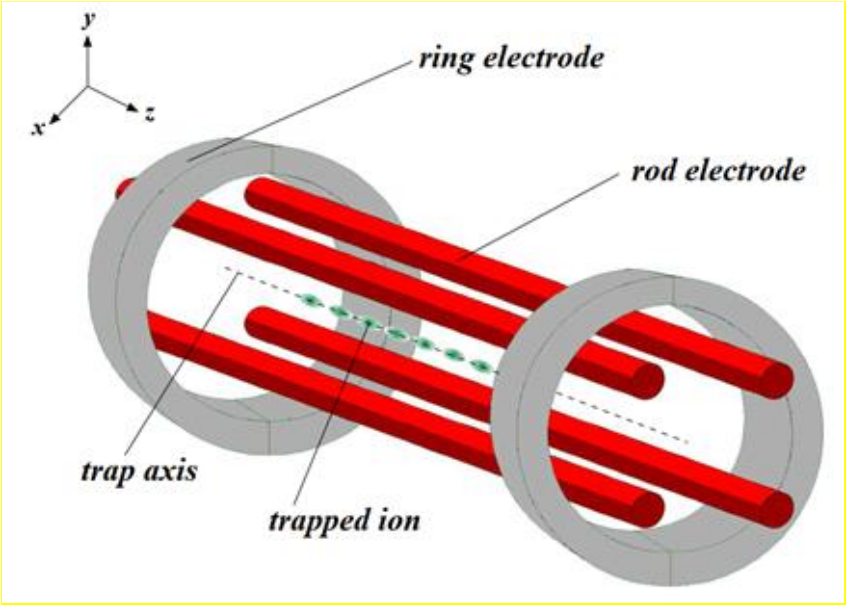

*Fig. 7 $l^{ine}$ Paul trap using two ring electrodes spaced by $2z_0$. [119].*

---

In Table 1 we summarize recommended reading on *qb* -$p^{hys}$.

| focus | references |
|---|---|
| superconducting qubits | [1-4,10-15, 125, 40-42,88, 70-89, 52,56,64] |
| electron spins in silicon | [16-21,] |
| quantum dots | [22-25, 37,90-111] |
| Trapped ions | [26-29,116-121] |
| ultracold atoms | [30-33] |
| polarized photons | [36-39] |

*Table1: Recommended reading on qb -$p^{hys}$*

## IV Q- $C^{omp}$ GATES LIBRARIES

The $c^{irc}$ -$m^{del}$ of a $C^{omp}$ is a representation of the $C^{omp}$ -$p^{ro}$ used in the design of $C^{omp}$ -$h^{rdw}$. In the $c^{irc}$ -$m^{del}$, any $C^{omp}$ represents the action ($a^{ct}$) of a $c^{irc}$ made of few discinct types of Boolean ($B^{ool}$) *lG* acting on an $i^{np}$ bit string. Each *lG* -$t^{rans}$ 's its $i^{np}$- bits into one or several $o^{ut}$- bits according to the function of the $g^{at}$. By $c^{omb}$ 'ing the $g^{at}$ 's in a graph ($g^{ph}$) such that the $o^{ut}$'s from preceding $g^{at}$'s feed into the $i^{np}$'s of succeeding $g^{at}$'s, any feasible $C^{omp}$ can be $p^{erf}$ 'ed. Here we will talk about the types of *lG* used within $c^{irc}$ 's and how the concept of *lG* is different in the *q*- context.

### *A. Q- $g^{at}$ 's- $l^{ibr}$*

*A.1. Classical lG- $l^{ibr}$:* The sets $o^{per}$ 'ions intersection, union, and complement can be defined in $t^{rm}$ 's of the $l^{gic}$ 'al connectives AND (∧), OR (∨), and NOT (¬), indicating close parallel between set $o^{per}$ 'ions and $l^{gic}$ 'al $o^{per}$ 'ions.

*Ir- $r^{vers}$ -$g^{at}$ 's AND and OR* : The $l^{gic}$ 'al connections AND (∧) and OR (∨) corresponds, to the notions of $l^{gic}$ 'al conjunction and disjunction respectively. The $a^{ct}$ of a *lG* is defined in $t^{rm}$ 's of its "truth table" showing all the possible $l^{gic}$ 'al values of the $i^{np}$'s together with their $c^{resp}$ 'ing $o^{ut}$'s. The AND $g^{at}$ is $l^{gic}$ 'ally *ir- $r^{vers}$*, since you cannot know unique $i^{np}$'s for all $o^{ut}$'s. For instance, if the $o^{ut}$ is 0, you cannot tell whether the $i^{np}$- values were 00, 01, or 10. It "erases" some $\mathcal{J}$ whenever its $o^{ut}$ is 0.

There is a variant of the OR $g^{at}$, referred to as exclusive-OR ( "XOR" or "⊕") which is like the OR $g^{at}$ except that it returns 0 when both its $i^{np}$'s are 1.

*Universal $g^{at}$ 's: NAND and NOR*: The *u*- $g^{at}$ 's, can express any desired $C^{omp}$. The existance of such *u*- $g^{at}$ 's makes possible optimization of modern $C^{omp}$ since $C^{omp}$ designers need only focus on optimizing a single type of $g^{at}$.

Any $B^{ool}$ -*F* can be expressed by using only ¬ and ∧ or only ¬ and ∨. So, any $B^{ool}$ -*F* can be $C^{omp}$ by using a $c^{irc}$ comprising NOT and AND $g^{at}$ 's, or NOT and OR $g^{at}$ 's. The design of large-scale $l^{gic}$ -$c^{irc}$ 's would be simpler if designers only had to use a single type of $g^{at}$. Such a $g^{at}$ is said to be "*u*-" since it can be used to derive any $B^{ool}$ - *F*. Restricting $c^{irc}$ 's to a single type of *u*- $g^{at}$ does not always lead to the smallest $c^{irc}$ for $C^{omp}$ a desired $B^{ool}$ - *F* but it does enable chip industry to improve the design and production $p^{ro}$ for the *u*- $g^{at}$, which, makes it easier to improve yield, reliability, and speed. Nowdays, the micro- $p^{ro}$ 'or industry favors this strategy by basing their $c^{irc}$ 's on the NAND ("NOT AND") $g^{at}$ 's. Formally, aNANDb ≡ ¬(a ∧ b) , commonly denoted as a|b, and is *u*- for *c*- ir- $r^{vers}$ -$C^{omp}$.

NAND:

| a | b | a\|b |
|---|---|---|
| 0 | 0 | 1 |
| 0 | 1 | 1 |
| 1 | 0 | 1 |
| 1 | 1 | 0 |

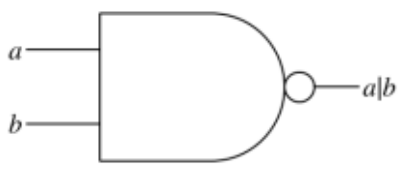

AND in terms of NAND:

| a | b | a\|b | (a\|b)\|(a\|b) | a ∧ b |
|---|---|---|---|---|
| 0 | 0 | 1 | 0 | 0 |
| 0 | 1 | 1 | 0 | 0 |
| 1 | 0 | 1 | 0 | 0 |
| 1 | 1 | 0 | 1 | 1 |

NOT in terms of NAND:

| a | a | a\|a | ¬a |
|---|---|---|---|
| 0 | 0 | 1 | 1 |
| 1 | 1 | 0 | 0 |

*Fig.8 Truth tables and Icons for NAND $g^{at}$ 's*

---

*EXAMPLE*: Operation of the NAND $g^{at}$ and the $c^{resp}$ 'ing $c^{irc}$ icon are shown in Fig.8. Since we can $C^{omp}$ any $B^{ool}$ - *F* in a $c^{irc}$ using only NOT and AND $g^{at}$ 's, to show that the NAND $g^{at}$ is *u*-, it is $s^{uff}$ to demonstrate that we can get NOT from NAND $g^{at}$ 's and AND from NAND $g^{at}$ 's. Two tables of Fig. 8 show how to get $\neg a$ from $a|a$ and $a \wedge b$ from two $a|b$ $g^{at}$'s. Since any logical operation can be written in $t^{rm}$ 's of only ¬ and ∧, and these two operation can, in turn, each be written in $t^{rm}$ 's of (NAND) this shows that any logical proposition can be written only in $t^{rm}$ 's of (NAND) $g^{at}$'s. This means that manufacturers need only perfect the $\mathcal{J}^{mpl}$ of just one type of $g^{at}$, the NAND $g^{at}$, to be sure that they can build a $c^{irc}$ that can $p^{erf}$ any feasible $C^{omp}$.

---

Other *u*- $g^{at}$ 's for *c*- ir- $r^{vers}$ -$C^{omp}$ include only the NOR $g^{at}$ ("NOT OR") and the NMAJORITY $g^{at}$ ("NOT MAJORITY"). The latter is $\mathcal{J}^{mpl}$ 'able in a new transistor design and leads to highly compact $c^{irc}$ 's.

*NOTE: logical ir- $r^{vers}$ 'ty comes at a price of $e^{nrgy}$ -$\mathcal{L}^{oss}$ 'es that occur when $\mathcal{J}$ is erased. Without a solution for this problem, further miniaturization of $C^{omp}$ technology might be jeopardized by the difficulty of removing this unwanted waste heat from deep within the ir -$r^{vers}$ -$c^{irc}$ -ry.*

*Reversible $g^{at}$ 's: NOT, $s^{wa}$, and $\bar{C}$*: One way to reduce the heat generated as a result of using ir -$r^{vers}$ -*lG* is to use only $r^{vers}$ -*lG*. In a $r^{vers}$ -*lG* a unique $i^{np}$- always corresponds to a unique $o^{ut}$- and vice versa. So $r^{vers}$ -$g^{at}$ 's do not erase any $\mathcal{J}$ when they act, so, a $C^{omp}$ based on $r^{vers}$ -$l^{gic}$ can be executed forward to get a result, the result copied, and after that the whole $C^{omp}$ undone to recover all the $e^{nrgy}$ used apart from the small amount used to copy the answer at the mid-way point.

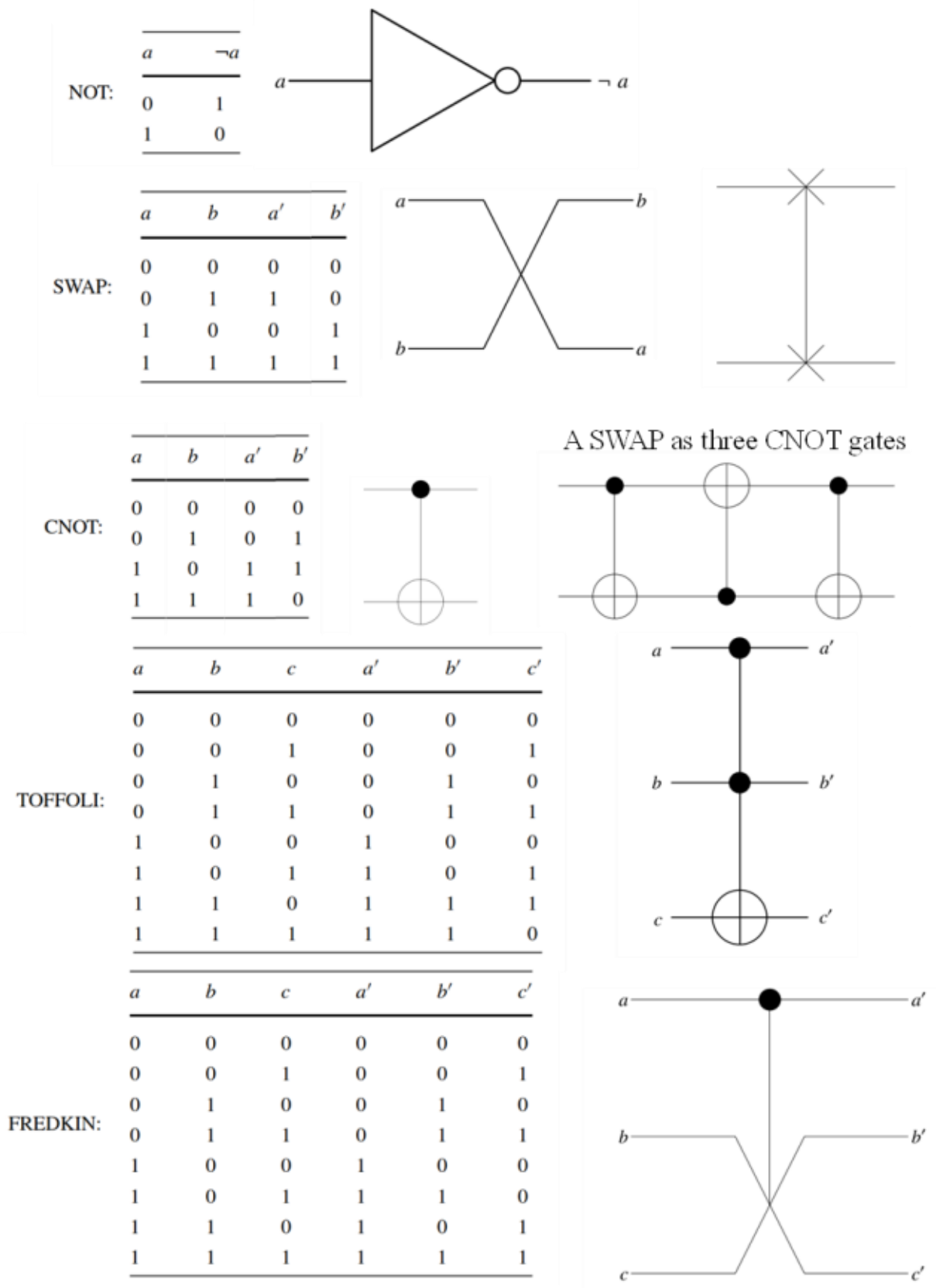

*Fig. 9 Truth tables and Icons for the logically $r^{vers}$ -$g^{at}$ 's*

---

*EXAMPLE*: The simplest $r^{vers}$ -*lG* is the NOT $g^{at}$. NOT is a 1- $i^{np}$/1- $o^{ut}$- $g^{at}$ that flips the $i^{np}$ bit. The truth tables and $c^{irc}$ icons for $r^{vers}$ -$g^{at}$ 's, are shown in Fig. 9. With NOT $g^{at}$, for a known $o^{ut}$- bit value, the $i^{np}$- bit value can be inferred and vice versa. The 2- $i^{np}$/2- $o^{ut}$- $s^{wa}$ -$g^{at}$ is a slightly more sophisticated example. SWAP exchanges the bit values it is handed. In *q- $C^{omp}$* a $c^{irc}$ may not have any $p^{hys}$ wires connecting the $g^{at}$ 's together. Instead, a $c^{irc}$ can be merely a visual specification of a $s^{equ}$ of $g^{at}$ -$o^{per}$ 'ions with time increasing from left to right in the $c^{irc}$ diagram as successive $g^{at}$ 's are $a^{pp}$ 'ed. So, in *q- $C^{omp}$* we sometimes use a different icon for a $s^{wa}$ -$g^{at}$ (showing in Fig. 9, that some $o^{per}$ 'ion , other than crossing wires, is needed to achieve the effect of a $s^{wa}$ -$o^{per}$ 'ion).
An important $r^{vers}$ -$g^{at}$ in *q- $C^{omp}$* is the 2-bit $C^{trol}$ 'ed-NOT ($\bar{C}$) $g^{at}$ which is flipping the bit value of the second bit iff the first bit is set to 1. In other words, the decision to flip or not flip the second bit is $C^{trol}$ 'ed by the value of the first bit. Hence, the name "$\bar{C}$". The $s^{wa}$ -$g^{at}$ can be obtained from a $s^{equ}$ of 3-$\bar{C}$- $g^{at}$ 's.

*Universal $r^{vers}$ -$g^{at}$ 's: FREDKIN ($F^{red}$) and TOFFOLI ($T^{off}$)*: Just as there can be *u-* $g^{at}$ 's for *c-* ir- $r^{vers}$ -$C^{omp}$, like NAND $g^{at}$ (2- $i^{np}$'s and 1- $o^{ut}$), there can be also *u-* $g^{at}$ 's for *c-* $r^{vers}$ -$C^{omp}$. Here, the smallest $g^{at}$ 's that are both $r^{vers}$ *and u-* need *3- $i^{np}$*'s and 3- $o^{ut}$'s. The $F^{red}$ ($C^{trol}$ 'ed $s^{wa}$) $g^{at}$ and the $T^{off}$ ($C^{trol}$ 'ed $\bar{C}$) $g^{at}$, whose truth tables and Icons are shown in Fig. 9, are such examples.

The $T^{off}$ -$g^{at}$ is also called the $C^{trol}$ 'ed- $\bar{C}$ -$g^{at}$ since it flips the third $i^{np}$- bit iff the first two $i^{np}$- bits are both 1.

Another well known $r^{vers}$ -$g^{at}$ is the $F^{red}$ ($C^{trol}$ 'ed $s^{wa}$) $g^{at}$. The $F^{red}$ -$g^{at}$ ( $C^{trol}$ 'ed $s^{wa}$ -$g^{at}$ ) $s^{wa}$ 's the values of the 2nd and 3rd bits, iff, the 1st is 1.

---

*Reversible $g^{at}$ 's Described by Permutation ($p^{erm}$) Matrices ($m^{tri}$'s):* Any $n$-bit $r^{vers}$ -$g^{at}$ must define mapping ($\mathcal{M}^{\leftrightarrows}$) of the $i^{np}$ *bit block* - into the $o^{ut}$ block of the same length where no two distinct $i^{np}$'s are allowed to be $\mathcal{M}^{\leftrightarrows}$ to the same $o^{ut}$ and vice versa. This makes the $\mathcal{M}^{\leftrightarrows}$ -$r^{vers}$. As a consequence, one can consider a $r^{vers}$ -$g^{at}$ as $e^{cod}$ 'ing , defining the rule how to permute the $2^n$ possible $i^{np}$ bit blocks expressible in $n$ bits.
In the case of the 2-bit $s^{wa}$ -$g^{at}$, for example, the four possible $i^{np}$-blocks are 00, 01, 10, 11 and these are $\mathcal{M}^{\leftrightarrows}$, respectively, into $00 \to 00, 01 \to 10, 10 \to 01, 11 \to 11$. For $\bar{C}$- $g^{at}$, the $i^{np}$'s 00, 01, 10, and 11 are $\mathcal{M}^{\leftrightarrows}$ into 00, 01, 11, and 10 respectively.
Thus, a way to represent an $n$-bit $r^{vers}$ -$g^{at}$ is as an array whose rows and columns are indexed by the $2^n$ possible bit blocks expressible in $n$ bits. The $(i, j)$-th element of this array is defined to be 1 if, and only if, the $i^{np}$- bit block $c^{resp}$ 'ing to the $i$-th row is $\mathcal{M}^{\leftrightarrows}$ to the $o^{ut}$- bit block $c^{resp}$ 'ing to the $j$-th column. The resulting array will contain a single 1 in each row and column and zeroes otherwise and will therefore be a $p^{erm}$ -$m^{tri}$. As arrays, the NOT, $s^{wa}$ and $\bar{C}$- $g^{at}$ 's would be described as follows:

$$\text{NOT}\ \begin{pmatrix} 0 & 1 \\ 1 & 0 \end{pmatrix};\quad \text{SWAP}\ \begin{pmatrix} 1 & 0 & 0 & 0 \\ 0 & 0 & 1 & 0 \\ 0 & 1 & 0 & 0 \\ 0 & 0 & 0 & 1 \end{pmatrix};\quad \text{CNOT}\ \begin{pmatrix} 1 & 0 & 0 & 0 \\ 0 & 1 & 0 & 0 \\ 0 & 0 & 0 & 1 \\ 0 & 0 & 1 & 0 \end{pmatrix}$$

TOFFOLI:

| | 000 | 001 | 010 | 011 | 100 | 101 | 110 | 111 |
|---|---|---|---|---|---|---|---|---|
| 000 | 1 | 0 | 0 | 0 | 0 | 0 | 0 | 0 |
| 001 | 0 | 1 | 0 | 0 | 0 | 0 | 0 | 0 |
| 010 | 0 | 0 | 1 | 0 | 0 | 0 | 0 | 0 |
| 011 | 0 | 0 | 0 | 1 | 0 | 0 | 0 | 0 |
| 100 | 0 | 0 | 0 | 0 | 1 | 0 | 0 | 0 |
| 101 | 0 | 0 | 0 | 0 | 0 | 1 | 0 | 0 |
| 110 | 0 | 0 | 0 | 0 | 0 | 0 | 0 | 1 |
| 111 | 0 | 0 | 0 | 0 | 0 | 0 | 1 | 0 |

FREDKIN:

| | 000 | 001 | 010 | 011 | 100 | 101 | 110 | 111 |
|---|---|---|---|---|---|---|---|---|
| 000 | 1 | 0 | 0 | 0 | 0 | 0 | 0 | 0 |
| 001 | 0 | 1 | 0 | 0 | 0 | 0 | 0 | 0 |
| 010 | 0 | 0 | 1 | 0 | 0 | 0 | 0 | 0 |
| 011 | 0 | 0 | 0 | 1 | 0 | 0 | 0 | 0 |
| 100 | 0 | 0 | 0 | 0 | 0 | 0 | 0 | 1 |
| 101 | 0 | 0 | 0 | 0 | 0 | 1 | 0 | 0 |
| 110 | 0 | 0 | 0 | 0 | 0 | 0 | 1 | 0 |
| 111 | 0 | 0 | 0 | 0 | 1 | 0 | 0 | 0 |

In fact, the $m^{tri}$*'s* -$c^{resp}$ 'ing to *c-* $r^{vers}$ -$g^{at}$ 's are always $p^{erm}$ -$m^{tri}$*'s*, i.e., 0/1 $m^{tri}$*'s* having a single 1 in each row and column, and $p^{erm}$ -$m^{tri}$*'s* are also always U -$m^{tri}$*'s*. To find out the effect of a $r^{vers}$ -$g^{at}$, e.g., the $F^{red}$ or $T^{off}$ -$g^{at}$, on an $i^{np}$- bit block, we prepare ($p^{rep}$) the column $\mathcal{V}$- $c^{resp}$ 'ing to that bit block, and then $p^{erf}$ the usual $m^{tri}$ -$\mathcal{V}$- $p^{duc}$ -$o^{per}$ 'ion. For instance, since the $F^{red}$ and $T^{off}$ -$g^{at}$ 's act on 3-bits, we can imagine a column $\mathcal{V}$ consisting of $2^3 = 8$ slots, one of which (the $i$-th say) contains a single 1, and all the other elements are 0.

$$000 \equiv \begin{pmatrix} 1 \\ 0 \\ 0 \\ 0 \\ 0 \\ 0 \\ 0 \\ 0 \end{pmatrix},\ 001 \equiv \begin{pmatrix} 0 \\ 1 \\ 0 \\ 0 \\ 0 \\ 0 \\ 0 \\ 0 \end{pmatrix},\ 010 \equiv \begin{pmatrix} 0 \\ 0 \\ 1 \\ 0 \\ 0 \\ 0 \\ 0 \\ 0 \end{pmatrix},\ \ldots\quad 111 \equiv \begin{pmatrix} 0 \\ 0 \\ 0 \\ 0 \\ 0 \\ 0 \\ 0 \\ 1 \end{pmatrix}$$

etc. We can find out the result of, e.g., applying the $T^{off}$ -$g^{at}$ on such an $i^{np}$- by $\mathcal{V}$- $m^{tri}$ product.

$$\text{TOFFOLI}|110\rangle = \begin{pmatrix} 1&0&0&0&0&0&0&0\\ 0&1&0&0&0&0&0&0\\ 0&0&1&0&0&0&0&0\\ 0&0&0&1&0&0&0&0\\ 0&0&0&0&1&0&0&0\\ 0&0&0&0&0&1&0&0\\ 0&0&0&0&0&0&0&1\\ 0&0&0&0&0&0&1&0 \end{pmatrix} \cdot \begin{pmatrix}0\\0\\0\\0\\0\\0\\1\\0\end{pmatrix} = \begin{pmatrix}0\\0\\0\\0\\0\\0\\0\\1\end{pmatrix} = |111\rangle$$

*$\mathcal{A}$'s in $r^{vers}$ -$C^{omp}$:* $\mathcal{A}$'s are important components in $c$- $r^{vers}$ -$C^{omp}$. Every $c^{irc}$ with more than 3 $i^{np}$'s over the NOT-$\bar{C}$- $T^{off}$ -$b^{as}$ does an *even* $p^{erm}$ -on the set of its $i^{np}$'s. So, to do an *odd* $p^{erm}$ on $\{0,1\}^n$, we need at least one $\mathcal{A}$ bit with fixed constant value in addition to the $n$ $v^{ria}$ -$i^{np}$'s. Toffoli has shown that one $\mathcal{A}$ bit is, actually, always $s^{uff}$.

The importance of $\mathcal{A}$'s can be also illustrated by considering $C^{omp}$ a $B^{ool}$ - F- $f : \{0,1\}^n \rightarrow \{0,1\}$ $r^{vers}$ 'ly. Every $r^{vers}$ -$c^{irc}$ on $m$ $i^{np}$'s, $C^{omp}$ -$f$ , has exactly $m$ -$o^{ut}$'s where one of them is $f$ . If $m = n$, i.e., there is no $\mathcal{A}$ bit, then every $o^{ut}$- $F$ must be a *balanced* $B^{ool}$ - F. So, if the simulated F is not balanced, $m > n$ is needed and there must therefore be at least one $\mathcal{A}$ bit.

---

*EXAMPLE:*

The $m^{del}$ described in Fig. 10 is used to define how a $r^{vers}$ -$c^{irc}$ -$C^{omp}$ a $F$- $f : \{0,1\}^n \rightarrow \{0,1\}$. In this $m^{del}$, we require that at the end of the $C^{omp}$ all $\mathcal{A}$'s are in their $i^{nit}$ states, except one $\mathcal{A}$ bit, the "answer" bit, that gives the value of the $F$.

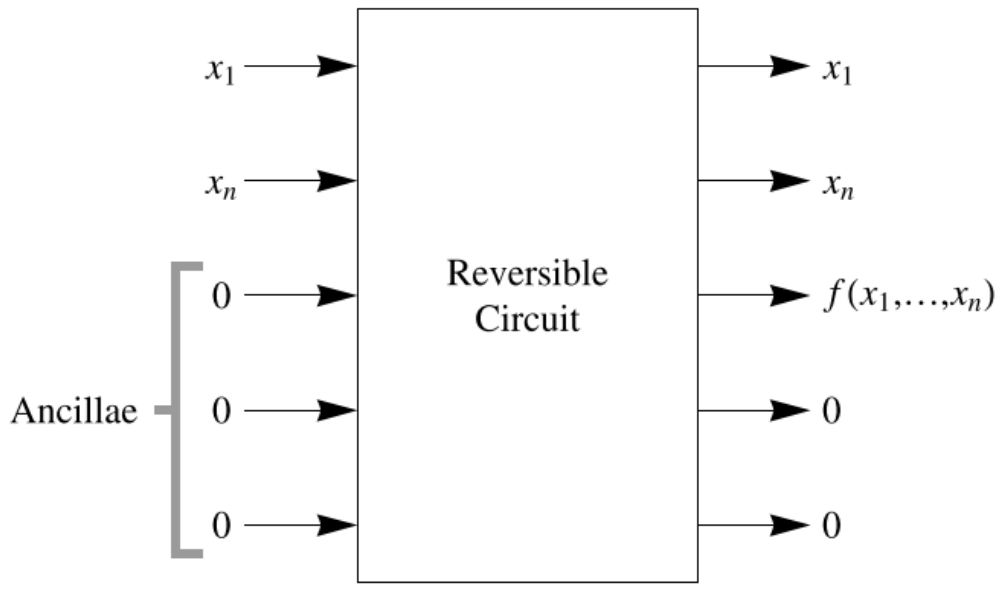

*Fig. 10 $C^{omp}$ a $B^{ool}$ - F using a $r^{vers}$ -$c^{irc}$*

---

*EXAMPLE:*

*Universal $r^{vers}$-$b^{as}$:* Here $r^{vers}$ -$c^{irc}$ 's over the NOT-$\bar{C}$- $T^{off}$ -$b^{as}$ are considered. Fig.11 defines the $a^{ct}$ and icons of these $g^{at}$ 's. While one can see that the $T^{off}$ -$g^{at}$ alone is $u$- for $r^{vers}$ -$C^{omp}$ so, the NOT and $\bar{C}$- $g^{at}$ 's are not needed the scheme, simplifies the constructions. Fig. 11 also shows how these $r^{vers}$ -$g^{at}$ 's can simulate the $c$- (ir-$r^{vers}$) standard $g^{at}$'s, in some cases with $\mathcal{A}$'s.

The action of reversible gates

| NOT | CNOT | TOFFOLI |
|---|---|---|
| $a \mapsto 1 \oplus a$ | $\begin{pmatrix}a\\b\end{pmatrix} \mapsto \begin{pmatrix}a\\a \oplus b\end{pmatrix}$ | $\begin{pmatrix}a\\b\\c\end{pmatrix} \mapsto \begin{pmatrix}a\\b\\c \oplus (a \cdot b)\end{pmatrix}$ |

The reversible basis

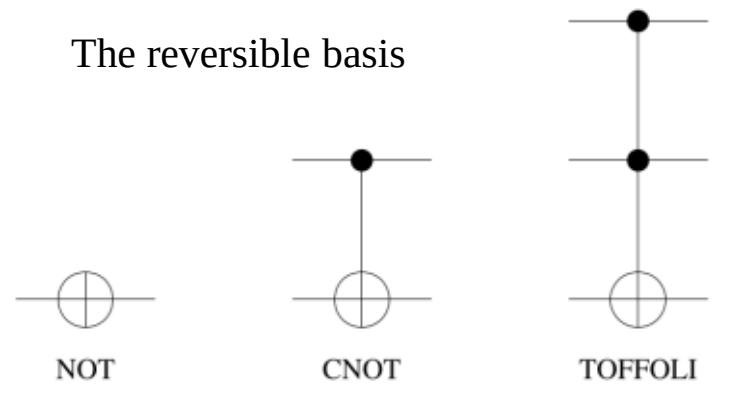

Reversible simulation of classical gates

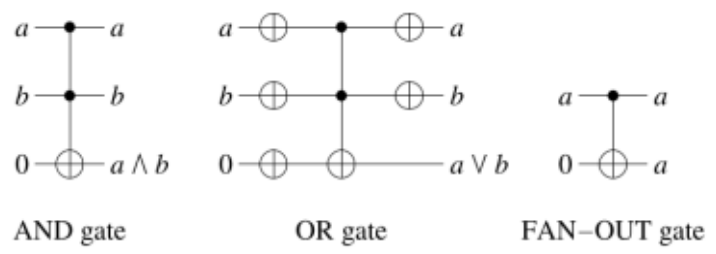

*Fig. 11 $r^{vers}$ -$c^{irc}$'s over the NOT-$\bar{C}$- $T^{off}$ -$b^{as}$*

---

*A.2. Q- lG -$l^{ibr}$*

Like in $c$-$C^{omp}$, any $q$- $C^{omp}$ can be represented as a $s^{equ}$ of $q$- $lG$ acting on a few $qb$ 's at a time. However, whereas $c$- $lG$'s manipulate the $c$- bit values, 0 or 1, $q$- $g^{at}$ 's can process multi ($m^{lti}$)-partite $q$- $s^{ta}$ including $S^{po}$ 's of the $C^{omp}$ -$b^{as}$ -$s^{ta}$, which may be also $\mathcal{E}$'ed. So, the $lG$ of $q$- $C^{omp}$ are more varied than the $lG$ of $c$- $C^{omp}$.

The $p^{hys}$ phenomena used to get the desired changes of a $q$- $s^{ta}$ can very a lot. For instance, if $qb$ 's are $e^{cod}$ 'ed in particles having $q$- $m^{ech}$ - $s^{pin}$, the $l^{gic}$ is affected by spin- changes performed by varying an $a^{pp}$ 'ed $m^{gn}$ -$f^{iel}$ at various orientations. On the other hand, if the $qb$ is $e^{cod}$ 'ed in an internal excitation $s^{ta}$ of an ion, the $g^{at}$ -$o^{per}$ 'ion can be obtained by changing the time a laser $b^{eam}$ can irradiate the ion or by changing the $w^{ve}$ -length of the laser light.

As any $q$- $g^{at}$ is $\mathcal{J}^{mpl}$ 'ed $p^{hys}$ 'ly as the $q$- $m^{ech}$ -$e^{vlt}$ of a given $q$- $s^{yst}$, the $t^{rans}$ is defined by $S^{chrö}$'s $eq$, $i\hbar\, \partial|\psi\rangle / \partial t = \mathcal{H}|\psi\rangle$, where $\mathcal{H}$ is the $\mathcal{H}$, defining the $p^{hys}$ -$f^{iel}$ 's and forces used. So, the U -$m^{tri}$'s specifying $q$- $g^{at}$ 's are given by $U = \exp(-i\mathcal{H}t/\hbar)$ . Here $\mathcal{H}$ is the $\mathcal{H}$ which specifies the $i^{ntac}$ 's that are present in the $p^{hys}$ -$s^{yst}$.

The $q$- $m^{ech}$ -$e^{vlt}$ defined by this $eq$ is U provided no $\mathcal{M}$'s are made, and no harmful $i^{ntac}$ 's occur with the $e^{vir}$. So, starting from some $i^{nit}$ -$s^{ta}$, $|\psi(0)\rangle$, the $q$- $s^{yst}$ will evolve, in time $t$ , into the $s^{ta}$ $|\psi(t)\rangle = \exp(-i\mathcal{H}t/\hbar)|\psi(0)\rangle = U|\psi(0)\rangle$ with $U$ being some U- $m^{tri}$. So, the $e^{vlt}$, in time $t$, of a $q$- $s^{yst}$ is defined by a U-$t^{rans}$ of an $i^{nit}$ -$s^{ta}$ $|\psi(0)\rangle$ to a final $s^{ta}$ $|\psi(t)\rangle = U|\psi(0)\rangle$. Therefore, a $q$- $lG$ operating on a given $q$- $C^{omp}$, will $t^{rans}$ that $s^{ta}$ unitarily up until the point at which a measurement is made. So, *q*- *lG* are specified, by U- $m^{tri}$*'s*, and their $a^{ct}$ is $r^{vers}$.

The early *q*- $C^{omp}$ researchers were already aware of the parallels between *c*- $r^{vers}$ -$g^{at}$ 's and *q*- $g^{at}$. They were aware that since the $m^{tri}$*'s* -$c^{resp}$ 'ing to $r^{vers}$ (*c*-) $g^{at}$ 's were $p^{erm}$ -$m^{tri}$*'s*, they were also U- $m^{tri}$*'s* and so could be seen as $o^{per}$ 'ors that evolved some $i^{nit}$ -*q*- $s^{ta}$ ($i^{np}$- to a $g^{at}$ ) into some final *q*- $s^{ta}$ ( its $o^{ut}$ ) following $S^{chrö}$'s *eq*. So, the closest *c*-analogs to *q*- *lG* are the *c*- $r^{vers}$ -$g^{at}$ 's like the NOT, $s^{wa}$, $\bar{C}$, $T^{off}$ and $F^{red}$. Whereas the repertoire of $g^{at}$ 's in *c*- $r^{vers}$ -$C^{omp}$ is limited to the U- $g^{at}$ 's whose $m^{tri}$ descriptions $c^{resp}$ to $p^{erm}$ -$m^{tri}$*'s*, in *q*- $C^{omp}$ any $g^{at}$ is allowed whose $m^{tri}$ is U whether or not it is also a $p^{erm}$ -$m^{tri}$.

*Properties of q- $g^{at}$ 's arising from unitarity*: The main $p^{pert}$ of *q*- *lG* follows from the fact that they are defined by U- $m^{tri}$*'s*. A $m^{tri}$, $U$, is U iff $U^{-1} = U^{\dagger}$. If $U$ is U we have: $U^{\dagger}$ *is* U, $U^{-1}$ *is* U, $U^{-1} = U^{\dagger}$ *(which defines unitarity)*, $U^{\dagger}U = 1, |\, det\ (U)| = 1$, *the columns (rows) of U form an orthonormal set of* $\mathcal{V}$*'s, for a given j*, $\sum_{i=1}^{2^n} | U_{ij}|^2 = 1$, *for a given i*, $\sum_{j=1}^{2^n} | U_{ij}|^2 = 1$, $U = exp\ (i\mathcal{H})$ *where* $\mathcal{H}$ *is an hermitian* $m^{tri}$, *i.e.*, $\mathcal{H} = \mathcal{H}^{\dagger}$.

Since, for any *q*- $g^{at}$ -$U$, $U^{\dagger}U = 1$ it is possible to undo the operation of a *q*- $g^{at}$, i.e., a *q*- $g^{at}$ is $r^{vers}$. Also, since for a given *j* $\sum_{i=1}^{2^n} | U_{ij}|^2 = 1$ and for a given $i$, $\sum_{j=1}^{2^n} | U_{ij}|^2 = 1$ , starting with a properly normalized *q*- $s^{ta}$ and applying on it a *q*- $g^{at}$, will result into a properly normalized *q*- $s^{ta}$. Since $|\det(U)|$ is constrained to be unity the constraint on the determinant can be satisfied with $\det(U) = \pm 1$ or $\pm i$. So, the entries of a *G*- U-$m^{tri}$ can be $c^{plex}$ -$n^{ber}$ 's.

*A.3.Qubit $g^{at}$ 's*

*Special 1- qb- $g^{at}$ 's*

$P^{au}$ - $s^{pin}$ -$m^{tri}$*'s*: Single- *qb* 's, "$P^{au}$ -$m^{tri}$*'s*" $(1, X, Y, Z)$ , are both hermitian and U, and any 1- *qb*- $\mathcal{H}$ can always be represented as a weighted sum of the $P^{au}$ -$m^{tri}$*'s*:

$$1 = \begin{pmatrix} 1 & 0 \\ 0 & 1 \end{pmatrix},\ X = \begin{pmatrix} 0 & 1 \\ 1 & 0 \end{pmatrix},\ Y = \begin{pmatrix} 0 & -i \\ i & 0 \end{pmatrix},\ Z = \begin{pmatrix} 1 & 0 \\ 0 & -1 \end{pmatrix}$$

Commonly used $\mathcal{H}$'s are $\mathcal{H} = Z^{(1)}Z^{(2)}$ (the Ising $i^{ntac}$) and $\mathcal{H} = X^{(1)} \otimes X^{(2)} + Y^{(1)} \otimes Y^{(2)}$ (the *XY* $i^{ntac}$) and $\mathcal{H} = 2X^{(1)} \otimes X^{(2)} + Y^{(1)} \otimes Y^{(2)}$ where the superscripts indicate which of two *qb* 's the $o^{per}$ 'or acts upon.

*NOT $g^{at}$:* The $P^{au}$-*X*-$m^{tri}$ is identical to the *c*- ($r^{vers}$) NOT $g^{at}$, i.e.,

$$X \equiv \text{NOT} = \begin{pmatrix} 0 & 1 \\ 1 & 0 \end{pmatrix}$$

Therefore $X$ negates the $C^{omp}$ -$b^{as}$ $s^{ta}$ $|0\rangle$ and $|1\rangle$, $c^{or}$ 'ly as these $c^{resp}$ to the *c*- bits, 0 and 1, respectively

$$X|0\rangle = \begin{pmatrix} 0 & 1 \\ 1 & 0 \end{pmatrix} \cdot \begin{pmatrix} 1 \\ 0 \end{pmatrix} = \begin{pmatrix} 0 \\ 1 \end{pmatrix} = |1\rangle$$

$$X|1\rangle = \begin{pmatrix} 0 & 1 \\ 1 & 0 \end{pmatrix} \cdot \begin{pmatrix} 0 \\ 1 \end{pmatrix} = \begin{pmatrix} 1 \\ 0 \end{pmatrix} = |0\rangle$$

$\sqrt{NOT}$ *$g^{at}$:* One simple example of 1- *qb* non- *c*- $g^{at}$ 's is a $f^{rct}$ 'al $p^{wer}$ of NOT $g^{at}$, such as $\sqrt{\text{NOT}}$:

$$\sqrt{\text{NOT}} = \begin{pmatrix} 0 & 1 \\ 1 & 0 \end{pmatrix}^{1/2} = \begin{pmatrix} \frac{1}{2} + \frac{i}{2} & \frac{1}{2} - \frac{i}{2} \\ \frac{1}{2} - \frac{i}{2} & \frac{1}{2} + \frac{i}{2} \end{pmatrix}$$

If a $g^{at}$ is $a^{pp}$ twice, i.e., $\sqrt{\text{NOT}} \cdot \sqrt{\text{NOT}}$, the result is the NOT -$o^{per}$ 'ion, but a single $a^{pp}$ gives a *q*- $s^{ta}$ that neither $c^{resp}$ 's to the *c*- bit 0, or the *c*- bit 1. So $\sqrt{\text{NOT}}$ is the truly non- *c*- $g^{at}$.

$$|0\rangle \xrightarrow{\sqrt{\text{NOT}}} \left(\frac{1}{2} + \frac{i}{2}\right)|0\rangle + \left(\frac{1}{2} - \frac{i}{2}\right)|1\rangle \xrightarrow{\sqrt{\text{NOT}}} |1\rangle$$

$$|1\rangle \xrightarrow{\sqrt{\text{NOT}}} \left(\frac{1}{2} - \frac{i}{2}\right)|0\rangle + \left(\frac{1}{2} + \frac{i}{2}\right)|1\rangle \xrightarrow{\sqrt{\text{NOT}}} |0\rangle$$

---

*EXAMPLE: $P^{au}$ -X is not a NOT -$g^{at}$ for qb 's:* The $P^{au}$ -*X* $g^{at}$ negates the $C^{omp}$ -$b^{as}$ -$s^{ta}$ -$c^{or}$ 'ly, and we are interested to see if it also acts as a true "NOT" $g^{at}$ when $a^{pp}$ 'ed on a *qb* in an *arbitrary q- $s^{ta}$*, i.e., a *qb* -$s^{ta}$ -$c^{resp}$ 'ing to any point on the Bloch sphere. Let us first specify what we expect a *q*- NOT $g^{at}$ to do, and then see whether $X$ can do it.

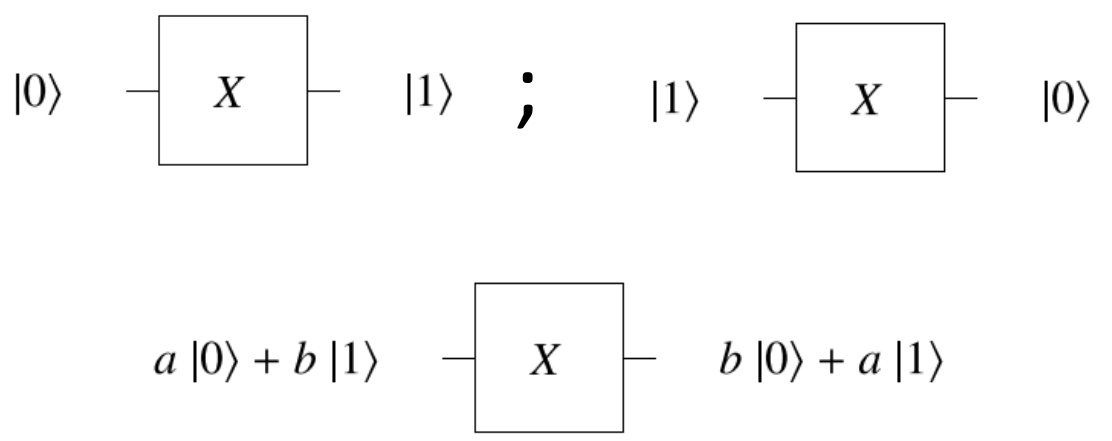

*Fig. 12 The effect of the $P^{au}$ -X- $g^{at}$ -$o^{per}$ 'ion on the $C^{omp}$ -$b^{as}$ $s^{ta}$ and an arbitrary $p^{ur}$ -$s^{ta}$ of a 1-qb. The $P^{au}$ -X- $g^{at}$ negates the $C^{omp}$ -$b^{as}$ -$s^{ta}$ -$c^{or}$ 'ly, but not an arbitrary $S^{po}$ -$s^{ta}$! So the $P^{au}$ -X -$g^{at}$ is not a u- NOT -$g^{at}$ for qb 's*

Since the NOT $g^{at}$ moves a $s^{ta}$ from the upper pole of the Bloch sphere to a $s^{ta}$ at the lower pole and vice versa, we redifine a NOT $g^{at}$ to be the $o^{per}$ 'ion that $\mathcal{M}^{\leftrightarrows}$ a *qb*, $|\psi\rangle$, lying at any point on the Bloch sphere, into its antipodal $s^{ta}$ , $|\psi^{\perp}\rangle$, on the opposite side of the Bloch sphere as shown in Fig.12.

---

The $|\psi^{\perp}\rangle$, is reached by drawing a straight line from the initial $s^{ta}$ through the center to arrive at the surface of the Bloch sphere on the opposite side. If starting $s^{ta}$ $|\psi\rangle$ is given by:

$$|\psi\rangle = \cos(\theta/2)|0\rangle + e^{i\varphi} \sin(\theta/2)|1\rangle$$

where $\theta$ is the "latitude" and φ the "longitude" angles of $|\psi\rangle$ on the Bloch sphere, to get $|\psi^{\perp}\rangle$ we move, to the equivalent latitude in the opposite hemisphere and shift the longitude by 180° (i.e., $\pi$ radians). Given the above $|\psi\rangle$, and by using the relations

$\cos\left(\frac{\pi-\theta}{2}\right) = \sin(\theta/2)$ and $\sin\left(\frac{\pi-\theta}{2}\right) = \cos(\theta/2)$ , the analytical form of the antipodal $s^{ta}$, $|\overline{\psi}\rangle$, must be:

$$|\psi^{\perp}\rangle = \cos\left(\frac{\pi-\theta}{2}\right)|0\rangle + e^{i(\varphi+\pi)} \sin\left(\frac{\pi-\theta}{2}\right)|1\rangle$$

$$= \cos\left(\frac{\pi-\theta}{2}\right)|0\rangle - e^{i\varphi} \sin\left(\frac{\pi-\theta}{2}\right)|1\rangle$$

$$= \sin(\theta/2)|0\rangle - e^{i\varphi}\,\varphi \cos(\theta/2)|1\rangle$$

Given the relationship between $|\psi\rangle$ and $|\psi^{\perp}\rangle$, we can now examine whether $X|\psi\rangle = |\psi^{\perp}\rangle$, and so, if $X$ is a true NOT $g^{at}$ for an arbitrary $qb$. By $d^{ir}$ check we have:

$$X|\psi\rangle = \begin{pmatrix} 0 & 1 \\ 1 & 0 \end{pmatrix} \cdot \begin{pmatrix} \cos(\theta/2) \\ \sin(\theta/2)e^{i\varphi} \end{pmatrix} = \begin{pmatrix} \sin(\theta/2)e^{i\varphi} \\ \cos(\theta/2) \end{pmatrix}$$

$$= e^{i\varphi} \sin(\theta/2)|0\rangle + \cos(\theta/2)|1\rangle$$

Since $s^{ta}$ that differ only in global $p^{has}$ cannot be distinguished, we can $m^{lti}$ 'ply by any overall $p^{has}$ factor we please. As the $a^{mpl}$ of the $|0\rangle$ component of the true $|\psi^{\perp}\rangle$ $s^{ta}$ is $\sin(\theta/2)$ , we $m^{lti}$ 'ply the above *eq* by $e^{-i\varphi}$ we get $X|\psi\rangle$ as:

$$X|\psi\rangle = \sin(\theta/2)|0\rangle + e^{-i\varphi} \cos(\theta/2)|1\rangle \neq |\psi^{\perp}\rangle$$

which is not $|\psi^{\perp}\rangle$. So in *c*- $C^{omp}$, the $g^{at}$ whose $m^{tri}$ is $\begin{pmatrix} 01 \\ 10 \end{pmatrix}$ is the "NOT" $g^{at}$, but it is not in the context of *q*- $C^{omp}$.

$\mathcal{H}^{mard}$ -$g^{at}$: *A* commonly used single $qb$ -$g^{at}$, is the $\mathcal{H}^{mard}$ -$g^{at}$, $H$, defined by the $m^{tri}$:

$$H = \frac{1}{\sqrt{2}} \begin{pmatrix} 1 & 1 \\ 1 & -1 \end{pmatrix}$$

It $\mathcal{M}^{\leftrightarrows}$'s (see Fig.13) $C^{omp}$ -$b^{as}$ -$s^{ta}$ into $S^{po}$- $s^{ta}$ and vice versa:

$$H|0\rangle = \frac{1}{\sqrt{2}}(|0\rangle + |1\rangle)$$

$$H|1\rangle = \frac{1}{\sqrt{2}}(|0\rangle - |1\rangle)$$

When the $\mathcal{H}^{mard}$ -$g^{at}$ $H$ operates on a $C^{omp}$ -$b^{as}$ -$s^{ta}$ $|x\rangle$ it $t^{rans}$ 's the $i^{np}$- according to $H|x\rangle = \frac{1}{\sqrt{2}}(|0\rangle + (-1)^{x}|1\rangle)$ .

*EXAMPLE:* ________________________________________

Although $\mathcal{H}^{mard}$ is simple looking $g^{at}$, it has a remarkable feature of vital importance to *q*- $C^{omp}$. If $n$ -$qb$ 's are $p^{rep}$'*ed* , each in the $s^{ta}$ $|0\rangle$ and its own $\mathcal{H}^{mard}$ -$g^{at}$ is $a^{pp}$ to each $qb$, in parallel ( as in Fig. 13) the $s^{ta}$ obtained is an equal $S^{po}$ of all the integers in the range 0 to $2^n - 1$.

$$H|0\rangle \otimes H|0\rangle \otimes \cdots \otimes H|0\rangle = \frac{1}{\sqrt{2^n}} \sum_{j=0}^{2^n-1} |j\rangle$$

where $|j\rangle$ is the $C^{omp}$ -$b^{as}$ -$s^{ta}$ indexed by the binary $n^{ber}$ that would $c^{resp}$ to the $n^{ber}$ -$j$ written in base-10. For instance, in a 7-$qb$ -$r^{eg}$ the $s^{ta}$ "|19⟩" $c^{resp}$ 's to the $C^{omp}$ -$b^{as}$ -$s^{ta}$ $|0010011\rangle$. The first two bits (00) are padding, and 10011 in base 2 $c^{resp}$ 's to 19 in base-10.The importance of the $\mathcal{H}^{mard}$ -$g^{at}$ comes from the fact that by $a^{pp}$'ing, in parallel, a $S^{epa}$ -$\mathcal{H}^{mard}$ -$g^{at}$ to each of $n$ $qb$ 's, each $i^{nit}$ 'ly in the $s^{ta}$ $|0\rangle$, we can create an $n$- $qb$ -$S^{po}$ containing $2^n$ component $e^{ig}$ - $s^{ta}$. These $e^{ig}$ - $s^{ta}$ represent all $n$ bits binary numbers.

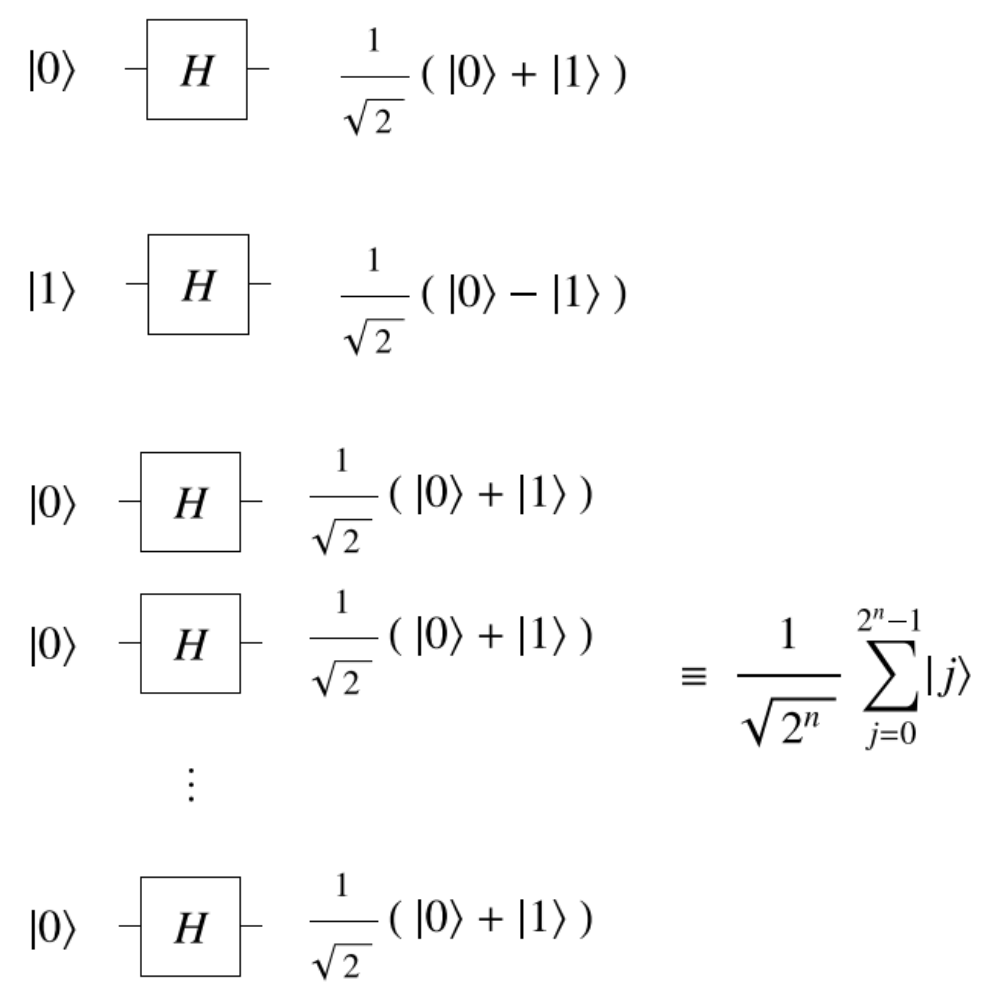

*Fig. 13 By $a^{pp}$ -n $\mathcal{H}^{mard}$ $g^{at}$ 's- $I^{dep}$ 'ly to n qb 's, all $p^{rep}$'d- $i^{nit}$ 'ly in $s^{ta}$ |0⟩, we can form an n- qb -$S^{po}$ whit its component $e^{ig}$ - $s^{ta}$ being the binary representation of all the integers in the range 0 to $2^n - 1$. Thus, a $S^{po}$ having exp 'ly many $t^{rm}$ 's can be $p^{rep}$ 'd using only a polynomial $n^{ber}$ of $o^{per}$ 'ions. This is used in a number of q- $a^{lg}$ 's to load a q- $m^{emo}$ -$r^{eg}$ -$e^{ffi}$ 'tly with an equally weighted $S^{po}$ of all the $n^{ber}$ 's it can contain $r^{ot}$ 's around the x-, y-, and z-axes*

This is one of the most important tricks of q- $C^{omp}$ as it enables loading exp 'ly many indices into a q- $C^{omp}$ using only polynomially many $o^{per}$ 'ions. Otherwise, if we needed to enter the different bit-blocks individually, as we do in c- $C^{omp}$, then q- $C^{omp}$ wouldn't have such potential for reduvcing $C^{omp}$ - $c^{plex}$ 'ity.

---

To generalize *q*- $g^{at}$ for a 1- $qb$, [137] first introduces the family of *q*- $g^{at}$ 's that $p^{erf}$ -$r^{ot}$ 's about the three axes of the Bloch sphere.

A 1- $qb$ -$p^{ur}$ -$s^{ta}$ is described by a point on the sphere. The result of a 1-$qb$ -$g^{at}$ acting on this $s^{ta}$ is to $\mathcal{M}^{\leftrightarrows}$ it to some other point on the sphere. The $g^{at}$ 's that rotate ($r^{ot}$) $s^{ta}$ around the *x*-, *y*-, and *z*-axes are important since enable us to $d^{comp}$ an arbitrary 1- $qb$ -*q*- $g^{at}$ into $s^{equ}$ 's of such $r^{ot}$ -$g^{at}$ 's.

We start by choosing *x*-, *y*-, and *z*-, or equivalently, three polar $c^{ord}$ 's. These two $c^{ord}$ -$s^{yst}$ 's are related via the *eq* 's:

$$x = r\sin(\theta)\cos(\varphi)\,;\; y = r\sin(\theta)\sin(\varphi)\,;\; z = r\cos(\theta)$$

The *q*- $g^{at}$ 's that $r^{ot}$ this $s^{ta}$ about the *x*-, *y*-, or *z*-axes can be built from the $P^{au}$ -*X*, *Y*, *Z*, $m^{tri}$'s, and in addition $P^{au}$ -$m^{tri}$, 1, can be utilized to grt a global overall $p^{has}$ shift. For this we define the U

-$m^{tri}$'s, $R_x(\theta)$, $R_y(\theta)$, $R_z(\theta)$ , and $Ph$ from $\mathcal{H}$'s as $P^{au}$ - $m^{tri}$'s, $X$, $Y$, $Z$, and $I$ (the $i^{d}$ - $m^{tri}$):

$$R_x(\alpha) = \exp(-i\alpha X/2) = \begin{pmatrix} \cos(\alpha/2) & -i\sin(\alpha/2) \\ -i\sin(\alpha/2) & \cos(\alpha/2) \end{pmatrix}$$

$$R_y(\alpha) = \exp(-i\alpha Y/2) = \begin{pmatrix} \cos(\alpha/2) & -\sin(\alpha/2) \\ \sin(\alpha/2) & \cos(\alpha/2) \end{pmatrix}$$

$$R_z(\alpha) = \exp(-i\alpha Z/2) = \begin{pmatrix} e^{-i\alpha/2} & 0 \\ 0 & e^{i\alpha/2} \end{pmatrix}$$

$$Ph(\delta) = e^{i\delta}\begin{pmatrix} 1 & 0 \\ 0 & 1 \end{pmatrix}$$

Consider the $g^{at}$ $R_z(\alpha)$ (see also illustration in Fig.14) and see how this $g^{at}$ -$t^{rans}$ 's an arbitrary single $qb$ -$s^{ta}$ $|\psi\rangle =$ $\cos(\theta/2)|0\rangle + e^{i\varphi}\sin(\theta/2)|1\rangle$.

$$R_z(\alpha)|\psi\rangle = \begin{pmatrix} e^{-i\alpha/2} & 0 \\ 0 & e^{i\alpha/2} \end{pmatrix} \cdot \begin{pmatrix} \cos(\theta/2) \\ \sin(\theta/2)e^{i\varphi} \end{pmatrix} = \begin{pmatrix} e^{-i\alpha/2}\cos(\theta/2) \\ \sin(\theta/2)e^{i\alpha/2}e^{i\varphi} \end{pmatrix}$$

$$= e^{-i\alpha/2}\cos(\theta/2)|0\rangle + e^{i\alpha/2}e^{i\varphi}\sin(\theta/2)|1\rangle$$

We can $m^{lti}$ 'ply this $s^{ta}$ by any overall $p^{has}$ factor since for any $q$-$s^{ta}$ $|\chi\rangle$, the $s^{ta}$ $|\chi\rangle$ and $e^{i\gamma}|\chi\rangle$ are indistinguishable. For an overall $p^{has}$ factor $\exp(i\alpha/2)$ , we get:

$$R_z(\alpha)|\psi\rangle \equiv \cos(\theta/2)|0\rangle + e^{i(\varphi+\alpha)}\sin(\theta/2)$$

where $\equiv$ means "equal up to an unimportant arbitrary overall $p^{has}$ factor". So, $a^{pp}$ -$R_z(\alpha)$ -$g^{at}$ on $|\psi\rangle$ advances the angle φ by $\alpha$ and hence $r^{ot}$ the $s^{ta}$ around the $z$-axis by angle $\alpha$. So, we call $R_z(\alpha)$ a $z$- $r^{ot}$ -$g^{at}$.

EXAMPLE: ____________________

It can be shown that $R_x(\alpha)$ and $R_y(\alpha)$ $r^{ot}$ the $s^{ta}$ around the $x$- and $y$-axes respectively, as demonstrated in Fig. 14. These $r^{ot}$ 's do not coincide with our intuitions about $r^{ot}$ 's base. As an example, a $r^{ot}$ of 360 $d^{gre}$ 's of an object around any axis, brings that object back to its $i^{nit}$ position which is not true of $r^{ot}$ 's on the Bloch sphere! If we $r^{ot}$ a $q$- $s^{ta}$ by 360 $d^{gre}$ 's on the Bloch sphere we don't bring it back to its $i^{nit}$ -$s^{ta}$. If we rotating an arbitrary 1-$qb$ -$p^{ur}$ -$s^{ta}$, $|\psi\rangle$ around the $z$-axis through $2\pi$ radians we get:

$$R_z(2\pi)|\psi\rangle = \begin{pmatrix} e^{-i\pi} & 0 \\ 0 & e^{i\pi} \end{pmatrix} \cdot \begin{pmatrix} \cos(\theta/2) \\ \sin(\theta/2)e^{i} \end{pmatrix} = \begin{pmatrix} -\cos(\theta/2) \\ -\sin(\theta/2)e^{i\varphi} \end{pmatrix} = -|\psi\rangle$$

which has an additional overall $p^{has}$ of $-1$. To bring a $s^{ta}$ back to its initial position we need to $r^{ot}$ it through $4\pi$ on the Bloch sphere.

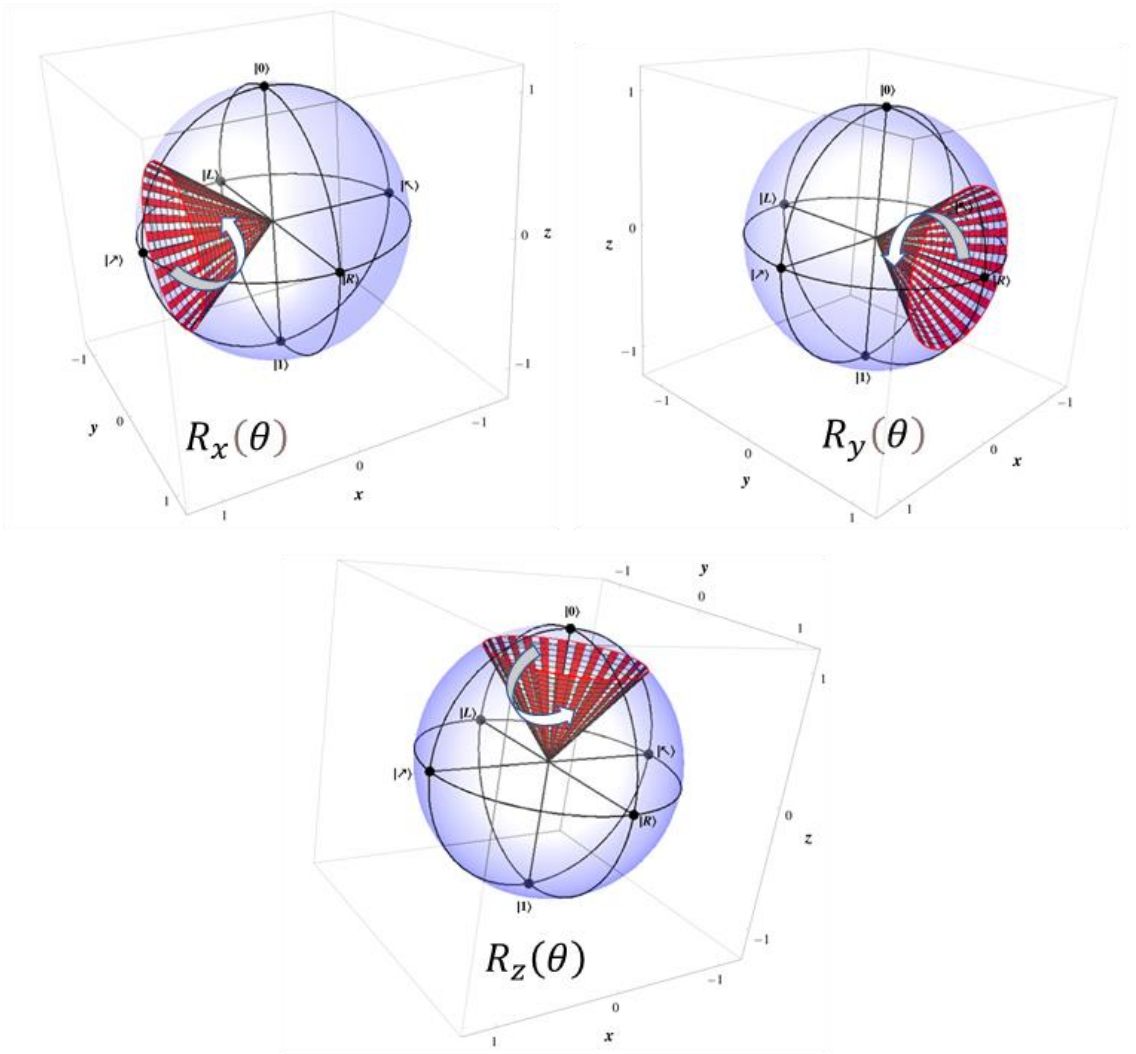

*Fig. 14 Illustrations of different rotating $\mathcal{M}^{\Leftarrow}$'s of a $s^{ta}$ -$|\psi\rangle$ on the surface of the Bloch sphere [137].*

---

*NOT*, $\sqrt{NOT}$, *and* $\mathcal{H}^{mard}$ -$g^{at}$ 's: can all be obtained via $s^{equ}$ 's of $r^{ot}$ -$g^{at}$ 's. For instance,

$$\text{NOT} \equiv R_x(\pi) \cdot Ph(\pi/2) \equiv R_y(\pi) \cdot R_z(\pi) \cdot Ph(\pi/2)$$

$$\sqrt{\text{NOT}} \equiv R_x\left(\frac{\pi}{2}\right) \cdot Ph\left(\frac{\pi}{4}\right) \equiv R_z\left(-\frac{\pi}{2}\right) \cdot R_y\left(\frac{\pi}{2}\right) \cdot R_z\left(\frac{\pi}{2}\right) \cdot Ph\left(\frac{\pi}{4}\right)$$

$$\mathcal{H}^{mard} \equiv R_x(\pi) \cdot R_y(\pi/2) \cdot Ph(\pi/2) \equiv R_y(\pi/2) \cdot R_z(\pi) \cdot Ph(\pi/2)$$

*Arbitrary* 1- *qb* -$g^{at}$ *'s - The* $P^{au}$ *-* $d^{comp}$*:* We have shown how *specific* 1- *qb*- $g^{at}$ 's can be $d^{comp}$ 'ed into $s^{equ}$ 's of $r^{ot}$ -$g^{at}$'s, i.e., $R_x(\cdot)$, $R_y(\cdot)$, $R_z(\cdot)$ and $p^{has}$-$g^{at}$ 's, i.e., $Ph(\cdot)$ . Let us now see how to $d^{comp}$ an arbitrary, $m^{ax}$ 'ly $G$, single- $qb$- $g^{at}$.

A $m^{ax}$ 'ly -$G$ single $qb$- $g^{at}$ will $c^{resp}$ to some $2 \times 2$ U $m^{tri}$, $U$. As $U$ is U we must have $|\det(U)| = 1$ that can be achieved by $\det(U)$ having any of the values $+1$, $-1$, $+i$, or $-i$. For $\det(U) = +1$ , $U$ is referred to as "special U". Otherwise, it can be written as $U = e^{i\delta}V$ where $V$ is a special U- $m^{tri}$, with, $\det(V) = +1$. Therefore, to get a $c^{irc}$ for the U- $m^{tri}$ $U$ it is $s^{uff}$ to get a $c^{irc}$ for the special U- $m^{tri}$ -$V$, since simply appending a $p^{has}$ shift $g^{at}$ $Ph(\delta)$ to the $c^{irc}$ for $V$ will give a $c^{irc}$ for $U$. This can be seen from

$$U = e^{i\delta}V = e^{i\delta}\begin{pmatrix} 1 & 0 \\ 0 & 1 \end{pmatrix} V = \begin{pmatrix} e^{i\delta} & 0 \\ 0 & e^{i\delta} \end{pmatrix} V = Ph(\delta) \cdot V$$

As $V$ is a $2 \times 2$ special U- $m^{tri}$ its rows and columns are orthonormal and, its entries, most $G$ 'ly, are $c^{plex}$ -$n^{ber}$ 's. So, $V$ is of the form:

$$V = \begin{pmatrix} \alpha & -\overline{\beta} \\ \beta & \overline{\alpha} \end{pmatrix}$$

where $\alpha$ and $\beta$ are arbitrary $c^{plex}$ -$n^{ber}$ ‘s with $\det(V) = \alpha\overline{\alpha} - \beta(-\overline{\beta}) = |\alpha|^2 + |\beta|^2 = 1$. This can be obtained by choosing $\alpha = e^{i\mu}\cos(\theta/2)$, and $\beta = e^{i\xi}\sin(\theta/2)$. So the $m^{tri}$ for $V$ can be also written as:

$$V = \begin{pmatrix} \alpha & -\overline{\beta} \\ \beta & \overline{\alpha} \end{pmatrix} \text{ with } \alpha \to e^{i\mu}\cos(\theta/2) \text{ and } \beta \to e^{i\xi}\sin(\theta/2)$$

$$= \begin{pmatrix} e^{i\mu}\cos(\theta/2) & -e^{-i\xi}\sin(\theta/2) \\ e^{i\xi}\sin(\theta/2) & e^{-i\mu}\cos(\theta/2) \end{pmatrix}$$

We can get this $m^{tri}$ as the $p^{duc}$ of the three $g^{at}$ ‘s $R_z(a)\cdot R_y(b)\cdot R_z(c)$ with $a \to -(\mu-\xi)$, $b \to \theta$, and $c \to -(\mu+\xi)$.

$$R_z(a)\cdot R_y(b)\cdot R_z(c) = \begin{pmatrix} e^{-i(a+c)/2}\cos(b/2) & -e^{-i(a-c)/2}\sin(b/2) \\ e^{i(a-c)/2}\sin(b/2) & e^{i(a+c)/2}\sin(b/2) \end{pmatrix}$$

with $a \to -(\mu-\xi)$, $b \to \theta$, and $c \to -(\mu+\xi)$

$$= \begin{pmatrix} e^{i\mu}\cos(\theta/2) & -e^{-i\xi}\sin(\theta/2) \\ e^{i\xi}\sin(\theta/2) & e^{-i\mu}\cos(\theta/2) \end{pmatrix} = V$$

---

*EXAMPLE*: Thus, any 1- *qb* special U- $g^{at}$ $-V$ can be $d^{comp}$ ‘ed into the form $R_z(a)\cdot R_y(b)\cdot R_z(c)$ and, any 1- *qb* -U- $g^{at}$, $U$ can be $d^{comp}$ ‘ed into the form $U \equiv R_z(a)\cdot R_y(b)\cdot R_z(c)\cdot Ph(d)$ as shown in Fig. 15.

---

*Decomposition of $R_x$ $g^{at}$*: It should be noticed that we could obtain an arbitrary 1- *qb*- $g^{at}$ without $p^{erf}$ ‘ing a $r^{ot}$ around the $x$-axis. *NOTE:* we can represent $r^{ot}$ ‘s around the $x$-axis $p^{ur}$ ‘ly in $t^{rm}$ ‘s of $r^{ot}$ ‘s around the $y$- and $z$-axe as

$$R_x(\theta) = \exp(-i\theta X/2) = \begin{pmatrix} \cos(\theta/2) & i\sin(\theta/2) \\ i\sin(\theta/2) & \cos(\theta/2) \end{pmatrix}$$

$$\equiv R_z(-\pi/2)\cdot R_y(\theta)\cdot R_z(\pi/2)$$

$$\equiv R_y(\pi/2)\cdot R_z(\theta)\cdot R_y(-\pi/2)$$

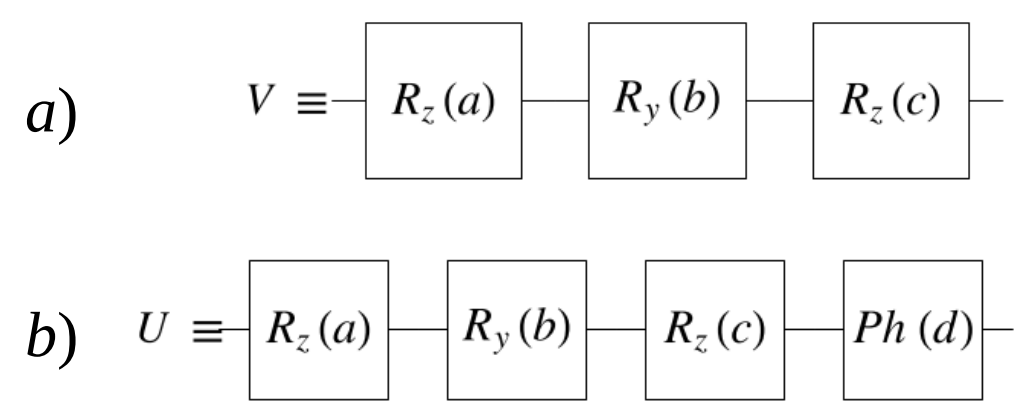

*Fig. 15 a) Decomposition of any 1- qb special U- $g^{at}$ and b) any 1- qb U- $g^{at}$*

*$C^{trol}$ ‘ed Q- $g^{at}$ ‘s*

To $p^{erf}$ non-trivial $C^{omp}$, we often need to change the $o^{per}$ ‘ion $a^{pp}$ ‘ed to one set of *qb*’s conditioned upon the values of some other set of qb’s. The $g^{at}$ ‘s that $\mathcal{J}^{mpl}$ these “if-then-else” type $o^{per}$ ‘ions are referred to as $C^{trol}$ *‘ed* $g^{at}$ ‘s. Some examples of $C^{trol}$ ‘ed $g^{at}$ ‘s discussed earlier in this section are $\bar{C}$ ($C^{trol}$ ‘ed-NOT), $F^{red}$ ($C^{trol}$ ‘ed- $s^{wa}$), and $T^{off}$ ($C^{trol}$ ‘ed- $\bar{C}$). The name for these $g^{at}$ ‘s “$C^{trol}$ ‘ed” comes from their effect on the $C^{omp}$ -$b^{as}$ -$s^{ta}$. For instance, $\bar{C}$ -$t^{rans}$ ‘s the $C^{omp}$ -$b^{as}$ -$s^{ta}$ such that the second *qb* is flipped iff the first *qb* is in $s^{ta}$ -$|1\rangle$.

$$|00\rangle \xrightarrow{\bar{C}} |00\rangle \ ; |01\rangle \xrightarrow{\bar{C}} |01\rangle;$$

$$|10\rangle \xrightarrow{\bar{C}} |11\rangle; \ |11\rangle \xrightarrow{\bar{C}} |10\rangle$$

Therefore, the value of the 2nd *qb* (the “$t^{rgt}$” *qb*) is $C^{trol}$ *‘ed* by the 1st *qb* (the “$C^{trol}$” -*qb*).

Similarly, under the $a^{ct}$ of the $F^{red}$ -$g^{at}$ the 2nd and 3rd *qb* ‘s are $s^{wa}$ ‘ped iff the 1st *qb* is in $s^{ta}$ -$|1\rangle$. So, the $F^{red}$ -$g^{at}$ -$p^{erf}$ ‘s a $C^{trol}$ ‘ed- $s^{wa}$ -$o^{per}$ ‘ion.

$$|000\rangle \xrightarrow{FREDKIN} |000\rangle ; |001\rangle \xrightarrow{FREDKIN} |001\rangle ;$$
$$|010\rangle \xrightarrow{FREDKIN} |010\rangle ; \ |011\rangle \xrightarrow{FREDKIN} |011\rangle ;$$

$$|100\rangle \xrightarrow{FREDKIN} |100\rangle \ ; |101\rangle \xrightarrow{FREDKIN} |110\rangle ;$$
$$|110\rangle \xrightarrow{FREDKIN} |101\rangle ; \ |111\rangle \xrightarrow{FREDKIN} |111\rangle$$

It is also possible to have $C^{trol}$ ‘ed $g^{at}$ ‘s with $m^{lti}$ ‘ple $C^{trol}$ -*qb* ‘s and $m^{lti}$ ‘ple $t^{rgt}$ -*qb* ‘s. The $a^{ct}$ of the $T^{off}$ -$g^{at}$ is to flip the third *qb* (i.e., the $t^{rgt}$ -*qb*) iff the first two *qb* ‘s (the $C^{trol}$ -*qb* ‘s) are in $s^{ta}$ -$|11\rangle$. Thus, the $T^{off}$ -$g^{at}$ has two $C^{trol}$ -*qb* ‘s and one $t^{rgt}$ -*qb*.

$$|000\rangle \xrightarrow{TOFFOLI} |000\rangle ; \ |001\rangle \xrightarrow{TOFFOLI} |001\rangle ;$$
$$|010\rangle \xrightarrow{TOFFOLI} |010\rangle ; \ |011\rangle \xrightarrow{TOFFOLI} |011\rangle$$

$$|100\rangle \xrightarrow{TOFFOLI} |100\rangle ; \ |101\rangle \xrightarrow{TOFFOLI} |101\rangle ;$$
$$|110\rangle \xrightarrow{TOFFOLI} |111\rangle ; \ |111\rangle \xrightarrow{TOFFOLI} |110\rangle$$

Can we say that $\bar{C}$, $F^{red}$ and $T^{off}$ are just *c*- $r^{vers}$ -$g^{at}$ ‘s? Well, yes, we can! They are also *q*- $g^{at}$ ‘s since the $t^{rans}$‘s they $p^{erf}$ (i.e., $p^{erm}$ ‘s of $C^{omp}$ -$b^{as}$ -$s^{ta}$) are also U. $C^{trol}$ ed *q*- $g^{at}$ ‘s can be far more sophisticated than $C^{trol}$ ‘ed *c*- $g^{at}$ ‘s. For instance, the natural *q*- $g^{ner}$ ‘ation of the $C^{trol}$ ‘ed-NOT $g^{at}$ is the $C^{trol}$ ‘ed-*U*- $g^{at}$:

$$\text{controlled-}U \equiv \begin{pmatrix} 1 & 0 & 0 & 0 \\ 0 & 1 & 0 & 0 \\ 0 & 0 & U_{11} & U_{12} \\ 0 & 0 & U_{21} & U_{22} \end{pmatrix}$$

where $U = \begin{pmatrix} U_{11} & U_{12} \\ U_{21} & U_{22} \end{pmatrix}$ is an arbitrary 1- *qb*- $g^{at}$. If we are using $\bar{C}$, $F^{red}$ or $T^{off}$ -$g^{at}$ ‘s within *c*- $r^{vers}$ -$C^{omp}$ their $i^{np}$’s are only *c*-bits. So, we can read each $C^{trol}$ bit to find what $a^{ct}$ to $p^{erf}$ on the $t^{rgt}$ -bit. But in the context of *q*- $C^{omp}$, where these $g^{at}$ ‘s may need to act on arbitrary $S^{po}$ -$s^{ta}$, the question is whether it makes sense to speak of “$C^{trol}$ ‘ed” $g^{at}$ ‘s since, in the *q*- case, the act of reading the $C^{trol}$ -*qb* will, in *G*, perturb it.

It turned out that it is not necessary to read $C^{trol}$ *-bits* during the $a^{pp}$ of a $C^{trol}$ 'ed *q- g^{at}*! Instead, if a $C^{trol}$ 'ed *q-* $g^{at}$ $a^{ct}$ 's on a $S^{po}$ -$s^{ta}$ *all* of the $C^{trol}$ -$a^{ct}$ 's are $p^{erf}$ 'ed in parallel to a $d^{gre}$ commensurate with the $a^{mpl}$ of the $c^{resp}$ 'ing $C^{trol}$ *qb* $e^{ig}$ - $s^{ta}$ within the $i^{np}$- $S^{po}$ -$s^{ta}$.

---

*EXAMPLE*: If $A$ and $B$ are a pair of U- $m^{tri}$*'s -*$c^{resp}$ 'ing to arbitrary 1- *qb -q-* $g^{at}$ 's. Then the $g^{at}$ defined by their $d^{ir}$ sum:

$$A \oplus B = \begin{pmatrix} A & 0 \\ 0 & B \end{pmatrix} = \begin{pmatrix} A_{11} & A_{12} & 0 & 0 \\ A_{21} & A_{22} & 0 & 0 \\ 0 & 0 & B_{11} & B_{12} \\ 0 & 0 & B_{21} & B_{22} \end{pmatrix}$$

$p^{erf}$ 's a "$C^{trol}$ 'ed" -$o^{per}$ 'ion as follows. If the first *qb* is in $s^{ta}$ -$|0\rangle$ then the $o^{per}$ 'ion $A$ is $a^{pp}$ 'ed to the second *qb*. As a consequence, if the first *qb* is in $s^{ta}$ -$|1\rangle$ then the $o^{per}$ 'ion $B$ is $a^{pp}$ 'ed to the second *qb*. And if the $C^{trol}$ '*qb* is some $S^{po}$ of $|0\rangle$ and $|1\rangle$ then both $C^{trol}$ -$a^{ct}$ 's are $p^{erf}$ 'ed to some $d^{gre}$. The *q-* $c^{irc}$ for such a $g^{at}$ is shown in Fig. 16

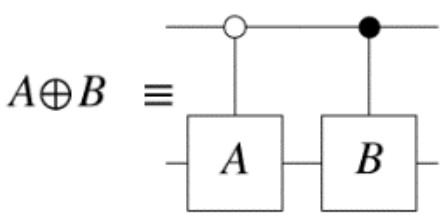

*Fig. 16 The q-* $c^{irc}$ *-*$c^{resp}$ *'ing to a* $g^{at}$ *that* $p^{erf}$ *'s different* $C^{trol}$ *-*$a^{ct}$ *'s according to whether the top qb is* $|0\rangle$ *or* $|1\rangle$

---

Here we note that if the first *qb* is in $s^{ta}$ -$|0\rangle$ the $i^{np}$ $s^{ta}$ can be written as $|0\rangle(a|0\rangle + b|1\rangle)$ , and if the first *qb* is in $s^{ta}$ -$|1\rangle$ we write the $i^{np}$ as $|1\rangle(a|0\rangle + b|1\rangle)$ . Therefore, for the first case, when the $g^{at}$ -$a^{ct}$'s, we get:

$$(A \oplus B)(|0\rangle \otimes (a|0\rangle + b|1\rangle)) =$$

$$\begin{pmatrix} A_{11} & A_{12} & 0 & 0 \\ A_{21} & A_{22} & 0 & 0 \\ 0 & 0 & B_{11} & B_{12} \\ 0 & 0 & B_{21} & B_{22} \end{pmatrix} \cdot \begin{pmatrix} a \\ b \\ 0 \\ 0 \end{pmatrix} = \begin{pmatrix} aA_{11} + bA_{12} \\ aA_{21} + bA_{22} \\ 0 \\ 0 \end{pmatrix}$$

$$= (aB_{11} + bB_{12})|10\rangle + (aB_{21} + bB_{22})|11\rangle$$

$$= |1\rangle \otimes B(a|0\rangle + b|1\rangle)$$

Similarly, in the second case, when the $g^{at}$ $a^{ct}$'s on an $i^{np}$ of the form $|1\rangle \otimes (a|0\rangle + b|1\rangle)$ we get:

$$(A \oplus B)(|1\rangle \otimes (a|0\rangle + b|1\rangle)) =$$

$$\begin{pmatrix} A_{11} & A_{12} & 0 & 0 \\ A_{21} & A_{22} & 0 & 0 \\ 0 & 0 & B_{11} & B_{12} \\ 0 & 0 & B_{21} & B_{22} \end{pmatrix} \cdot \begin{pmatrix} 0 \\ 0 \\ a \\ b \end{pmatrix} = \begin{pmatrix} 0 \\ 0 \\ aB_{11} + bB_{12} \\ aB_{21} + bB_{22} \end{pmatrix}$$

$$= (aB_{11} + bB_{12})|10\rangle + (aB_{21} + bB_{22})|11\rangle$$

$$= |1\rangle \otimes B(a|0\rangle + b|1\rangle)$$

*EXAMPLE:* ___

So, when the 2- *qb* $C^{trol}$ 'ed $g^{at}$ $(A \oplus B)$ acts on a $G$ two- *qb* -$S^{po}$ -$s^{ta}$ $|\psi\rangle = a|00\rangle + b|01\rangle + c|10\rangle + d|11\rangle$ the $C^{trol}$ -*qb* is no longer $p^{ur}$ 'ly $|0\rangle$ or $p^{ur}$ 'ly $|1\rangle$. Here, due to the $l^{ine}$ 'ity of *q-* mechanics the $c^{or}$ -$C^{trol}$ -$a^{ct}$ 's are $p^{erf}$ 'ed, in the $c^{or}$ proportions, on the $t^{rgt}$ -*qb*.

$$(A \oplus B)|\psi\rangle = |0\rangle \otimes A(a|0\rangle + b|1\rangle) + |1\rangle \otimes B(c|0\rangle + d|1\rangle)$$

$C^{trol}$ 'ed $g^{at}$ 's can be $g^{ner}$ 'ed to have $m^{lti}$ 'ple $C^{trol}$ 's as shown in Fig. 17

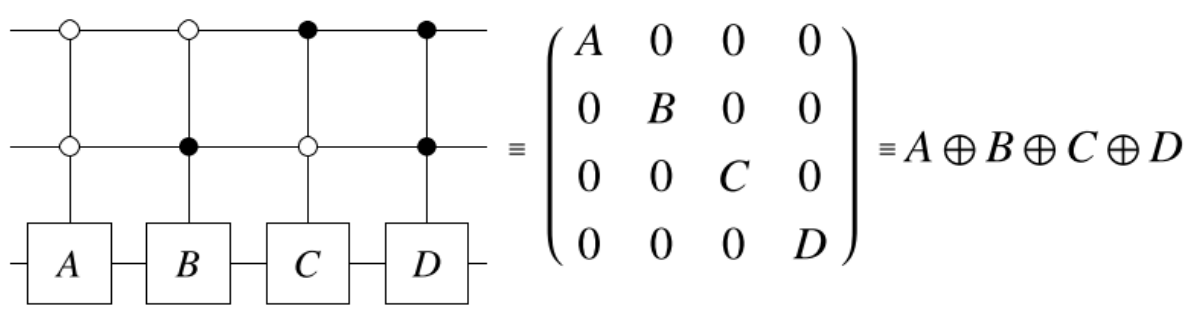

*Fig. 17 The q-* $c^{irc}$ *-*$c^{resp}$ *'ing to a* $g^{at}$ *that* $p^{erf}$ *'s different* $C^{trol}$ *-*$a^{ct}$ *'s according to whether the top two qb 's are* $|00\rangle$, $|01\rangle$, $|10\rangle$, *or* $|1\rangle$

---

Here a distinct $o^{per}$ 'ion is $p^{erf}$ 'ed on the third *qb* conditioned on the $s^{ta}$ of the top two *qb* 's. Such $m^{lti}$ 'ply-$C^{trol}$ 'ed *q-* $g^{at}$ 's are often used in practical *q-* $c^{irc}$ 's. It should be noted, that the $n^{ber}$ of distinct $s^{ta}$ of the $C^{trol}$ 's grows *exp* 'ly with the $n^{ber}$ of $C^{trol}$ 's. Therefore, it is more difficult to build $m^{lti}$ 'ply-$C^{trol}$ 'ed -$g^{at}$ 's beyond just a few $C^{trol}$ -*qb* 's.

$c^{irc}$ *for* $C^{trol}$ *'ed-U*: $I^{dep}$ of when *qb* 's should be read, we need to know how to $d^{comp}$ these $C^{trol}$ 'ed $g^{at}$ 's into a less complex set of standard $g^{at}$ s. Factoring a $C^{trol}$ 'ed $g^{at}$ as in $A \oplus B = (1 \otimes A) \cdot (1 \otimes A^{-1} \cdot B)$ where $1 = \begin{pmatrix} 1 0 \\ 0 1 \end{pmatrix}$, it can be seen that the core "$C^{trol}$ 'ed" component of the $g^{at}$ is really a $g^{at}$ of the form:

$$\text{controlled-}U \equiv$$

$$\begin{pmatrix} 1 & 0 & 0 & 0 \\ 0 & 1 & 0 & 0 \\ 0 & 0 & U_{11} & U_{12} \\ 0 & 0 & U_{21} & U_{22} \end{pmatrix}$$

with $U_{ij}$ being the entries of an arbitrary 1- *qb*- $g^{at}$ $U = A^{-1} \cdot B$.

We refer to a 2- *qb*- $g^{at}$ of the form $\begin{pmatrix} 1 & \hat{0} \\ \hat{0} & U \end{pmatrix}$ as a $C^{trol}$ 'ed-*U*- $g^{at}$.

A *q-* $c^{irc}$ for a 2- *qb*- $C^{trol}$ 'ed-*U* $g^{at}$ can be built in $t^{rm}$ 's of $\bar{C}$- $g^{at}$ 's and 1- *qb*- $g^{at}$ 's as follows. For an arbitrary 1- *qb*- $g^{at}$ -*U* with a single *qb* ($P^{au}$) $d^{comp}$ of the form $U = e^{ia} R_z(b) \cdot R_y(c) \cdot R_z(d)$ , the $a^{ct}$ of the $C^{trol}$ 'ed-*U* -$g^{at}$ is not to $a^{ct}$ on the $t^{rgt}$ -*qb* when the $C^{trol}$ -*qb* is $|0\rangle$ and to $a^{pp}$ -*U* to the $t^{rgt}$ -*qb* when the $C^{trol}$ -*qb* is $|1\rangle$. The result of "doing nothing" is the same as $a^{pp}$'ing the $i^{d}$ -$g^{at}$ to the $t^{rgt}$. So given the *q-* $c^{irc}$ -$d^{comp}$ for $C^{omp}$ $U$, what is a *q-* $c^{irc}$ that $C^{omp}$ -$C^{trol}$ 'ed-*U*?

By definition, there exist angles $a$, $b$, $c$, and $d$ such that:

$$U = e^{ia} R_z(b) \cdot R_y(c) \cdot R_z(d)$$

Given these angles and using notation $\xi = (d - b)/2$ and $\zeta = (d + b)/2$ , define $m^{tri}$*'s* $A$, $B$, $C$ as follows:

$$A = R_z(\xi) \;\; ; B = R_y(-c/2) \cdot R_z(-\zeta) \;\; ; C = R_z(b) \cdot R_y(c/2) \, ;$$
$$\Delta = \mathrm{diag}\big(1, e^{ia}\big)$$

*EXAMPLE:* ___

The $c^{irc}$ shown in Fig. 18 $C^{omp}$ -$C^{trol}$ 'ed-$U$. When the $C^{trol}$ -$qb$ is in $s^{ta}$ -$|0\rangle$ the $\Delta$ -$g^{at}$ does not alter it since $\Delta|0\rangle = |0\rangle$ (with no $p^{has}$ addition).

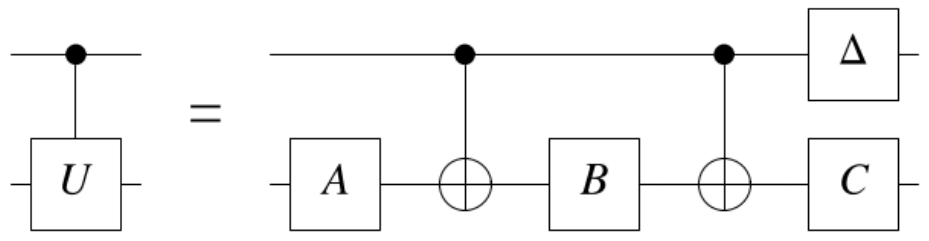

*Fig. 18 A q- $c^{irc}$ for a $C^{trol}$ 'ed-U $g^{at}$, where U is an arbitrary 1- qb- $g^{at}$*

The $C^{trol}$ -$qb$ -s of the $\bar{C}$- $g^{at}$ 's are also $|0\rangle$ and so the $\bar{C}$'s do not $a^{ct}$ on the $t^{rgt}$ -$qb$. So, when the $C^{trol}$ -$qb$ in the $c^{irc}$ is $|0\rangle$ the $t^{rans}$ of the $t^{rgt}$ -$qb$ will be defined by $C \cdot B \cdot A$. Here, if the $A$ -$g^{at}$ $a^{ct}$ 's first, then the $B$ -$g^{at}$, and then the $C$ -$g^{at}$, the $m^{tri}$'s should be processed in the $o^{rde}$ $C \cdot B \cdot A$ since when this object $a^{ct}$'s in an $i^{np}$- $s^{ta}$ $|\psi\rangle$ we want the $\mathcal{G}r$ 'ing to be $\Big(C \cdot \big(B \cdot (A|\psi\rangle)\big)\Big)$ ($g^{at}$ -$A$ first then $g^{at}$ -$B$ then $g^{at}$ -$C$). It can be shown that the overall result of these three $o^{per}$ 'ions is the $i^{d}$ .

$$C \cdot B \cdot A \equiv R_z(b) \cdot R_y(c/2) \cdot R_y(-c/2) \cdot R_z(-\zeta) \cdot R_z(\xi)$$
$$= \begin{pmatrix} 1 & 0 \\ 0 & 1 \end{pmatrix}$$

___

In the sequel, we check up the situation when the $C^{trol}$ -$qb$ is in $s^{ta}$ -$|1\rangle$. Now the $C^{trol}$ -$qb$ first acts on a $p^{has}$ component because $\Delta|1\rangle = e^{ia}|1\rangle$. All $C^{trol}$ -$qb$ 's of the $\bar{C}$- $g^{at}$ 's will be set to $|1\rangle$, and so they will $a^{pp}$ a NOT -$g^{at}$ (implemented as $P^{au}$ -$X$ -$g^{at}$) to the $t^{rgt}$ -$qb$ when the $\bar{C}$- $g^{at}$ -$a^{ct}$'s. So, the $t^{rans}$ to which the $t^{rgt}$ -$qb$ will be subject when the $C^{trol}$ -$qb$ is $|1\rangle$ is $e^{ia}C \cdot X \cdot B \cdot X \cdot A$. Here, we should notice that $X \cdot R_y(\theta) \cdot X \equiv R_y(-\theta)$ and $X \cdot R_z(\theta) \cdot X \equiv R_z(-\theta)$ giving:

$$C \cdot X \cdot B \cdot X \cdot A = R_z(b) \cdot R_y(c/2) \cdot X \cdot$$
$$\cdot R_y(-c/2) \cdot R_z(-\zeta) \cdot X \cdot R_z(\xi)$$
$$= R_z(b) \cdot R_y(c/2) \cdot X \cdot R_y(-c/2) \cdot X \cdot X \cdot R_z(-\zeta) \cdot X \cdot R_z(\xi)$$
$$= R_z(b) \cdot R_y(c/2) \cdot X \cdot R_y(-c/2) \cdot X \cdot X \cdot R_z(-\zeta) \cdot X \cdot R_z(\xi)$$
$$= R_z(b) \cdot R_y(c/2) \cdot R_y(c/2) \cdot R_z(\zeta) \cdot R_z(\xi)$$
$$= R_z(b) \cdot R_y(c) \cdot R_z(d)$$

Hence the $c^{irc}$ for $C^{trol}$ 'ed-$U$ =$U^{c}$ , $p^{erf}$ 's as follows:

$$U^{c}|0\rangle(a|0\rangle + b|1\rangle) =$$
$$= |0\rangle \otimes C \cdot B \cdot A(a|0\rangle + b|1\rangle) =$$
$$|0\rangle \otimes (a|0\rangle + b|1\rangle)$$

$$U^{c}|1\rangle(a|0\rangle + b|1\rangle) =$$
$$= e^{ia}|1\rangle \otimes C \cdot X \cdot B \cdot X \cdot A(a|0\rangle + b|1\rangle)$$
$$= |1\rangle \otimes e^{ia}C \cdot X \cdot B \cdot X \cdot A(a|0\rangle + b|1\rangle)$$
$$= |1\rangle \otimes U(a|0\rangle + b|1\rangle)$$

Thus $U$ is $a^{pp}$ 'ed to the $t^{rgt}$ -$qb$ iff the $C^{trol}$ -$qb$ is $|1\rangle$.

*Flipping the $C^{trol}$ and $t^{rgt}$ -$qb$ 's*: It is not necessary that the $C^{trol}$ -$qb$ is the topmost $qb$ in a $q$- $c^{irc}$. With notation $U^{sw}$ = $s^{wa}$, an *upside down* $U^{c}$ -$g^{at}$ ($U^{udc}$) would be given by $U^{udc}$ = $U^{sw}$ $\cdot U^{c}\cdot$ $U^{sw}$ as shown in Fig.19.

$$U^{udc} = U^{sw} \cdot U^{c} \cdot U^{sw} = \begin{pmatrix} 1 & 0 & 0 & 0 \\ 0 & U_{11} & 0 & U_{12} \\ 0 & 0 & 1 & 0 \\ 0 & U_{21} & 0 & U_{22} \end{pmatrix}$$

The second $qb$ is now the $C^{trol}$ -$qb$ and the first $qb$ the $t^{rgt}$ -$qb$. The result is the $m^{tri}$ -$c^{resp}$ 'ing to a 2- $qb$- $C^{trol}$ 'ed -$q$- $g^{at}$ inserted into a $c^{irc}$ "upside down".

*EXAMPLE:* ___

*Control- on-* $|0\rangle$ $q$- $g^{at}$'s: In a $C^{trol}$ 'ed $q$- $g^{at}$ it is not necessary that the value that controls $a^{pp}$ of a special $a^{ct}$ is $|1\rangle$. A 2- $qb$- $q$- $g^{at}$ with the $a^{ct}$ dependent on the first $qb$ being $|0\rangle$ is related to the usual $C^{trol}$ 'ed $g^{at}$ as follows:

$$controlled[1]\text{-}U = U^{c[1]} = \begin{pmatrix} 1 & 0 & 0 & 0 \\ 0 & 1 & 0 & 0 \\ 0 & 0 & U_{11} & U_{12} \\ 0 & 0 & U_{21} & U_{22} \end{pmatrix}$$

$$C^{trol}\ \text{'ed}[0]\text{-}U = U^{c[0]} = (\mathrm{NOT} \otimes 1_2) \cdot U^{c[1]} \cdot (\mathrm{NOT} \otimes 1_2)$$
$$= \begin{pmatrix} U_{11} & U_{12} & 0 & 0 \\ U_{21} & U_{22} & 0 & 0 \\ 0 & 0 & 1 & 0 \\ 0 & 0 & 0 & 1 \end{pmatrix}$$

as illustrated in Fig. 19.

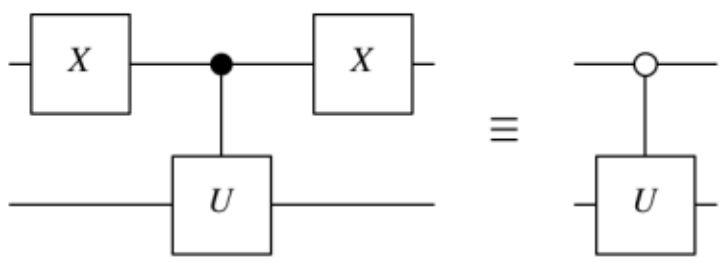

*Fig. 19 A q- $c^{irc}$ for a $C^{trol}$ 'ed q- $g^{at}$ that acts when its $C^{trol}$ -qb is in $s^{ta}$ -$|0\rangle$ (as indicated by the open circle on the $C^{trol}$ -qb) rather than $s^{ta}$ -$|1\rangle$*

___

*$c^{irc}$ for $C^{trol}$ 'ed- $C^{trol}$ 'ed-U ($U^{cc}$)*: We can continue in a similar way by, e.g., using $m^{lti}$ 'ple $C^{trol}$ -$qb$ 's and/or $t^{rgt}$ -$qb$ 's.

___

*EXAMPLE*: We have interpreted earlier the $T^{off}$ -$g^{at}$ as a $C^{trol}$ 'ed- $\bar{C}$ -$g^{at}$ which is suggesting to consider a $C^{trol}$ 'ed- $C^{trol}$ 'ed-$U$ ($U^{cc}$) -$g^{at}$, where $U$ is an arbitrary 1- *qb*- $g^{at}$. As a $m^{tri}$, the $C^{trol}$ ‘ed-$C^{trol}$ ‘ed-$U$- $g^{at}$ has the form:

$$U^{cc} \equiv \begin{pmatrix} 1 & 0 & 0 & 0 & 0 & 0 & 0 & 0 \\ 0 & 1 & 0 & 0 & 0 & 0 & 0 & 0 \\ 0 & 0 & 1 & 0 & 0 & 0 & 0 & 0 \\ 0 & 0 & 0 & 1 & 0 & 0 & 0 & 0 \\ 0 & 0 & 0 & 0 & 1 & 0 & 0 & 0 \\ 0 & 0 & 0 & 0 & 0 & 1 & 0 & 0 \\ 0 & 0 & 0 & 0 & 0 & 0 & U_{11} & U_{12} \\ 0 & 0 & 0 & 0 & 0 & 0 & U_{21} & U_{22} \end{pmatrix}$$

We can $d^{comp}$ a $U^{cc}$ -$g^{at}$ into a $c^{irc}$ constructed from only $\bar{C}$- $g^{at}$ ‘s and 1- *qb*- $g^{at}$ -s as shown in Fig. 20 where $V = U^{1/2}$. The $o^{per}$ ‘ion of this $c^{irc}$ can be explained by analyzing what it does to the eight possible $C^{omp}$ -$b^{as}$ -$s^{ta}$ of a 3- *qb*- $s^{yst}$.

$$|000\rangle \xrightarrow{U^{cc}} |000\rangle\,;\;\; |001\rangle \xrightarrow{U^{cc}} |001\rangle$$

$$|010\rangle \xrightarrow{U^{cc}} |01\rangle \otimes \left(V^{\dagger}\cdot V|0\rangle\right) = |010\rangle\,;\;\; |011\rangle \xrightarrow{U^{cc}} |01\rangle \otimes \left(V^{\dagger}\cdot V|1\rangle\right) = |011\rangle$$

$$|100\rangle \xrightarrow{U^{cc}} |10\rangle \otimes \left(V\cdot V^{\dagger}|0\rangle\right) = |100\rangle\,;\;\; |101\rangle \xrightarrow{U^{cc}} |10\rangle \otimes \left(V\cdot V^{\dagger}|1\rangle\right) = |101\rangle$$

$$|110\rangle \xrightarrow{U^{cc}} |11\rangle \otimes V^2|0\rangle = |11\rangle \otimes U|0\rangle\,;\; |111\rangle \xrightarrow{U^{cc}} |11\rangle \otimes V^2|1\rangle = |11\rangle \otimes U|1\rangle \text{ (since } V^2 = U)$$

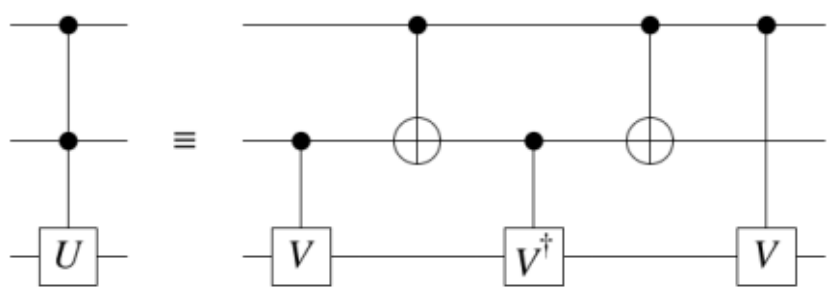

*Fig. 20 Q- $c^{irc}$ for the $U^{cc}$ -$o^{per}$ ‘ion. Here $V$ is any q- $g^{at}$ such that $V^2 = U$*

---

*Universal Q-$g^{at}$‘s:* A set of $g^{at}$ ‘s, $\mathcal{S}$, is *u*- if an arbitrary $C^{omp}$ can be done in a $c^{irc}$ composed only of $g^{at}$ ‘s from $\mathcal{S}$. The special *u*-sets of $g^{at}$ ‘s are those composed of a single $g^{at}$. The NAND $g^{at}$, the NOR $g^{at}$, and the NMAJORITY=$U^{nm}$ $g^{at}$, are all, *u*- for *c*-ir’$r^{vers}$ -$C^{omp}$. Similarly, the $T^{off}$ =$U^{t}$ and $F^{red}$ =$U^{f}$ $g^{at}$ ‘s are each *u*- for *c*- $r^{vers}$ -$C^{omp}$. What about *u*- $g^{at}$ ‘s for *q*- $C^{omp}$? How many *qb* ‘s does the *smallest u*- *q*- $g^{at}$ must have?

The fact that the closest *c*- $g^{at}$ ‘s to the *q*- $g^{at}$ ‘s are the *c*- $r^{vers}$ -$g^{at}$ ‘s, and these need at least 3 bits to be *u*-, suggesting the smallest *u*- *q*- $g^{at}$ will be a 3- *qb*- $g^{at}$ too. Well, it is true, that there is a 3- *qb*- $g^{at}$ that is *u*- for *q*- $C^{omp}$, known as a DEUTSCH=$U^{d}$ -$g^{at}$, and an arbitrary *q*- $C^{omp}$ can be achieved in a $c^{irc}$ built only from $U^{d}$ -$g^{at}$‘s $a^{ct}$’ing on various triplets of *qb* ‘s. The $g^{at}$ can be represented as:

$$U^{d} = \begin{pmatrix} 1 & 0 & 0 & 0 & 0 & 0 & 0 & 0 \\ 0 & 1 & 0 & 0 & 0 & 0 & 0 & 0 \\ 0 & 0 & 1 & 0 & 0 & 0 & 0 & 0 \\ 0 & 0 & 0 & 1 & 0 & 0 & 0 & 0 \\ 0 & 0 & 0 & 0 & 1 & 0 & 0 & 0 \\ 0 & 0 & 0 & 0 & 0 & 1 & 0 & 0 \\ 0 & 0 & 0 & 0 & 0 & 0 & i\cos(\theta) & \sin(\theta) \\ 0 & 0 & 0 & 0 & 0 & 0 & \sin(\theta) & i\cos(\theta) \end{pmatrix}$$

where $\theta$ is an arbirtary constant angle such that $2\theta/\pi$ is an irrational $n^{ber}$. However, $c^{irc}$ ‘s for an arbitrary $2^n \times 2^n$ U- $m^{tri}$ constructed from this $g^{at}$ are typically very in’ $e^{ffi}$ in $g^{at}$ count.

$U^{d}$ -$g^{at}$ is not the minimum option. D. DiVincenzo and J.Smolin proved that $U^{d}$ -$g^{at}$ could be constructed by using only 2- *qb*- $g^{at}$ ‘s , and A. Barenco by a single type of 2- *qb*-$g^{at}$ -the BARENCO= $U^{b}$ -$g^{at}$, which can be expressed as:

$$U^{b} = \begin{pmatrix} 1 & 0 & 0 & 0 \\ 0 & 1 & 0 & 0 \\ 0 & 0 & e^{i\alpha}\cos(\theta) & -ie^{i(\alpha-\phi)}\sin(\theta) \\ 0 & 0 & -ie^{i(\alpha+\phi)}\sin(\theta) & e^{i\alpha}\cos(\theta) \end{pmatrix}$$

where $\phi$, $\alpha$ and $\theta$ are fixed irrational $m^{lti}$ ‘ples of $\pi$ and each other. Therefore, *q*- $g^{at}$ ‘s are different from *c*- $g^{at}$ ‘s in $t^{rm}$ ‘s of *u*’ity. While in *c*- $r^{vers}$ -$C^{omp}$ there is no 2-bit $g^{at}$ that is both $r^{vers}$ and *u*-, in *q*- $C^{omp}$ *almost all* 2- *qb*- $g^{at}$ ‘s are *u*-. This suggests that certain *c*- $r^{vers}$ -$C^{omp}$ (which are described by $p^{erm}$ -$m^{tri}$’*s* and are, therefore, U) can potentially be $\mathcal{J}^{mpl}$ ‘ed more $e^{ffi}$ ‘tly using *q*- $g^{at}$ ‘s than using only *c*- $r^{vers}$ -$g^{at}$ ‘s. This suggests that one of the nearest $t^{rm}$ large scale $a^{pp}$ ‘s of *q*- $g^{at}$ ‘s migh be in $\mathcal{J}^{mpl}$ of “*c*-” $r^{vers}$ -$C^{omp}$ for fast, low -$p^{wer}$, $r^{vers}$ micro-$p^{ro}$ ‘ors. The main reason to study *u*- $g^{at}$ ‘s is to make $e^{xp}$’*s* simpler. If all *q*- $C^{omp}$ can be made from the same type of $g^{at}$, then an $e^{xp}$ ‘ist need only focus on how to design that $g^{at}$ in $o^{rde}$ to make any *q*- $C^{omp}$ possible. On the other hand, it is difficult to use in practice the $U^{b}$ -$g^{at}$ as a primitive $g^{at}$ as it needs a 2- *qb* -$\mathcal{H}$ having 3 “tunable” $p^{met}$ ‘s, $\phi$, $\alpha$ and $\theta$.

But the $U^{b}$ -$g^{at}$ is clearly a $C^{trol}$ ‘ed-$U$ -$g^{at}$ and can be further $d^{comp}$ ‘ed into a $s^{equ}$ of 1- *qb*- $g^{at}$ ‘s and one 2- *qb*- $g^{at}$ like $\bar{C}$. So, the collection of $g^{at}$ ‘s $\mathcal{S} = \{R_x(\alpha), R_y(\beta), R_z(\gamma), Ph(\delta)\,, \bar{C}\}$ must be a *u*- set of $g^{at}$ ‘s for *q*- $C^{omp}$. Actually, the set of all 1- *qb*- $g^{at}$ ‘s and $\bar{C}$ is the preferred set of $g^{at}$ ‘s used in building *q*- $c^{irc}$ ‘s. Other *u*- $g^{at}$ collections are also available, see Table 2, that involve only fixed-angle $g^{at}$ ‘s, although, they do not necessary lead to $e^{ffi}$ -*q*- $c^{irc}$ ‘s since they need to repeat fixed angle $r^{ot}$ ‘s a number of times to $a^{prox}$ a specific 1- *qb*- $g^{at}$ to required accuracy. However, even if a given set of $g^{at}$ ‘s is *u*-, in practice, certain $C^{omp}$ can be done more $e^{ffi}$ ‘tly if an “over-complete” family of *u*- $g^{at}$ ‘s is deployed.

*Selected 2- qb- $g^{at}$ ‘s -$l^{ibr}$’s*

Using the collection of all 1- *qb*- $g^{at}$ ‘s and $\bar{C}$ as the *u*- $l^{ibr}$ of $g^{at}$ ‘s, is not necessarily the optimum option for all types of *q*- $C^{omp}$ hardware ($h^{rdw}$). Different variants of *q*- $C^{omp}$ -$h^{rdw}$ are used with different $\mathcal{H}$'s. So, while a $\bar{C}$ -$g^{at}$ for example may be easy to obtain in one implementation, it might not be easy in another.

*Table2: u- $g^{at}$ Families*

| |
|---|
| $\{R_x, R_y, R_z, Ph, \text{CNOT}\}$ |
| $\text{BARENCO}(\phi, \alpha, \theta)$ |
| $\{H, S, T, \text{CNOT}\}$ where $H = \frac{1}{\sqrt{2}}\begin{pmatrix}1 & 1\\ 1 & -1\end{pmatrix}$ is the Walsh-Hadamard gate, $S = \begin{pmatrix}1 & 0\\ 0 & i\end{pmatrix}$ is the "phase gate", and $T = \begin{pmatrix}1 & 0\\ 0 & \exp(i\pi/4)\end{pmatrix}$ is the "$\pi/8$ gate" |

For this reason, in the sequel several different $l^{ibr}$ 's of 1- *qb* and 2- *qb*- $g^{at}$ 's are described which are more "natural" for a given types of *q*- $C^{omp}$ -$h^{rdw}$. The rules are provided for inter-changing between these types of 2- *qb*- $g^{at}$ 's so that designers can look at a *q*- $c^{irc}$ expressed using one $g^{at}$ -$l^{ibr}$ and $\mathcal{M}^{\leftrightharpoons}$ it into another, perhaps easier to attain, $l^{ibr}$.

$$\text{CSIGN} = \begin{pmatrix}1&0&0&0\\0&1&0&0\\0&0&1&0\\0&0&0&-1\end{pmatrix},$$

$$\text{SWAP}^{\alpha} = \begin{pmatrix}1&0&0&0\\0&\frac{1}{2}(1+e^{i\pi\alpha})&\frac{1}{2}(1-e^{i\pi\alpha})&0\\0&\frac{1}{2}(1-e^{i\pi\alpha})&\frac{1}{2}(1+e^{i\pi\alpha})&0\\0&0&0&1\end{pmatrix}$$

$$\text{iSWAP} = \begin{pmatrix}1&0&0&0\\0&0&i&0\\0&i&0&0\\0&0&0&1\end{pmatrix},$$

$$B = \begin{pmatrix}\cos(\frac{\pi}{8})&0&0&i\sin(\frac{\pi}{8})\\0&\cos(\frac{3\pi}{8})&i\sin(\frac{3\pi}{8})&0\\0&i\sin(\frac{3\pi}{8})&\cos(\frac{3\pi}{8})&0\\i\sin(\frac{\pi}{8})&0&0&\cos(\frac{\pi}{8})\end{pmatrix}$$

*CSIGN, $SWAP^{\alpha}$, iSWAP, Berkeley B*: The $p^{hys}$ -$i^{ntac}$ 's available within different types of *q*- $C^{omp}$ -$h^{rdw}$ leads to a number of "natural" 2- *qb*- $g^{at}$ 's such as is$^{wa}$ , SWAP$^{\alpha}$, CSIGN etc. They are easier to implement than $\bar{C}$ in the given $p^{hys}$ setup, and if $m^{ax}$ 'ly $\mathcal{E}$'ing, provide no less $e^{ffi}$ -$d^{comp}$ 's of arbitrary 2- *qb* -$o^{per}$ 'ions. The four alternatives to $\bar{C}$ are shown below.

---

*EXAMPLE:* For the $g^{at}$ 's shown in Fig. 21, CSIGN are used in $l^{ine}$ -$o^{pti}$ 'al -Q- $C^{omp}$ (LOQC), is$^{wa}$ in $s^{con}$- *q*- $C^{omp}$ via $\mathcal{H}$'s -$\mathcal{J}^{mpl}$ 'ing the so-called XY $m^{del}$, $\sqrt{\text{SWAP}}$ in $s^{pin}$ -tronic *q*- $C^{omp}$ as that approach uses the "exchange $i^{ntac}$" and SWAP$^{\alpha}$ in $s^{pin}$ -tronic *q*- $C^{omp}$. The duration of the exchange $o^{per}$ 'ion defines $\alpha$ in SWAP$^{\alpha}$. For the Berkeley B $g^{at}$ -$\mathcal{H}$ is $\mathcal{H} = \frac{\pi}{8}(2X \otimes X + Y \otimes Y)$ and $g^{at}$ is $U = \exp(i\mathcal{H})$ .

---

$$\text{iSWAP} = \begin{pmatrix}1&0&0&0\\0&0&i&0\\0&i&0&0\\0&0&0&1\end{pmatrix} \quad \sqrt{\text{SWAP}} = \begin{pmatrix}1&0&0&0\\0&\frac{1}{2}+\frac{i}{2}&\frac{1}{2}-\frac{i}{2}&0\\0&\frac{1}{2}-\frac{i}{2}&\frac{1}{2}+\frac{i}{2}&0\\0&0&0&1\end{pmatrix} \quad \text{SWAP}^{\alpha} = \begin{pmatrix}1&0&0&0\\0&\frac{1}{2}(1+e^{i\pi\alpha})&\frac{1}{2}(1-e^{i\pi\alpha})&0\\0&\frac{1}{2}(1-e^{i\pi\alpha})&\frac{1}{2}(1+e^{i\pi\alpha})&0\\0&0&0&1\end{pmatrix}$$

The Berkeley B Gate: Hamiltonian is $\mathcal{H} = \frac{\pi}{8}(2X \otimes X + Y \otimes Y)$. Gate is $U = \exp(i\mathcal{H})$.

$$B = e^{i\frac{\pi}{8}(2X\otimes X + Y\otimes Y)} = \begin{pmatrix}\cos(\frac{\pi}{8})&0&0&i\sin(\frac{\pi}{8})\\0&\cos(\frac{3\pi}{8})&i\sin(\frac{3\pi}{8})&0\\0&i\sin(\frac{3\pi}{8})&\cos(\frac{3\pi}{8})&0\\i\sin(\frac{\pi}{8})&0&0&\cos(\frac{\pi}{8})\end{pmatrix} = \frac{\sqrt{2-\sqrt{2}}}{2}\begin{pmatrix}1+\sqrt{2}&0&0&i\\0&1&i(1+\sqrt{2})&0\\0&i(1+\sqrt{2})&1&0\\i&0&0&1+\sqrt{2}\end{pmatrix}$$

*Fig. 21Matrix representation of the four $g^{at}$ 's*

*Interrelationships Between the Members of 2- qb- $g^{at}$ 's -$l^{ibr}$*: In $e^{xp}$ 'al *q*- $C^{omp}$ one has to work with the $p^{hys}$ -$i^{ntac}$ 's imposed by Nature. In *G*, we would not expect that the most accessible and $C^{trol}$ 'lable $p^{hys}$ -$i^{ntac}$ 's permit a *q*- $m^{ech}$ -$e^{vlt}$ that can be interpreted as a $\bar{C}$- $g^{at}$.

However, from the $\mathcal{H}$'s available in different types of $p^{hys}$ -$s^{yst}$ 's one can always find 2- *qb*- $g^{at}$ 's from which we can, in conjunction with 1- *qb*- $g^{at}$ 's, build $\bar{C}$- $g^{at}$ 's. In the sequel we show how to build $\bar{C}$- $g^{at}$ 's out of the kinds of 2-body $i^{ntac}$ 's that are commonly available in real $p^{hys}$ -$s^{yst}$ 's.

---

*EXAMPLE*: Getting $\bar{C}$ *from CSIGN* is shown in Fig. 22, *from* $\sqrt{SWAP}$ in Fig. 23, *from is$^{wa}$ and one s$^{wa}$ in Fig. 24 and from two is$^{wa}$ 's in Fig. 25*.

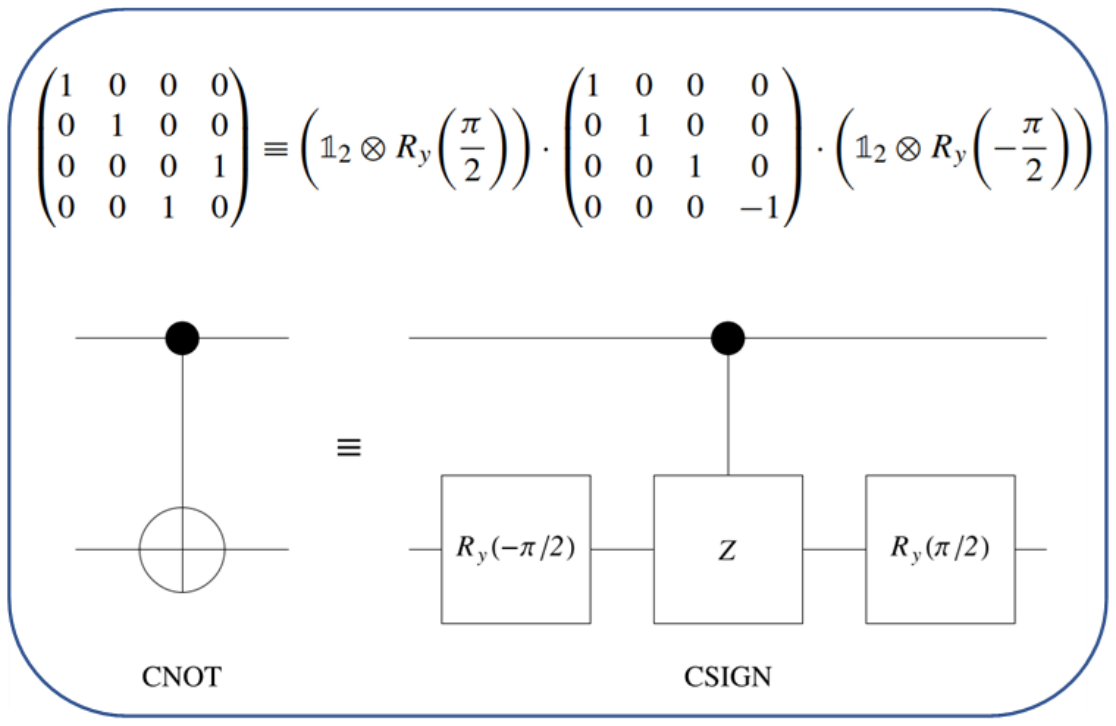

*Fig. 22 Q- $c^{irc}$ for getting a $\bar{C}$- $g^{at}$ given the ability to achieve 1- qb- $g^{at}$ 's and CSIGN*

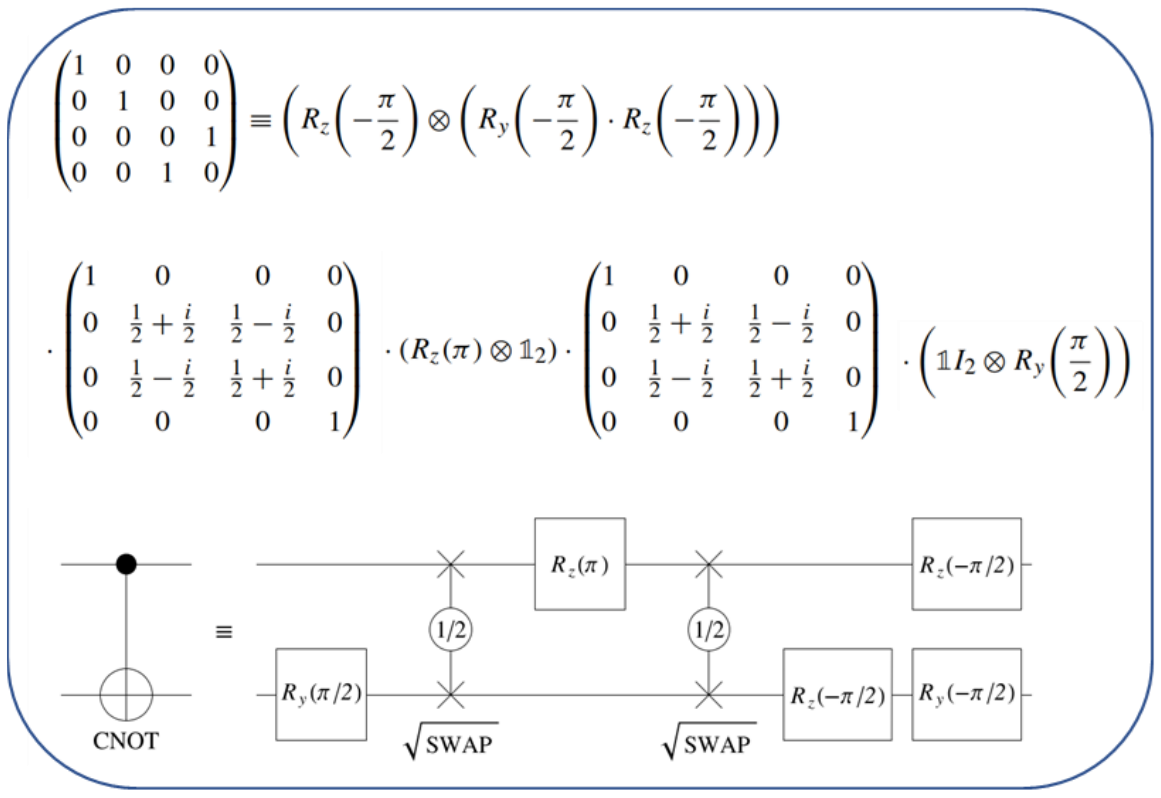

*Fig. 23 Q- $c^{irc}$ for getting a $\bar{C}$- $g^{at}$ given the ability to achieve 1- qb- $g^{at}$ 's and $\sqrt{SWAP}$*

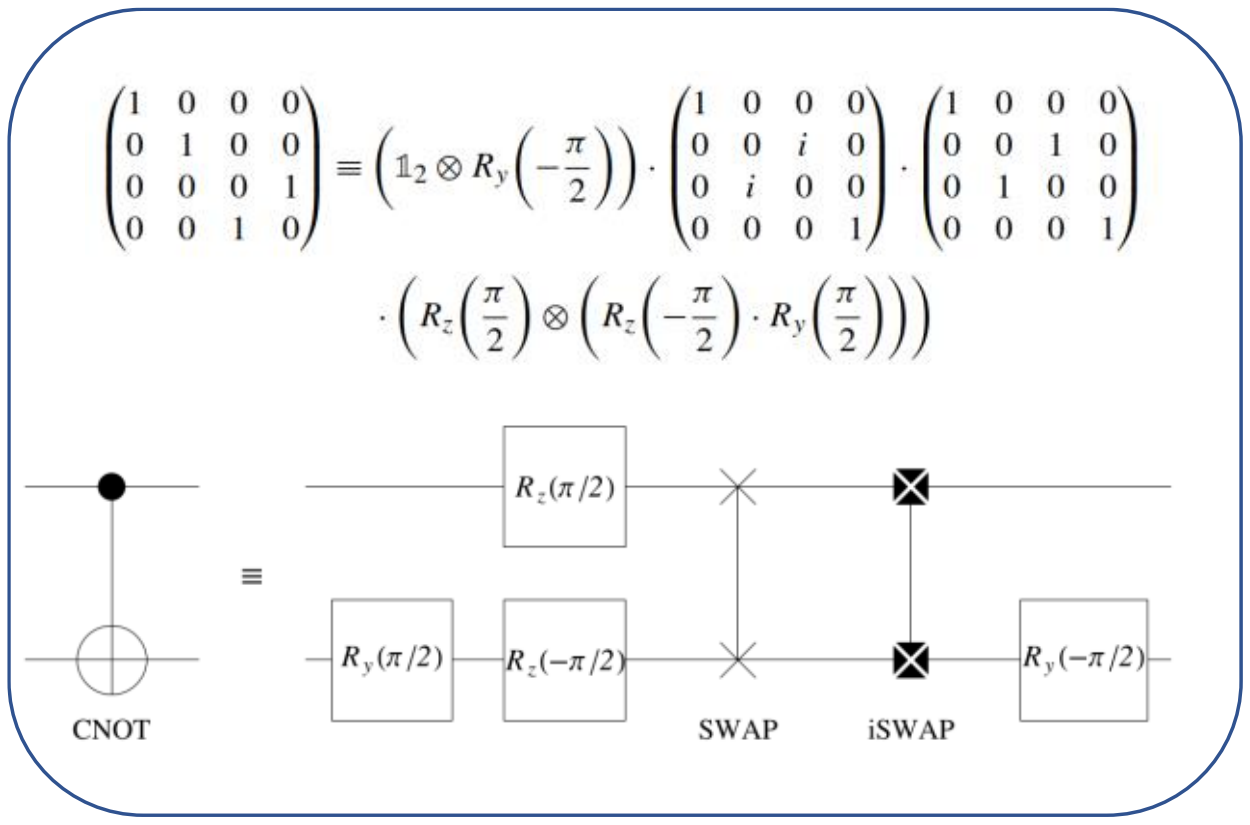

*Fig. 24 Q- $c^{irc}$ for getting a $\bar{C}$- $g^{at}$ given the ability to achieve 1- qb- $g^{at}$ 's, $is^{wa}$, and $s^{wa}$*

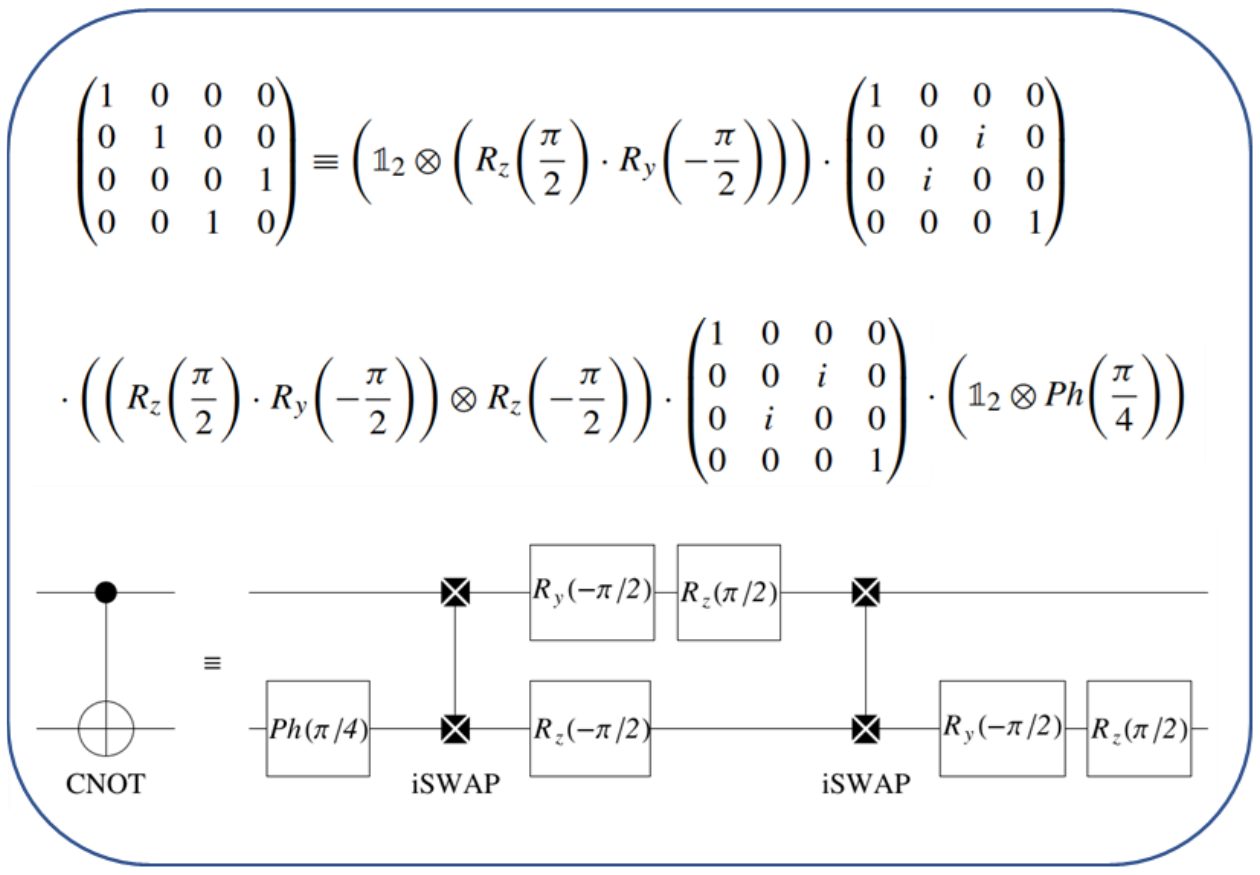

*Fig. 25 Q- $c^{irc}$ for getting a $\bar{C}$- $g^{at}$ given the ability to achieve 1- qb- $g^{at}$ 's and $is^{wa}$*

*$\mathcal{E}$'ing $p^{wer}$ of Q- $g^{at}$ 's*

A set $S$ of $qb$'s is *$\mathcal{E}$'ed* if the $o^{per}$ 'ions $p^{erf}$ 'ed on one subset ($S_\wedge$) of $qb$'s have impact on the remaining $S_\wedge$ of $qb$'s, even though those $qb$'s are not $o^{per}$ 'ed upon $d^{ir}$ 'ly. For instance, suppose a set of $n$ $qb$'s $\mathcal{S}$ is partitioned into two $S_\wedge$'s $\mathcal{A} \subset \mathcal{S}$ and $\mathcal{B} = \mathcal{S} \setminus \mathcal{A}$. If $o^{per}$ 'ions $p^{erf}$ 'ed on the $qb$'s in $\mathcal{A}$ have impact on the $s^{ta}$ of the $qb$ 's in $\mathcal{B}$ then the $qb$ 's in $\mathcal{A}$ and those in $\mathcal{B}$ are entangled. If so, the $s^{ta}$ of $\mathcal{S}$ cannot be written as the $d^{ir}$ -$p^{duc}$ of $s^{ta}$'s for $\mathcal{A}$ and $\mathcal{B}$. Such $\mathcal{E}$ is unmediated and undiminished by distance and gives rise to so-called "non-local" effects.

The main difference between $q$- $lG$ and $c$- $lG$ is that $q$- $lG$ can cause the $qb$ 's upon which they act to become more or less $\mathcal{E}$'ed, whereas $c$- $g^{at}$ 's cannot. Actually, $\mathcal{E}$ is not present in $c$- $C^{omp}$ at all and $c$- $g^{at}$ 's can neither $\mathcal{E}$ nor dis-$\mathcal{E}$ the bits upon which they act. Thus, $\mathcal{E}$ is an unique $q$- resource that is only available to $q$- $C^{omp}$. As a result, $\mathcal{E}$ is essential in achieving the *exp* speedups seen in $q$- $a^{lg}$ 's without other $C^{omp}$ -$r^{eso}$ 's, such as space ($m^{emo}$), time and $e^{nrgy}$, increasing *exp* 'ly.

It is important to know if all 2- $qb$- $g^{at}$ 's are equally good at generating $\mathcal{E}$ ? So far we know that some 2- $qb$- $g^{at}$ 's, such as those built as the $d^{ir}$ -$p^{duc}$ of two 1- $qb$- $g^{at}$ 's, cannot $g^{ner}$ any $\mathcal{E}$. But $g^{at}$ 's, such as $\bar{C}$, seem able to $\mathcal{M}^{\leftrightarrows}$ un'$\mathcal{E}$'ed $i^{np}$'s into maximally ($m^{ax}$'*ly*) $\mathcal{E}$'ed - $o^{ut}$'s. So, there are different options how the 2- $qb$- $g^{at}$ 's can $g^{ner}$ -$\mathcal{E}$. Here, we need to quantify the $d^{gre}$ of $\mathcal{E}$ within a $s^{ta}$ (*$\mathcal{E}$- measure*), and we should define an ensemble of $i^{np}$- $s^{ta}$ over which we would like to average this $\mathcal{E}$-measure. If we take a set of $i^{nit}$ 'ly un-$\mathcal{E}$'ed $i^{np}$'s, i.e., $p^{duc}$ -$s^{ta}$, then we want to quantify how successful a given $g^{at}$ is at producing $\mathcal{E}$ i.e. how $\mathcal{E}$'ed, on average, its $o^{ut}$'s will be upon acting on $i^{nit}$ 'ly un-$\mathcal{E}$'ed $i^{np}$'s. This brings us to the notion of the "$\mathcal{E}$'ing $p^{wer}$" of a $q$- $g^{at}$. We would expect, the more the $o^{ut}$ is $\mathcal{E}$'ed, the higher the $\mathcal{E}$'ing $p^{wer}$ of the $g^{at}$.

"*Tangle*" *as a measure of the $\mathcal{E}$ within a $s^{ta}$*: There are different ways to characterize the $d^{gre}$ of $\mathcal{E}$ within a 2- $qb$- $q$- $s^{ta}$. For 2- $qb$- $s^{ta}$, all the different $\mathcal{E}$ measures are equivalent to one another. This is not the case for $\mathcal{E}$ measures of $n$- $qb$- $s^{ta}$. Since here we are interested only in the $\mathcal{E}$'ling $p^{wer}$ of 2- $qb$- $g^{at}$'s, any of the equivalent 2- $qb$- $\mathcal{E}$ measures can be used.

---

*EXAMPLE*: We define, the *tangle as the square of the concurrence* ($c^{crc}$). For a 2- $qb$ -$p^{ur}$ -$s^{ta}$, $|\psi\rangle$, the $c^{crc}$ is defined as: $c^{crc}(|\psi\rangle) = \mathcal{C}(|\psi\rangle) = |\langle\psi|\tilde{\psi}\rangle|$ with $|\tilde{\psi}\rangle$ being the $s^{pin}$ -flipped version of $|\psi\rangle$. This is given by $|\tilde{\psi}\rangle = (Y \otimes Y)|\psi^*\rangle$, with $Y$ being the $P^{au}$ -$Y$ -$m^{tri}$, and $|\psi^*\rangle$ is $|\psi\rangle$ with its amplitudes ($a^{mpl}$'*s*)

---

$c^{plex}$ conjugated. Thus, if $|\psi\rangle = a|00\rangle + b|01\rangle + c|10\rangle + d|11\rangle$, then $|\psi^*\rangle = a^*|00\rangle + b^*|01\rangle + c^*|10\rangle + d^*|11\rangle$ and $|\tilde{\psi}\rangle = -d^*|00\rangle + c^*|01\rangle + b^*|10\rangle - a^*|11\rangle$. So, the $c^{crc}$ of a general (G) 2- $qb$- $s^{ta}$ $|\psi\rangle$ is given by: $\mathcal{C}(a|00\rangle + b|01\rangle + c|10\rangle + d|11\rangle) = |2b^*c^* - 2a^*d^*|$ .

The "$s^{pin}$ -flip" $t^{rans}$ $\mathcal{M}^{\leftrightharpoons}$'s the $s^{ta}$ of an individual $qb$ into its orthogonal ($o^{rtg}$) $s^{ta}$. So, the $t^{rans}$ is not U and cannot, be $p^{erf}$ 'ed formally by any individual $q$- $s^{yst}$. Therefore, there is no an ideal $s^{pin}$ -flip "$g^{at}$" as such. (If there were it would be a $u$- NOT $g^{at}$.) Still, this $t^{rans}$ is a valid analytical description of a $t^{rans}$. One of the $p^{pert}$ of the $t^{rans}$ is that, if the 2- $qb$- $s^{ta}$ $|\psi\rangle$ is a $p^{duc}$ -$s^{ta}$ (i.e., an un-$\mathcal{E}$'ed- $s^{ta}$) its $|\tilde{\psi}\rangle$, will be $o^{rtg}$ to $|\psi\rangle$. So, the overlap $\langle\psi|\tilde{\psi}\rangle$ will be zero and therefore the $c^{crc}$ of $s^{ta}$ $|\psi\rangle$ will be zero. So un-$\mathcal{E}$'ed- $s^{ta}$ have a $c^{crc}$ of zero.

---

On the other hand, for this $t^{rans}$-$m^{ax}$ 'ly -$\mathcal{E}$'ed -$s^{ta}$, such as $B^{ell}$ -$s^{ta}$, remain unchanged except for an overall $p^{has}$. With, the four $B^{ell}$ -$s^{ta}$

$$|\beta_{00}\rangle = \frac{1}{\sqrt{2}}(|00\rangle + |11\rangle);\ |\beta_{01}\rangle = \frac{1}{\sqrt{2}}(|01\rangle + |10\rangle)$$

$$|\beta_{10}\rangle = \frac{1}{\sqrt{2}}(|00\rangle - |11\rangle);\ |\beta_{11}\rangle = \frac{1}{\sqrt{2}}(|01\rangle - |10\rangle)$$

the $s^{pin}$ -flip $t^{rans}$ gives:

$$|\beta_{00}\rangle \overset{sf}{\to} -|\beta_{00}\rangle;\ |\beta_{01}\rangle \overset{sf}{\to} |\beta_{01}\rangle;\ |\beta_{10}\rangle \overset{sf}{\to} |\beta_{10}\rangle;\ |\beta_{11}\rangle \overset{sf}{\to} -|\beta_{11}\rangle$$

where "*sf*" stands for $s^{pin}$ *-flip*. So, the overlap between a $m^{ax}$ 'ly -$\mathcal{E}$'ed -$s^{ta}$ and its $s^{pin}$ -flipped version is unity, which is the $m^{ax}$ possible value, implying that $m^{ax}$ 'ly- $\mathcal{E}$'ed- $s^{ta}$ have a unit $c^{crc}$. So, the tangle, as defined earlier, is a quantitative measure for the $d^{gre}$ of $\mathcal{E}$ within a $p^{ur}$- 2- $qb$- $s^{ta}$.

"*$\mathcal{E}$'ing $p^{wer}$" as the Mean Tangle $g^{ner}$'ed by a $g^{at}$*: We can now quantify the $d^{gre}$ of $\mathcal{E}$ which different $g^{at}$ 's -$g^{ner}$ when acting upon $i^{nit}$ 'ly un'$\mathcal{E}$'ed -$i^{np}$'s. For this we can define the *$\mathcal{E}$'ing $p^{wer}$* of a 2- $qb$- $g^{at}$ -$U$, $\mathrm{EP}(U)$ , as the average tangle that $U$ $g^{ner}$ 's averaged over all $i^{np}$- $p^{duc}$ -$s^{ta}$ -$i^{np}$'s sampled uniformly on the Bloch sphere, defined as:

$$\mathrm{EP}(U) = \langle E(U|\psi_1\rangle \otimes |\psi_2\rangle)\rangle_{|\psi_1\rangle,|\psi_2\rangle}$$

Here $E(\cdot)$ is the tangle of any other 2- $qb$- $\mathcal{E}$ measure such as the $l^{ine}$ entropy (since all the 2- $qb$- $\mathcal{E}$ measures are equivalent), and $|\psi_1\rangle$ and $|\psi_2\rangle$ are single $qb$- $s^{ta}$ sampled uniformly on the Bloch sphere.

*$\bar{C}$ from any $m^{ax}$'ly -$\mathcal{E}$'ing -$g^{at}$*: In $e^{xp}$ 'al $q$- $C^{omp}$, we need to find a way to get a $\bar{C}$- $g^{at}$ from whatever $p^{hys}$'ly feasible 2- $qb$ -$i^{ntac}$, is available. How easy we can get a $\bar{C}$ from the $p^{hys}$ 'ly available 2- $qb$- $g^{at}$, $U$, depends on the $\mathcal{E}$'ing $p^{wer}$ of $U$. If $\mathrm{EP}(U) = \frac{2}{9}$ , i.e., $m^{ax}$, but $U$ itself is not a $\bar{C}$- $g^{at}$, then we can get a $\bar{C}$- $g^{at}$ by $a^{pp}$'s of $U$ via a $d^{comp}$ of the form: $\bar{C} \equiv (A_1 \otimes A_2) \cdot U \cdot (H \otimes 1_2) \cdot U$ where $H$ is the $\mathcal{H}$- $g^{at}$ and $A_1$ and $A_2$ are 1- $qb$- $g^{at}$'s. This is a useful result to designers since sometimes it is not possible to get a $\bar{C}$ -$g^{at}$ -$d^{ir}$ 'ly from $i^{ntac}$ -$\mathcal{H}$ within some $p^{hys}$ context. However, once it is understood how a $m^{ax}$ 'ly $\mathcal{E}$'ing -$o^{per}$ 'ion can be obtained from the available $i^{ntac}$ -$\mathcal{H}$'s, then we can use the above result, in conjunction with 1- $qb$- $g^{at}$ 's, to achieve a $\bar{C}$.

*The Magic $b^{as}$ and Its Effect on $\mathcal{E}$'ing $p^{wer}$*: We already know that, a $q$- $g^{at}$ with U- $m^{tri}$ -$U$ in the $C^{omp}$ -$b^{as}$ can be seen as the $m^{tri}$ $V \cdot U \cdot V^\dagger$ in the "V- $b^{as}$". For 2- $qb$- $g^{at}$ 's a specific $b^{as}$, referred to as the *magic* $b^{as}$, has several useful $p^{pert}$. The "magic $b^{as}$" is a collection of 2- $qb$- $s^{ta}$'s that are $p^{has}$ shifted versions of the $B^{ell}$ $s^{ta}$ defined as

$$|00\rangle \overset{\mathcal{M}}{\to} |\mathcal{M}_1\rangle = |\beta_{00}\rangle\ ;\ |01\rangle \overset{\mathcal{M}}{\to} |\mathcal{M}_2\rangle = i|\beta_{10}\rangle$$

$$|10\rangle \overset{\mathcal{M}}{\to} |\mathcal{M}_3\rangle = i|\beta_{01}\rangle\ ;\ |11\rangle \overset{\mathcal{M}}{\to} |\mathcal{M}_4\rangle = |\beta_{11}\rangle$$

where $|\beta_{00}\rangle$, $|\beta_{01}\rangle$, $|\beta_{10}\rangle$, and $|\beta_{11}\rangle$ are the $B^{ell}$ -$s^{ta}$ defined by:

$$|\beta_{00}\rangle = \frac{1}{\sqrt{2}}(|00\rangle + |11\rangle);\ |\beta_{01}\rangle = \frac{1}{\sqrt{2}}(|01\rangle + |10\rangle)$$

$$|\beta_{10}\rangle = \frac{1}{\sqrt{2}}(|00\rangle - |11\rangle);\ |\beta_{11}\rangle = \frac{1}{\sqrt{2}}(|01\rangle - |10\rangle)$$

So, the $m^{tri}$, $\mathcal{M}$, which $\mathcal{M}^{\leftrightharpoons}$'s the $C^{omp}$ -$b^{as}$ into the "magic" $b^{as}$ is:

$$\mathcal{M} = |\mathcal{M}_1\rangle\langle 00| + |\mathcal{M}_2\rangle\langle 01| + |\mathcal{M}_3\rangle\langle 10| + |\mathcal{M}_4\rangle\langle 11|$$

$$= \frac{1}{\sqrt{2}}\begin{pmatrix} 1 & i & 0 & 0 \\ 0 & 0 & i & 1 \\ 0 & 0 & i & -1 \\ 1 & -i & 0 & 0 \end{pmatrix}$$

This $b^{as}$ is known as the "magic $b^{as}$" since any partially or $m^{ax}$ 'ly $\mathcal{E}$'ing 2- $qb$- $g^{at}$, defined by a $p^{ur}$ *'ly real* U- $m^{tri}$, $U$, becomes a *non-$\mathcal{E}$'ing* $g^{at}$ in the "magic" $b^{as}$, i.e., regardless of how $\mathcal{E}$'ing $U$ may be, $\mathcal{M} \cdot U \cdot \mathcal{M}^\dagger$ is always a non-$\mathcal{E}$'ing -$g^{at}$, and therefore $\mathrm{EP}(\mathcal{M} \cdot U \cdot \mathcal{M}^\dagger) = 0$.

---

*EXAMPLE*: This fact is used to get a $c^{irc}$ for an arbitrary 2- $qb$- $g^{at}$ defined by a $p^{ur}$'ly real U- $m^{tri}$, $U$, since either $\mathcal{M} \cdot U \cdot \mathcal{M}^\dagger = A \otimes B$ (the first non-$\mathcal{E}$'ing -$c^{irc}$ option) or is related to a single $s^{wa}$ -$g^{at}$ (the second non-$\mathcal{E}$'ing -$g^{at}$ option). It can be easily seen which is the case. So, if we know the simplest $q$- $c^{irc}$ -$\mathcal{J}^{mpl}$ 'ing the magic $b^{as}$ -$t^{rans}$, we can then use one of the above two options to find a $c^{irc}$ for $U$. It is easy to find a $q$- $c^{irc}$ for the magic $b^{as}$ -$t^{rans}$. A simple $q$- $c^{irc}$-$\mathcal{J}^{mpl}$'ing the magic $b^{as}$-$g^{at}$ is shown in Fig.26.

---

The magic $b^{as}$ -$t^{rans}$ can be also used to relate a given $p^{ur}$'ly real U, via a manipulation involving $\mathcal{M}$, to $g^{at}$ that is certainly $m^{ax}$ 'ly $\mathcal{E}$'ing! For any $p^{ur}$ 'ly real $4 \times 4$ U- $m^{tri}$, $U$, then, $I^{dep}$ of its $\mathcal{E}$'ing $p^{wer}$, the $\mathcal{E}$'ing $p^{wer}$ of the $g^{at}$ defined by $\mathcal{M} \cdot U \cdot \mathcal{M}$ is $m^{ax}$, i.e., $\frac{2}{9}$ .

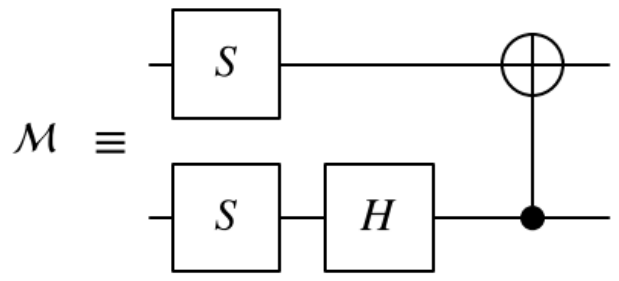

*Fig. 26 Q- $c^{irc}$ that $\mathcal{J}^{mpl}$ 's the magic $b^{as}$ -$t^{rans}$ , where $S = Ph(\pi/4) \cdot R_z(\pi/2)$ and $H = ZR_y(-\pi/2)$*

*B. Depth-$\mathcal{O}$- Q- $c^{irc}$ 's*

In the past a lot of effort has been invested into synthesizing $\mathcal{O}$ -$c^{irc}$ 's for c-, specifically, $r^{vers}$ -$B^{ool}$ - $F$ 's. The $s^{ynth}$ of 3-bit $r^{vers}$ -$l^{gic}$ -$c^{irc}$ 's using NOT, $\bar{C}$, and $T^{off}$ -$g^{at}$ 's, by generating $c^{irc}$ -$l^{ibr}$ 's, then iteratively searching through them is discussed in [139]. The $s^{ynth}$ of $\mathcal{O}$ -4-bit $r^{vers}$ -$c^{irc}$ 's $c^{omp}$ 'ed with NOT, $\bar{C}$, $T^{off}$, and the 4-bit $T^{off}$ -$g^{at}$ 's was considered in [140]. In their paper, very $e^{ffi}$ -$p^{erf}$ *is described*. While the speed of $c^{irc}$ from [140] is not matched in this work, [138] contends that $s^{ynth}$ of U- $c^{irc}$ 's is a more $C^{omp}$ intensive $p^{ro}$ than $r^{vers}$ -$c^{irc}$ 's, making $d^{ir}$ comparison difficult. Even so, [138] adapts many of the search $t^{chn}$ 's described in [140] to attain fast $p^{erf}$.

A $p^{blem}$ somewhat closer to that of *q- $c^{irc}$ -$s^{ynth}$* was discussed in [141]. In the paper a method for $C^{omp}$ -$\mathcal{O}$ cost $d^{comp}$ 's of $r^{vers}$ -$l^{gic}$ into NOT, $\bar{C}$, and the *q- $C^{trol}$'ed -$\sqrt{X}$- $g^{at}$ was developed*. By $a^{pp}$ -$t^{chn}$ 's from formal verification, minimum cost *q- $c^{irc}$ -$\mathcal{J}^{mpl}$* of various *lG*, including the $T^{off}$, $F^{red}$, and Peres $g^{at}$ 's was found. Here, a restricted *q- $c^{irc}$ -$m^{del}$* was used (for instance, $C^{trol}$ 's are required to remain $B^{ool}$ ), and only a finite $S^{\wedge}$ of *q- $c^{irc}$* 's on $n$- *qb* -s can be described using four valued -$l^{gic}$. The work presented in [138], $\mathcal{O}$ over all *q- $c^{irc}$* 's using any $g^{at}$ set, while at the same time generating $c^{irc}$ 's for all *q- $g^{at}$* 's, not just $B^{ool}$ - ones, and enabling $\mathcal{O}$ over other cost $F$ 's. The $s^{ynth}$ of 3-bit $c^{irc}$ 's over this $g^{at}$ set, was discussed also in [142], using a pruned breadth-first search similar to [143] rather than formal methods.

The studies of the $p^{blem}$ of *$\mathcal{O}$- q- $c^{irc}$ -$s^{ynth}$* are rather limited and mostly focused on finding $a^{prox}$ 's in small $s^{ta}$ spaces. Here we focus on finding exact $d^{comp}$ 's for various $l^{gic}$ 'al $g^{at}$ 's, although the $a^{lg}$ may be extended to find $a^{prox}$ -$g^{at}$ $s^{equ}$ 's as well. An $a^{lg}$ for $C^{omp}$ $\varepsilon$- $a^{prox}$ 's of 1- *qb*- $g^{at}$ 's in time $O(\log^{2.71}(1/\varepsilon))$ , along with a $g^{ner}$ 'ation to $m^{lti}$ - *qb* cases is discussed in [144]. The $a^{lg}$, quickly finds a logarithmic (in the precision) $d^{pth}$ -$\varepsilon$- $a^{prox}$ for a given U, though the $c^{irc}$ produced may be far from $\mathcal{O}$ [145-147]. It recursively $C^{omp}$ -$a^{prox}$ 's of U's, which in the base case reduces to searching through sets of previously $g^{ner}$ 'ed U's for a basic $a^{prox}$. The $a^{lg}$ finds a minimal $d^{pth}$ -$c^{irc}$.

A similar work, [148] describes an *exp* time $a^{lg}$ for finding $d^{pth}$ -$\mathcal{O}$- $\varepsilon$- $a^{prox}$ 's of single *qb*- $g^{at}$ 's. The $a^{lg}$ uses previously $C^{omp}$ knowledge of equivalent sub-$s^{equ}$ 's to remove entire $s^{equ}$ 's from the search. Authors in [149] developed a $d^{pth}$ -$\mathcal{O}$ form for 1- *qb*- $c^{irc}$ 's over the $g^{at}$ set $\{H, T\}$. They provide a speed up over brute force searching that finds $d^{pth}$ -$\mathcal{O}$- $\varepsilon$- $a^{prox}$ 's by searching through $d^{ta}$ - $b^{as}$ of canonical $c^{irc}$ 's. The work presented in [138] is somewhat similar, though *$a^{pp}$'s* their canonical form only to 1 -*qb*- $c^{irc}$ 's over $\{H, T\}$, while method of [138] $a^{pp}$ to $n$- *qb*- $c^{irc}$ 's over any $g^{at}$ set and requires significantly less RAM at the expense of slower searches. Authors in [138] focus on the $s^{ynth}$ of small $\mathcal{O}$- $m^{lti}$ - *qb*- $c^{irc}$ 's, that *$\mathcal{O}$- 1- qb -$s^{ynth}$ -$a^{lg}$* 's such as [149,147] are unable to deal with.

In the $c^{irc}$ -$m^{del}$ of *q- $C^{omp}$*, *wires* carry *qb's* to $g^{at}$ 's, which $t^{rans}$ their $s^{ta}$. Given that the $s^{ta}$ of a $s^{yst}$ of $n$ -*qb* 's is typically described as a $\mathcal{V}$ in a $\mathcal{D}^{2^n}$ -$c^{plex}$ -$\mathcal{V}$ space $\mathcal{H}$, *q- $g^{at}$* 's can be considered as $l^{ine}$ -$o^{per}$ 'ors on $\mathcal{H}$. We focus on U $o^{per}$ 'ors, i.e. $o^{per}$ 'ors $U$ such that $UU^\dagger = U^\dagger U = I$, where $U^\dagger$ stands for the conjugate-transpose of $U$ and $I$ for the $i^d$ - $o^{per}$ 'or. The $l^{ine}$ -$o^{per}$ 'or $p^{erf}$ 'ed by the $c^{irc}$ is then given as the sequential $c^{omp}$ 'ition of the individual $g^{at}$ 's within the $c^{irc}$, and it is easily verified that this $l^{ine}$ -$o^{per}$ 'or is itself U.

A given $g^{at}$ acts only on a $S^{\wedge}$ of the *qb* 's in a given $s^{yst}$. It is handy to define the U- $p^{erf}$ 'ed by the $g^{at}$ as the tensor $p^{duc}$ of the non-trivial U on a smaller $s^{ta}$ space $c^{resp}$ 'ing to the affected *qb* 's, and the $i^d$ - $o^{per}$ 'or on the remaining *qb* 's. This disciption of $g^{at}$ 's also displays the parallel nature of $c^{irc}$ 's: a sequential $c^{irc}$ -$c^{omp}$ 'ed of two $g^{at}$ 's $g_1$, $g_2$ represented by U's $(g_1 \otimes I)$ and $(I \otimes g_2)$ can be rewritten in parallel as $g_1 \otimes g_2$.

The $d^{pth}$ of a $c^{irc}$ , as the main $\mathcal{O}$ criteria for $c^{irc}$ 's used in [138], is defined as the length of any critical $\mathcal{P}^{th}$ through the $c^{irc}$.

*EXAMPLE:* __________________________________

Desciring a $c^{irc}$ as a $d^{ir}$ 'ed acyclic $g^{ph}$ with $n^{od}$ 's -$c^{resp}$ 'ing to the $c^{irc}$'s $g^{at}$ 's and edges $c^{resp}$ 'ing to $g^{at}$ $i^{np}$'s/ $o^{ut}$'s, a critical $\mathcal{P}^{th}$ is a $\mathcal{P}^{th}$ of $m^{ax}$ length from an $i^{np}$ of the $c^{irc}$ to an $o^{ut}$ (see Fig. 27).

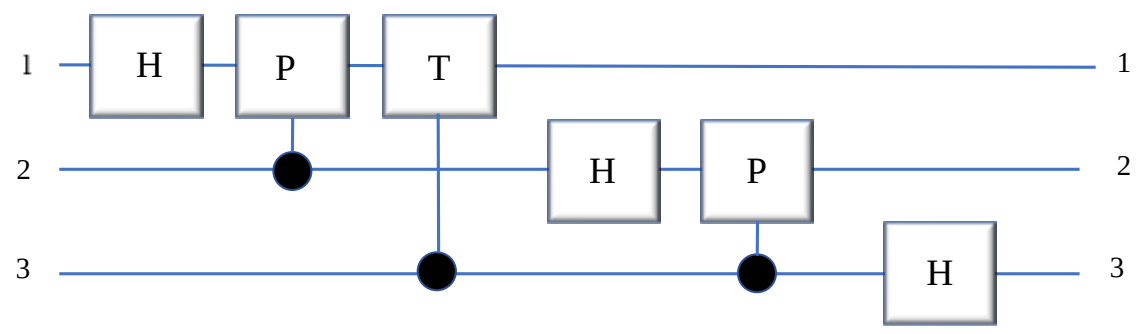

*Fig. 27A q- $c^{irc}$ -$p^{erf}$ 'ing the q- Fourier -$t^{rans}$, up to $p^{erm}$ of the $o^{ut}$'s. This $c^{irc}$ has $d^{pth}$ -5, with two critical $\mathcal{P}^{th}$ 's flowing from $i^{np}$- 1 to $o^{ut}$- 3.*

The $p^{blem}$ of $q$- $c^{irc}$ -$s^{ynth}$ consists of finding some $c^{irc}$ composed only of $g^{at}$ ‘s from a fixed set $p^{erf}$ ‘ing the required U. This set is called an *instruction set* ( $^{i}S$), and it must contain the inverse ($i^{nv}$) of each $g^{at}$ in the set. An $n$- $qb$- $c^{irc}$ over $^{i}S$- $\mathcal{G}$ is then the $c^{omp}$ ‘ition of individual $g^{at}$ ‘s $a^{pp}$ ‘ed to non-empty $S^{\wedge}$’s of $n$- $qb$ ‘s, tensored with the $i^{d}$ on the remaining $qb$ ‘s.

We may use $g^{at}$ ‘s from an $^{i}S$, each $a^{ct}$’ing on distinct $qb$ ‘s, to get $c^{irc}$ ‘s of $d^{pth}$ one over $n$ $qb$ ‘s. As such $c^{irc}$ ‘s will be integral to our $a^{lg}$, we define $\mathcal{V}_{n,\mathcal{G}}$, the set of all U’s- $c^{resp}$ ‘ing to $d^{pth}$ one $n$- $qb$- $c^{irc}$ ‘s over the $^{i}S$- $\mathcal{G}$. An $n$ $qb$- $c^{irc}$ $C$ over the $^{i}S$- $\mathcal{G}$ then has $d^{pth}$ at most $m$ if $C$ $c^{resp}$ ‘s to some $s^{equ}$ of U’s $U_1U_2\cdots U_m$ where $U_1$, $U_2$, $U_m \in \mathcal{V}_{n,\mathcal{G}}$. We say that $C$- $\mathcal{J}^{mpl}$ ‘s a U- $U \in U(2^n)$ if $U_1U_2\cdots U_m = U$.

In $G$, many distinct $c^{irc}$’s may $\mathcal{J}^{mpl}$ the same U. Often we don’t distinguish a $c^{irc}$ from the U it $\mathcal{J}^{mpl}$ ‘s. Stil here we point out that by “$c^{irc}$” we mean a given $s^{equ}$ of $g^{at}$ ‘s, rather than the resulting U -$t^{rans}$. As a consequence $c^{irc}$ ‘s are written in $t^{rm}$ ‘s of $o^{per}$ ‘or $c^{omp}$ ‘ition, so U’s are $a^{pp}$ ‘ed right to left — in a $c^{irc}$ diagram, while, $g^{at}$ ‘s are $a^{pp}$ ‘ed left to right.

In light of these definitions, authors in [138] describe an $a^{lg}$ that, given an $^{i}S$- $\mathcal{G}$ and U -$t^{rans}$ $U \in U(2^n)$ , determines whether $U$ can be $\mathcal{J}^{mpl}$ ‘ed by a $c^{irc}$ over $\mathcal{G}$ of $d^{pth}$ at most $l$ in time $O\left(|\mathcal{V}_{n,\mathcal{G}}|^{\lceil l/2\rceil}\log\left(|\mathcal{V}_{n,\mathcal{G}}|^{\lceil l/2\rceil}\right)\right)$ .

Furthermore, if $U$ can be $\mathcal{J}^{mpl}$ ‘ed with a $c^{irc}$ of $d^{pth}$ at most $l$, the $a^{lg}$ returns a $c^{irc}$ $\mathcal{J}^{mpl}$ ‘ing $U$ in $m^{ini}$ -$d^{pth}$ over $\mathcal{G}$. Ref. [138] also presents a C/C++ $\mathcal{J}^{mpl}$ of the $a^{lg}$ and describes better $c^{irc}$ ‘s, $g^{ner}$ ‘ed with this $\mathcal{J}^{mpl}$, than existing ones.

This $a^{lg}$ provides a $p^{erf}$ enhancement over the brute force $O\left(|\mathcal{V}_{n,\mathcal{G}}|^{l}\right)$ $a^{lg}$. By careful selection of $d^{ta}$ -structures running times close to $\Theta\left(|\mathcal{V}_{n,\mathcal{G}}|^{\lceil l/2\rceil}\right)$ are achievable. Even so, we should point out, that the runtime is still *exp*, since $|\mathcal{V}_{n,\mathcal{G}}| \geq k^n$ for any $^{i}S$- $\mathcal{G}$ with $k$ single $qb$- $g^{at}$ ‘s, making it only practical for small $n^{ber}$ ‘s of $qb$ ‘s.

Encouraged by results in fault tolerance, Ref. [4] uses the $^{i}S$ ($g^{at}$ -$l^{ibr}$) ($H$, $P$, $\bar{C}$, $T$, $P^\dagger$,$T^\dagger$) comprising the $\mathcal{H}$ -$g^{at}$ $H = \frac{1}{\sqrt{2}}\begin{pmatrix}1 & 1\\ 1 & -1\end{pmatrix}$, $p^{has}$ -$g^{at}$ $P = \begin{pmatrix}1 & 0\\ 0 & i\end{pmatrix}$, $C^{trol}$ ‘ed-NOT

$$CNOT = \begin{pmatrix}1 & 0 & 0 & 0\\ 0 & 1 & 0 & 0\\ 0 & 0 & 0 & 1\\ 0 & 0 & 1 & 0\end{pmatrix},$$

and $T = \begin{pmatrix}1 & 0\\ 0 & e^{\frac{i\pi}{4}}\end{pmatrix}$, along with $P^\dagger$ and $T^\dagger$. The set of $c^{irc}$ ‘s -$c^{omp}$ ‘ed from these $g^{at}$ ‘s forms a $S^{\wedge}$ of $2^n \times 2^n$ U- $m^{tri}$*’s* over the ring $\mathbb{Z}\left[\frac{1}{\sqrt{2}}, i\right]$ defined as

$$\mathbb{Z}\left[\frac{1}{\sqrt{2}}, i\right] = \left\{ \frac{a + be^{i\frac{\pi}{4}} + ce^{i\frac{\pi}{2}} + de^{i\frac{3\pi}{4}}}{\sqrt{2^n}} \;\middle|\; \begin{matrix} a, b, c, d, n \in \\ \mathbb{Z}, \\ n \geq 0 \end{matrix} \right\}.$$

Two important $\mathcal{G}r$ ‘s of $m^{tri}$*’s* over this ring are also identified: the $P^{au}$ -$\mathcal{G}r$ on $n$- $qb$ ‘s, $\mathcal{P}_n$ , defined as the set of all $n$-fold tensor $p^{duc}$ ‘s of the $P^{au}$ - $m^{tri}$*’s* $I = \begin{pmatrix}1 & 0\\ 0 & 1\end{pmatrix}$ , $X = \begin{pmatrix}0 & 1\\ 1 & 0\end{pmatrix}$ , $Y = \begin{pmatrix}0 & -i\\ i & 0\end{pmatrix}$ , $Z = \begin{pmatrix}1 & 0\\ 0 & -1\end{pmatrix}$, and the Clifford $\mathcal{G}r$ on $n$- $qb$ ‘s, $\mathcal{C}_n$, defined as the normalizer ($2\mathcal{C}_n = \{U \in U(2^n) | U\mathcal{P}_n U^{-1} \subseteq \mathcal{P}_n\}$) of $\mathcal{P}_n$ in $U(2^n)$ . U’s computable by $c^{irc}$ ‘s -$c^{omp}$ ‘ed with $I$, $X$, $Y$, $Z$, $H$, $P$, and $\bar{C}$ are elements of the Clifford $\mathcal{G}r$, while the $T$ -$g^{at}$ does not belong to the Clifford $\mathcal{G}r$.

$\mathcal{C}_n$ together with any one U- $U \notin \mathcal{C}_n$ forms a set dense in $U(2^n)$ [145]. Since $\{H, P, \bar{C}\}$ $g^{ner}$ ‘s the Clifford $\mathcal{G}r$ up to global $p^{has}$, the $^{i}S$ comprising $\{H, P, P^\dagger, \bar{C}, \mathrm{T}, T^\dagger\}$ is $u$- for $q$- $C^{omp}$.

*Meet-in-the-Middle (mm) Search Algorithm [138]*

Here we briefly describe the $a^{lg}$ used to $C^{omp}$ -$\mathcal{O}$- $c^{irc}$ ‘s. For more details on exact $\mathcal{J}^{mpl}$ of U and its $a^{prox}$ ‘ing $c^{irc}$ ‘s see [138]. The $a^{lg}$ enable us to search for $c^{irc}$ ‘s of $d^{pth}$ -$l$ by only generating $c^{irc}$ ‘s of $d^{pth}$ at most $\lceil l/2\rceil$, hence the name *mm* $a^{lg}$ [138].

*Lemma 1. [138] Let $S_i \subset U(2^n)$ be the set of all* U’s- $\mathcal{J}^{mpl}$ *‘able in $d^{pth}$ -i over the $g^{at}$ set $\mathcal{G}$. Given a* U *-$U$, there exists a $c^{irc}$ over $\mathcal{G}$ of $d^{pth}$ -l -$\mathcal{J}^{mpl}$ ‘ing U iff $S^\dagger_{\lceil l/2\rceil}U \cap S_{\lceil l/2\rceil} \neq \emptyset$.*

For the proof see [138]. Ref. [138] uses this lemma to design a simple $a^{lg}$ to find whether there exists a $c^{irc}$ over $\mathcal{G}$ of $d^{pth}$ at most $l$ $\mathcal{J}^{mpl}$ ‘ing U- $U$, and if so return a $m^{ini}$ -$d^{pth}$ -$c^{irc}$ -$\mathcal{J}^{mpl}$ ‘ing $U$.

*EXAMPLE*

Given an $^{i}S$- $\mathcal{G}$ and U- $U$, $c^{irc}$ ‘s of increasing $d^{pth}$ are repeatedly $g^{ner}$ ‘ed, then used to search for $c^{irc}$ ‘s $\mathcal{J}^{mpl}$ ‘ing $U$ with up to twice the $d^{pth}$ (Fig. 28).

```
function   mm-FACTOR(G, U, l)

  S_0 := {I}
  i := 1
  for i ≤ ⌈l/2⌉ do
     S_i := V_{n,G} S_{i-1}
     if S†_{i-1} U ∩ S_i ≠ ∅ then
        return any circuit VW s.t.
            V ∈ S_{i-1}, W ∈ S_i, V†U = W
     else if S†_i U ∩ S_i ≠ ∅ then
        return any circuit VW s.t.
            V, W ∈ S_i, V†U = W
     end if
     i := i + 1
  end for
end function
```

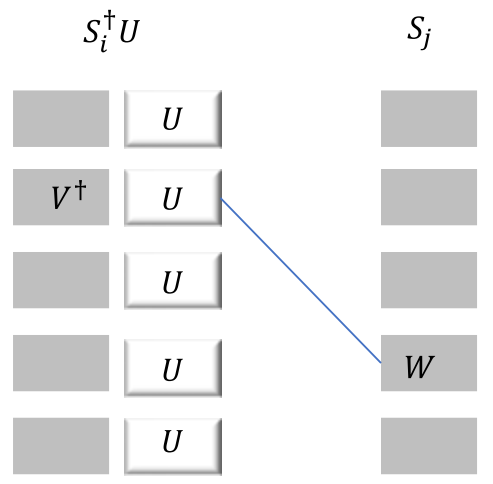

*Fig. 28: For each $V \in S_i$ we construct $W = V^{\dagger}U$ and $p^{erf}$ a logarithmic-time search for $W$ in $S_j$.*

At each round all $d^{pth}$ -$i$ -$c^{irc}$ ‘s -$S_i$ are $g^{ner}$ ‘ed by extending the $d^{pth}$ –$(i-1)$ -$c^{irc}$ ‘s with one more level of $d^{pth}$, then the sets $S_{i-1}^{\dagger}U$ and $S_i^{\dagger}U$ are $C^{omp}$ and checked if there are any collisions with $S_i$. By Lemma 1, there exists a $c^{irc}$ of $d^{pth}$ –$(2i-1)$ or $2i$ $\mathcal{J}^{mpl}$ ‘ing $U$ iff $S_{i-1}^{\dagger}U \cap S_i \neq \emptyset$ or $S_i^{\dagger}U \cap S_i \neq \emptyset$, respectively, so the $a^{lg}$ stops at the smallest $d^{pth}$ less than or equal to $l$ for which there exists a $c^{irc}$ $\mathcal{J}^{mpl}$ ‘ing $U$. If $U$ can be $\mathcal{J}^{mpl}$ ‘ed in $d^{pth}$ less than $l$, the $a^{lg}$ returns one such $c^{irc}$ of $m^{ini}$ -$d^{pth}$.

To observe the claimed $O\left(|\mathcal{V}_{n,\mathcal{G}}|^{\lceil l/2\rceil}\log\left(|\mathcal{V}_{n,\mathcal{G}}|^{\lceil l/2\rceil}\right)\right)$ runtime, a strict lexicographic $o^{rde}$ is imposed on U’s — as a simple example two U- $m^{tri}$’s can be $o^{rde}$ ‘ed according to the first element on which they differ. The set $S_i$ can then be sorted with respect to this $o^{rde}$ in $O(|S_i|\log(|S_i|))$ time, so that searching for each element of $S_{i-1}^{\dagger}U$ and $S_i^{\dagger}U$ in $S_i$ can be $p^{erf}$ ‘ed in time $O(|S_{i-1}|\log(|S_i|))$ and $O(|S_i|\log(|S_i|))$ , respectively. As $|S_i| \leq |\mathcal{V}_{n,\mathcal{G}}|^i$, the $i$-th iteration thus takes time with $b^{\wedge}$ $2|\mathcal{V}_{n,\mathcal{G}}|^i\log\left(|\mathcal{V}_{n,\mathcal{G}}|^i\right)$ .

$$\sum_{i=1}^{\lceil l/2\rceil}|\mathcal{V}_{n,\mathcal{G}}|^i\log\left(|\mathcal{V}_{n,\mathcal{G}}|^i\right) \leq \sum_{i=1}^{\lceil l/2\rceil}|\mathcal{V}_{n,\mathcal{G}}|^i\log\left(|\mathcal{V}_{n,\mathcal{G}}|^{\lceil l/2\rceil}\right) \text{ and}$$

$$\sum_{i=1}^{\lceil l/2\rceil}|\mathcal{V}_{n,\mathcal{G}}|^i \leq |\mathcal{V}_{n,\mathcal{G}}|^{\lceil l/2\rceil}\left(1+\frac{1}{|\mathcal{V}_{n,\mathcal{G}}|^{\lceil l/2\rceil}-1}\right),$$

Since we see that the $a^{lg}$ runs in $O\left(|\mathcal{V}_{n,\mathcal{G}}|^{\lceil l/2\rceil}\log\left(|\mathcal{V}_{n,\mathcal{G}}|^{\lceil l/2\rceil}\right)\right)$ time. It can also be noted that $\mathcal{V}_{n,\mathcal{G}} \in O(|\mathcal{G}|^n)$ , so the runtime is in $O\left(|\mathcal{G}|^{\lceil n\cdot l/2\rceil}\log\left(|\mathcal{G}|^{\lceil n\cdot l/2\rceil}\right)\right)$ .

---

*C. Exact $m^{ini}$ of Q- $c^{irc}$ ‘s*

Since non-permutative $q$- $g^{at}$ ‘s ($npqg$), such as $C^{trol}$ ‘ed-square-root-of-NOT $g^{at}$ ‘s ($CV = CV^{+}$ $g^{at}$ ‘s), have more $c^{plex}$ rules than permutative $q$- $g^{at}$ ‘s ($pqg$), it is very hard to synthesize them.

An $n$-letter $p^{erm}$ admits a $n \times n$ binary $m^{tri}$ format with only one entry of 1 in each row and each column and 0s elsewhere. “Magic” $p^{erm}$ -$m^{tri}$’s show one entry of 1 on their main diagonals. Some well-known $p^{erm}$ -$m^{tri}$’s / $g^{at}$ ‘s are the $P^{au}$ -$g^{at}$

$$X = \begin{pmatrix}01\\10\end{pmatrix} \equiv (2,1), I \otimes X \equiv (2,1)(4,3) ,$$

$$\bar{C} = \begin{pmatrix}1000\\0100\\0001\\0010\end{pmatrix} \equiv (1,2)(4,3) \;\; ; \;\; C\bar{C} \equiv (1,2,3,4,5,6)(8,7) ,$$

that $a^{ct}$ ‘s on one, two, or three $qb$ ‘s, respectively. The $p^{erm}$ -$g^{at}$ ‘s may $a^{ct}$ on qudits as the shif $g^{at}$

$$X = \begin{pmatrix}010\\001\\100\end{pmatrix} \equiv (2,3,1)$$

$a^{ct}$ ‘ing on qutrits.

In the $e^{ffi}$ -$s^{ynth}$ -$a^{lg}$, $d^{ir}$ use of $npqg$ should be avoided. Instead, the key method is to use $q$- $g^{at}$ ‘s to create new $pqg$ to replace $npqg$. This assumes the $l^{ibr}$ of $q$- $g^{at}$ primitives is constructed so as to have the lowest possible $q$- cost. Authors in [150] first discuss some new $CV = CV^{+}$-like $g^{at}$ ‘s, i.e. $C^{trol}$ ‘ed-$k$th-root-of-NOT $g^{at}$ ‘s where $k$ =2,4,8 . . ., and give all $c^{resp}$ ‘ing $m^{tri}$ ‘s. They also describe a generic method to quickly and $d^{ir}$ ‘ly construct this $\mathcal{O}$- $q$- $lG$ -$l^{ibr}$ using $\bar{C}$ and these $npqg$. The method gives new means to find $pqg$ with lower $q$- cost.

As already pointed out in the previous sections, $r^{vers}$ -$l^{gic}$ synthesis ($s^{ynth}$) has a major role in $q$- $\mathcal{J}$; small $n^{ber}$ ‘s of elementary $q$- $g^{at}$ ‘s, each with carefully $\mathcal{O}$’ed $q$- cost, can build an arbitrary $q$- $l^{gic}$ -$c^{irc}$. A $c^{irc}$ with lower $q$- cost is faster and cheaper, but also generate fewer errors ($\varepsilon$’s). The $s^{ynth}$ of $r^{vers}$ -$l^{gic}$ (permutative $q$-) $c^{irc}$ ‘s using elementary $q$- $g^{at}$ ‘s is different from $c$- (non- $r^{vers}$) $l^{gic}$ -$s^{ynth}$. In addition to the discussion presented in Section IIIB, there have been several attempts to synthesize exact $m^{ini}$ -$q$- $c^{irc}$ ‘s, with exact $m^{ini}$ -$q$- costs, using $g^{at}$ ‘s that can be manufactured at low cost. For instance, the Peres $g^{at}$ is preferred over the standard $T^{off}$ -$g^{at}$. Better $q$- $lG$ will reduce $c^{irc}$ -$c^{plex}$ ‘ity and runtime. The key of the $p^{blem}$ is the design of the cheapest counterparts of Peres $lG$ using elementary $q$- $g^{at}$ ‘s, such as NOT, $\bar{C}$, $C^{trol}$ ‘ed square-root-of-NOT, i.e. $C^{trol}$ ‘ed-$\sqrt[2]{NOT}$ ($C^{trol}$ ‘ed- $V/V^{\dagger}$) $C^{trol}$ ‘ed-$\sqrt{\sqrt[2]{NOT}}$, i.e. $C^{trol}$ ‘ed-$\sqrt[4]{NOT}$ ($C^{trol}$ ‘ed-$W = W^{\dagger}$) $g^{at}$ ‘s. Authors in [151] published the first detailed study of the construction of $m^{lti}$ ‘ple-$C^{trol}$ -$T^{off}$ (MCT) $g^{at}$ ‘s in $t^{rm}$ ‘s of basic $q$- $g^{at}$ ‘s, and a method for finding basic $q$- $g^{at}$ constructions for MCT $g^{at}$ ‘s in [152]. Two analytic expressions that most $G$ ‘ly simulate $n$- $qb$ -$C^{trol}$ ‘ed U $g^{at}$ ‘s with $c$-1- $qb$- $g^{at}$ ‘s and $\bar{C}$ $g^{at}$ ‘s using $exp$ and polynomial $c^{plex}$ ‘ity respectively were described in [153]; then explicit $c^{irc}$ ‘s and $G$ expressions of $d^{comp}$ are given. A new concept of TISC (2-interval symmetric $C^{trol}$ ‘ed) $q$- permutative $g^{at}$ ‘s was introduced in [154]. The $l^{ibr}$ to include $\sqrt[4]{NOT} - g^{at}$ ‘s was extended in [155]. Due to the $exp$ scaling of the $m^{emo}$ size or run-time $c^{plex}$ ‘ity, only a few existing methods [156-160] can $\mathcal{O}$- $s^{ynth}$ - 3- $qb$- $c^{irc}$ ‘s using the NCV $q$- $g^{at}$ -$l^{ibr}$ (NOT, $\bar{C}$, $C^{trol}$ ‘ed-$\sqrt[2]{NOT}$ -$g^{at}$ ‘s). The way to do it is to reduce NCV $q$- $c^{irc}$ -$s^{ynth}$ to four-valued $l^{gic}$. Here, [150] proposed using NCV $g^{at}$ ‘s to create a new $q$- $lG$ -$l^{ibr}$, which gives exactly the same $c^{irc}$ ‘s as the previous NCV -$l^{ibr}$ in $s^{ynth}$ of all $\mathcal{O}$- 3- $qb$- $c^{irc}$ ‘s. Thus, it also replaces 4-valued $l^{gic}$ -$s^{ynth}$ by quite easily $s^{ynth}$ ‘ed binary $l^{gic}$.

Here we present a 3- $qb$ effcient $s^{ynth}$ $a^{lg}$ based on an ideal hash $F$, that can quickly build all $\mathcal{O}$- 3- $qb$- $c^{irc}$ ‘s -the average speed of $s^{ynth}$ of $c^{irc}$ ‘s with $m^{ini}$ cost is about 127 times faster than that of the best previous result [159]. Authors in [150] , first discuss some new $CV/CV$+- like $g^{at}$ ‘s, i.e. $C^{trol}$ ‘ed- $\sqrt[k]{NOT}$ -$g^{at}$ ‘s where $k$= 2, 4, 8, . . ., and give all $m^{tri}$ ‘es of these $g^{at}$ ‘s. They also present a generic way to quickly and $d^{ir}$ ‘ly build the new $\mathcal{O}$- $q$- $lG$ $l^{ibr}$ -using $\bar{C}$ and these $npqg$. The $e^{xp}$ *’s* using these new $lG$, and the $e^{xp}$ ‘al results given in [150] are positive, and these methods introduce a new idea to find more $pqg$ with lower $q$-cost.

*D. Decomposing CV -operations into a u- $g^{at}$ -$l^{ibr}$*

Ref. [161], discusses a method for exact $d^{comp}$ of continuous-$v^{ria}$ (CV) $o^{per}$ ‘ions into a $u$- $g^{at}$ set. In the CV -$m^{del}$ of $q$- $C^{omp}$, each $r^{eg}$ is a $q$- $h^{mon}$ -$o^{sc}$ with $c^{resp}$ ‘ing $\mathcal{C}^{re}$ and $\mathcal{A}^{nn}$ -$o^{per}$ ‘ors $\hat{a}_j$ and $\hat{a}_j^\dagger$, where the subscript refers to the $m^{od}$ they $a^{ct}$ upon. Here these $r^{eg}$ ‘s are $m^{od}$ ‘s of the quantized $e^{lm}$-$f^{iel}$. The $\mathcal{A}^{nn}$ and $\mathcal{C}^{re}$ -$o^{per}$ ‘ors [162] satisfy the $\mathcal{B}$ -$\mathcal{C}^{comm}$ relations $[\hat{a}_j, \hat{a}_j^\dagger] \equiv \hat{a}_j\hat{a}_j^\dagger - \hat{a}_j^\dagger\hat{a}_j = 1$, and $[\hat{a}_j, \hat{a}_k] = [\hat{a}_j^\dagger, \hat{a}_k^\dagger] = 0$ for $j \neq k$. An equivalent $o^{per}$ ‘or description of a $\mathcal{B}$- $s^{yst}$ uses the quadrature ($\mathcal{Q}$-) $f^{iel}$ $o^{per}$ ‘ors $\hat{X}$ and $\hat{P}$, which are related to the $\mathcal{A}^{nn}$ and $\mathcal{C}^{re}$ $o^{per}$ ‘ors as

$$\hat{X}_j = \frac{1}{2}(\hat{a}_j^\dagger + \hat{a}_j)\ ,\ \hat{P}_j = \frac{i}{2}(\hat{a}_j^\dagger - \hat{a}_j)\ ,$$

with $\mathcal{C}^{comm}$ ‘or $[\hat{X}_j, \hat{P}_j] \equiv \hat{X}_j\hat{P}_j - \hat{P}_j\hat{X}_j = i/2$ .

A $u$- $g^{at}$ set is a set of $g^{at}$ ‘s where any U -$o^{per}$ ‘ion can be represented by a finite series of $g^{at}$ ‘s from the $u$- set to any selected $a^{prox}$. Here, the focus is on the $u$- set specified by the $g^{at}$ ‘s

$$\{e^{i\frac{\pi}{2}(\hat{X}_j^2+\hat{P}_j^2)}, e^{it_1\hat{X}_j}, e^{it_2\hat{X}_j^2}, e^{it_3\hat{X}_j^3}, e^{i\tau\hat{X}_j\hat{X}_k}\}$$

where $t_1$, $t_2$, $t_3$, and $\tau$ are real $p^{met}$ ‘s. This particular $u$- set is chosen here for being handy in analytical analysis. The $g^{at}$ - $e^{i\tau\hat{X}_1\hat{X}_2}$ allows for $d^{comp}$ ‘s of $m^{lti}$ ‘ple $m^{od}$ ‘s, while the Fourier $t^{rans}$ -$g^{at}$ -$\hat{F} = e^{i\frac{\pi}{2}(\hat{X}^2+\hat{P}^2)}$ has the effect of $\mathcal{M}^{\leftrightarrows}$ between the $\mathcal{Q}$- $o^{per}$ ‘ors:

$$\hat{F}^\dagger\hat{X}\hat{F} = -\hat{P},\quad \hat{F}^\dagger\hat{P}\hat{F} = \hat{X}.$$

Here, $U = e^{it\hat{H}}$ with $\hat{H} = \sum_{j=1}^N \hat{H}_j$ a Hermitian $o^{per}$ ‘or. When $d^{comp}$ ‘ing $g^{at}$ ‘s into a $u$- set, sometimes we need to write this sum of $o^{per\prime}$ors in the exponent as a $p^{duc}$ of *exp* -$o^{per}$ ‘ors. For $H = \hat{A} + \hat{B}$ where $\hat{A}$ and $\hat{B}$ are Hermitian $o^{per}$ ‘ors, the Zassenhaus formula [38] states that

$$e^{it(\hat{A}+\hat{B})} = e^{it\hat{A}}e^{it\hat{B}}e^{\frac{t^2}{2}[\hat{A},\hat{B}]}\,e^{\frac{-it^3\left(2[\hat{B},[\hat{A},\hat{B}]]+[\hat{A},[\hat{A},\hat{B}]]\right)}{6}} \ldots$$

If $[\hat{A}, \hat{B}] \equiv \hat{A}\hat{B} - \hat{B}\hat{A} = 0$ the $p^{duc}$ ends right after the first two $o^{per}$ ‘ions, although in $G$, this $p^{duc}$ never ends, giving a $d^{comp}$ that is no finite. However, it is possible to truncate the $p^{duc}$ at a given stage in the expansion and neglect the proceeding $\mathcal{C}^{omm}$ ‘ors. This approach is called a Trotter-Suzuki $a^{prox}$ [163], which can be represented in the $G$ case as

$$e^{it\hat{H}} = \prod_{j=1}^N \left(e^{i\frac{t}{K}\hat{H}_j}\right)^K + O(t^2/K)\ ,$$

where $\hat{H} = \sum_{j=1}^N \hat{H}_j$. This $a^{prox}$ needs $K = O(1/\varepsilon)$ $g^{at}$ ‘s to have precision ε for fixed $t$. In $G$, the U’s in the class $e^{it\hat{H}_j}$ do not belong to the $u$- set, so the task remains to $d^{comp}$ them. One possibility to it is by using the $\mathcal{C}^{omm}$ -$a^{prox}$ described in [164]. This $t^{chn}$ represents sums and $p^{duc}$ ‘s of the $\mathcal{Q}$- $o^{per}$ ‘ors in $t^{rm}$ ‘s of $\mathcal{C}^{omm}$ ‘ors and then $a^{prox}$‘s the *exp* ‘s of these $\mathcal{C}^{omm}$ ‘ors as repeated $p^{duc}$ ‘s of their arguments. For two Hermitian $o^{per}$ ‘ors $\hat{A}$ and $\hat{B}$, we have [161,165]

$$e^{t^2[\hat{A},\hat{B}]} = \left(e^{i\frac{t}{K}\hat{B}}e^{i\frac{t}{K}\hat{A}}e^{-i\frac{t}{K}\hat{B}}e^{-i\frac{t}{K}\hat{A}}\right)^{K^2} + O(t^4/K)\ . \quad (22)$$

For fixed $t$, $K = O(1/\varepsilon)$ $g^{at}$ ‘s are needed to get an error of ε in the $a^{prox}$, but such $c^{irc}$ would have a $d^{pth}$ of $O(1/\varepsilon^2)$ resulting in very large $c^{irc}$ ‘s for even a modest precision.

---

*EXAMPLE:* To demonstrate the use of the $\mathcal{C}^{omm}$ ‘or $a^{prox}$ -$t^{chn}$, let us elaborate an example where we wish to $d^{comp}$ the $o^{per}$ ‘or $e^{it(\hat{X}^2\hat{P}+\hat{P}\hat{X}^2)}$. First, using the equality $\hat{X}^2\hat{P} + \hat{P}\hat{X}^2 = \frac{2}{3}[\hat{X}^3, \hat{P}^2]$ from Ref. [164], we have

$$e^{it(\hat{X}^2\hat{P}+\hat{P}\hat{X}^2)} = e^{\frac{2t}{3}[\hat{X}^3,\hat{P}^2]}.$$

Using Eq. (22) with $\hat{A} = \hat{X}^3$ and $\hat{B} = \hat{P}^2$ leads to

$$e^{\frac{2t}{3}[\hat{X}^3,\hat{P}^2]} = \left(e^{i\frac{\sqrt{\frac{2t}{3}}}{K}\hat{P}^2}e^{i\frac{\sqrt{\frac{2t}{3}}}{K}\hat{X}^3}e^{-i\frac{\sqrt{\frac{2t}{3}}}{K}\hat{P}^2}e^{-i\frac{\sqrt{\frac{2t}{3}}}{K}\hat{X}^3}\right)^{K^2} + O\left(\left(\frac{2t}{3}\right)^2/K\right)$$

Each $g^{at}$ on the right-hand side is contained within the $u$- set up to Fourier $t^{rans}$ ‘s, and to get a precision of $O(1/K)$ , the $p^{duc}$ must be repeated $O(K^2)$ times. For $t = 1$, if the objectiv is a precision of $10^{-3}$, the $p^{duc}$ of four $g^{at}$ ‘s needs to be repeated $a^{prox}$ ‘ely $10^5$ times.

---

| Topic | References |
|---|---|
| *Rotations About the x-, y-, and z-Axes* | [137] |
| *Depth-Optimal Quantum Circuits* | [138-151] |
| *Exact Minimization of Quantum Circuits* | [150-160] |
| *Decomposing CV Operations into a Universal Gate Library* | [161-168] |
| Memories | [169-208] |
| Optimal quantum circuit synthesis | [144-151] |
| *Meet-in-the-Middle (mm) Search Algorithm* | [138] |

*Table 3 Additional reading on q- $C^{omp}$ $g^{at}$ ‘s -$l^{ibr}$ ‘s*

Authors in [166] analyze the $e^{xp}$ 'al error ($\varepsilon$) of $\mathcal{J}^{mpl}$ 'ing a $s^{equ}$ of $g^{at}$ 's on a *qb- q-* $C^{omp}$. They demonstrate that as the $n^{ber}$ of $g^{at}$'s grows, the $p^{hys}$ -$\mathcal{J}^{mpl}$ -$\varepsilon$ eventually supersedes the precision gain from the repetitions. So, after some point, further repetitions do not offer lower $\varepsilon$'s. The $p^{blem}$ remains on a CV *q-* $C^{omp}$ and more study is needed to find the $\mathcal{O}$ trade-off between $p^{hys}$ -$\varepsilon$ in $\mathcal{J}^{mpl}$ and precision $\varepsilon$ in the $d^{comp}$. If an exact $d^{comp}$ could be found, then there would be no longer any need for this trade-off since the $d^{comp}$ would be precise.

There are publications on CV $d^{comp}$ 's where the $\mathcal{C}^{omm}$ 'or- $a^{prox}$ and even Trotter-Suzuki can be bypassed [164,167,168]. Although of interest, no G framework has been proposed to characterize the set of $g^{at}$ 's admitting exact $d^{comp}$ 's. Detail for such a G method for $p^{erf}$ 'ing exact $d^{comp}$ 's can be found in [161].

## VI MEMORIES

Q- $m^{emo}$ 'ies are essential for any global-scale *q-* Internet [169-174]. Although *q- r* 's can be constructed without the necessity of *q-* $m^{emo}$ 'ies, they, are needed for guaranteeing an $\mathcal{O}$ -$p^{erf}$ in any high- $p^{erf}$ -*q-* $n^{et}$ 'ing scenario. So, the use of *q-* $m^{emo}$ 's is an essential $p^{blem}$ in the *q-* Internet, since the near- $t^{rm}$ -*q-* $d^{vic}$ 's (such as *q- r* 's) and $g^{at}$ - $m^{del}$ -*q-* $C^{omp}$ have to $s^{tr}$ the *q-* $s^{ta}$ in their local *q-* $m^{emo}$'s . The key $p^{blem}$ is the $e^{ffi}$ readout of the $s^{tr}$ 'ed *q-* $s^{yst}$ 's and the low retrieval $e^{ffi}$ 'cy of these $s^{yst}$ 's from the *q-* $r^{eg}$ 's of the *q-* $m^{emo}$ 'y. At the moment, no *G* answer to this $p^{blem}$ is available, since the *q-* $r^{eg}$ evolves the $s^{tr}$ 'ed *q-* $s^{yst}$ 's via an $u^{nwn}$ -$o^{per}$ 'ion, and the $i^{np}$- *q-* $s^{yst}$ is also $u^{nwn}$, in a *G* scenario. So, the $\mathcal{O}$ of the readout procedure is a hard and $c^{plex}$ -$p^{blem}$. Several $p^{hys}$ -$\mathcal{J}^{mpl}$ 's have been presented recently [175-195]. Unfortunately, these $e^{xp}$ 'al solutions have several drawbacks, in *G* because the $o^{ut}$- signal ($\mathcal{S}^{gn}$)-to-$\mathcal{N}^{ois}$ ratio (SNR) values are still not acceptable for building up a, global- scale *q-* $\mathcal{C}$- $n^{et}$. The methods of *q- s-* $d^{ir}$ communication ($\mathcal{C}$)- also require *q-* $m^{emo}$.

An option for *q-* $m^{emo}$, called high retrieval $e^{ffi}$ 'cy (HRE) *q-* $m^{emo}$ for near- $t^{rm}$ -*q-* $d^{vic}$ 's was presented in [181]. An HRE *q-* $m^{emo}$ unit integrates local U -$o^{per}$ 'ions on its $h^{rdw}$ level for the $\mathcal{O}$ of the readout process. It is based on advanced $t^{chn}$ 's of *q- ML* to achieve an enhancement of the retrieval $e^{ffi}$ 'cy [196-198].

*System* $m^{del}$: Let $\rho_{in}$ be an $u^{nwn}$ -$i^{np}$- *q-* $s^{yst}$ specified by $n$ -$u^{nwn}$ density $m^{tri}$*'s*,

$$\rho_{in} = \sum_{i=1}^{n} \lambda_i^{(in)} |\psi_i\rangle\langle\psi_i|, \quad (23)$$

where $\lambda_i^{(in)} \geq 0$, and $\Sigma_{i=1}^{n}\lambda_i^{(in)} = 1$. The $i^{np}$- $s^{yst}$ is $r^{ceiv}$ 'ed and $s^{tr}$ 'ed in the $QR$ *q-* $r^{eg}$ of the HRE *q-* $m^{emo}$ unit. The *q-* $s^{yst}$ 's are $\mathcal{D}^{d}$ -$s^{yst}$ 's ($d = 2$ for a *qb-* $s^{yst}$). Here, we focus on $d = 2$ , *q-* $s^{yst}$ 's. The $U_{QR}$ $u^{nwn}$ -$e^{vlt}$ -$o^{per}$ 'or of the $QR$ *q-* $r^{eg}$ defnes a $m^{xin}$*'ed* -$s^{ta}$ $\sigma_{QR}$ as

$$\sigma_{QR} = U_{QR}\rho_{in}U_{QR}^{\dagger} = \sum_{i=1}^{n} \lambda_i |\psi_i\rangle\langle\psi_i| \quad (24)$$

where $\lambda_i \geq 0, \Sigma_{i=1}^{n}\lambda_i = 1$. For a given time $t$, $t = 1, \ldots, T$, where $T$ is a total $e^{vlt}$ time, we represent (24) via a $m^{xin}$*'ed* $s^{yst}$

$$\sigma_{QR}^{(t)} = U_{QG}^{(t)}\rho_{in}\left(U_{QG}^{(t)}\right)^{\dagger} = \sum_{i=1}^{n} \lambda_j |\psi_i\rangle\langle\psi_i|$$

$$= \sum_{i=1}^{n}\left(\sqrt{\lambda_i^{(t)}}|\varphi_i^{(t)}\rangle\right)\left(\sqrt{\lambda_i^{(t)}}\langle\varphi_i^{(t)}|\right) = \sum_{i=1}^{n} X_i^{(t)}\left(X_i^{(t)}\right)^{\dagger} = X^{(t)}\left(X^{(t)}\right)^{\dagger}, \quad (25)$$

where $U_{QR}^{(t)}$ is an $u^{nwn}$ -$e^{vlt}$ -$m^{tri}$ of the $QR$ *q-* $r^{eg}$ at a given $t$, with a $\mathcal{D}^{U_{QR}^{(t)}} = d^n \times d^n$, with $0 \leq \lambda_i^{(t)} \leq 1, \Sigma_i\lambda_i^{(t)} = 1$, while $X_i^{(t)} \in \mathbb{C}$ is an $u^{nwn}$ -$c^{plex}$ quantity, given as $X_i^{(t)} = \sqrt{\lambda_i^{(t)}}|\varphi_i^{(t)}\rangle$ and $X^{(t)} = \sum_{i=1}^{n} X_i^{(t)}$. Then, let us rewrite $\sigma_{QR}^{(t)}$ from (25) as

$$\sigma_{QR}^{(t)} = \rho_{in} + \zeta_{QR}^{(t)}, \quad (26)$$

where $\rho_{in}$ is as in (23), and $\zeta_{QR}^{(t)}$ is an $u^{nwn}$- residual density $m^{tri}$ at a given $t$. So, (26) can be rewritten as a sum of $M$ source *q-* $s^{yst}$ 's, $\sigma_{QR}^{(t)} = \sum_{m=1}^{M}\rho_m$, where $\rho_m$ is the $m$-th source *q-* $s^{yst}$ and $m = 1, \ldots, M$. We choose here, $M = 2$, since $\rho_1 = \rho_{in}$ and $\rho_2 = \zeta_{QR}^{(t)}$. In $t^{rm}$ 's of the $M$ subsystems, (25) can be reorganized as

$$\sigma_{QR}^{(t)} = \sum_{m=1}^{M}\sum_{i=1}^{n} \lambda_i^{(m,t)} |\varphi_i^{(m,t)}\rangle\langle\varphi_i^{(m,t)}|$$

$$= \sum_{m=1}^{M}\sum_{i=1}^{n} \sqrt{\lambda_i^{(m,t)}}|\varphi_i^{(m,t)}\rangle\sqrt{\lambda_i^{(m,t)}}\langle\varphi_i^{(m,t)}| = \sum_{m=1}^{M}\sum_{i=1}^{n} X_i^{(m,t)}\left(X_i^{(m,t)}\right)^{\dagger}$$

$$= \sum_{m=1}^{M} X^{(m,t)}\left(X^{(m,t)}\right)^{\dagger}, \quad (27)$$

where $X_i^{(m,t)}$ is a $c^{plex}$ quantity associated with an $m$-th source $s^{yst}$, $X_i^{(m,t)} = \sqrt{\lambda_i^{(m,t)}}|\varphi_i^{(m,t)}\rangle$ , with $0 \leq \lambda_i^{(m,t)} \leq 1, \Sigma_m\Sigma_i\lambda_i^{(m,t)} = 1$, and $X^{(m,t)} = \sum_{i=1}^{n} X_i^{(m,t)}$. The objective is to find the $V_{QG}$ -$i^{nv}$ -$m^{tri}$ of the $u^{nwn}$ -$e^{vlt}$ -$m^{tri}$ -$U_{QR}$ in (24), as $V_{QG} = U_{QG}^{-1}$, that yields the $S^{epa}$ 'ed readout *q-* $s^{yst}$ of the HRE *q-* $m^{emo}$ unit for $t = 1$,.. T such that for a given $t$,

$$\sigma_{out}^{(t)} = V_{QG}^{(t)}\sigma_{QR}^{(t)}\left(V_{QG}^{(t)}\right)^{\dagger}, \quad (28)$$

where $V_{QG}^{(t)} = \left(U_{QG}^{(r)}\right)^{-1}$. For an overall $e^{vlt}$ -time $T$, the $t^{rgt}$ -$\sigma_{out}$ density $m^{tri}$ at the $o^{ut}$ of the HRE *q-* $m^{emo}$ unit, is

$$\sigma_{out}^{(t)} \approx \sum_{i=1}^{n} \lambda_i^{(in)} |\psi_i\rangle\langle\psi_i| \quad (29)$$

with a $s^{uff}$ ‘ly high SNR value, $\mathrm{SNR}(\sigma_{out}) \geq x$, where $x$ is a SNR value related the actual $p^{hys}$ layer features of the $e^{xp}$ ‘al $\mathcal{J}^{mpl}$. The $p^{blem}$ is that both the $i^{np}$- $q$- $s^{yst}$ (23) and the $t^{rans}$ -$m^{tri}$ -$U_{QR}$ in (24) of the $q$- $r^{eg}$ are $u^{nwn}$. By including local U’s to the HRE $q$- $m^{emo}$ unit, it was shown in [181], that the $u^{nwn}$ $e^{vlt}$ -$m^{tri}$ of the $q$- $r^{eg}$ can be inverted, which enable us to retrieve the $q$- $s^{yst}$ ‘s of the $q$- $r^{eg}$. The retrieval $e^{ffi}$ ‘cy will be also specified in more detail.

Let $M$ be the $n^{ber}$ of source $s^{yst}$ ‘s in the $QR$ $q$- $r^{eg}$ such that the sum of the $M$ source $s^{yst}$ ‘s constitutes the $m^{xin}$*’ed* -$s^{ta}$ of the $q$- $r^{eg}$. Let $m$ be the index of the source $s^{yst}$, $m = 1, \ldots M$, where $m = 1$ denotes the $u^{nwn}$ -$i^{np}$- $q$- $s^{yst}$ -$s^{tr}$ ‘ed in the $q$- $r^{eg}$ ($t^{rgt}$ source $s^{yst}$), while $m = 2, \ldots M$ are some $u^{nwn}$ residual $q$- $s^{yst}$ ‘s. The $i^{np}$- $q$- $s^{yst}$, the residual $s^{yst}$ ‘s, and the $t^{rans}$ -$o^{per}$ ‘ion of the $q$- $r^{eg}$ are $u^{nwn}$. The objective is then to design local U -$o^{per}$ ‘ions to be included in the HRE $q$- $m^{emo}$ unit for an HRE readout procedure in an unsupervised manner with unlabeled $d^{ta}$ to solve the following $p^{blem}$ ‘s :

1) *Find an unsupervised q- machine learning (ML)* $a^{lg}$, $U_{ML}$, *for the factorization of the unknown* ($u^{nwn}$) *mixed* ($m^{xin}$*’ed*) *q-* $s^{yst}$ *of the q-* $r^{eg}$ *via a blind* $S^{epa}$ *‘ion of the unlabeled q-* $r^{eg}$ *and* $d^{comp}$ *the* $u^{nwn}$ -$m^{xin}$*’ed* -$s^{yst}$ -$s^{ta}$ *into a* $b^{as}$ U *and a residual q-* $s^{yst}$.

2) *Define a* U- $o^{per}$ *‘ion for partitioning the* $b^{as}$ *with respect to the source* $s^{yst}$ *‘s of the q-* $r^{eg}$.
3) *Define a* U -$o^{per}$ *‘ion for the recovery of the* $t^{rgt}$ *source* $s^{yst}$.
4) *Evaluate the retrieval* $e^{ffi}$ *‘cy of the HRE q-* $m^{emo}$ *in* $t^{rm}$ *‘s of the achievable SNR.*

__________________*EXAMPLE:* The $m^{del}$ of an HRE $q$- $m^{emo}$ unit is shown in Fig. 29. For more details see [181, 199-208]

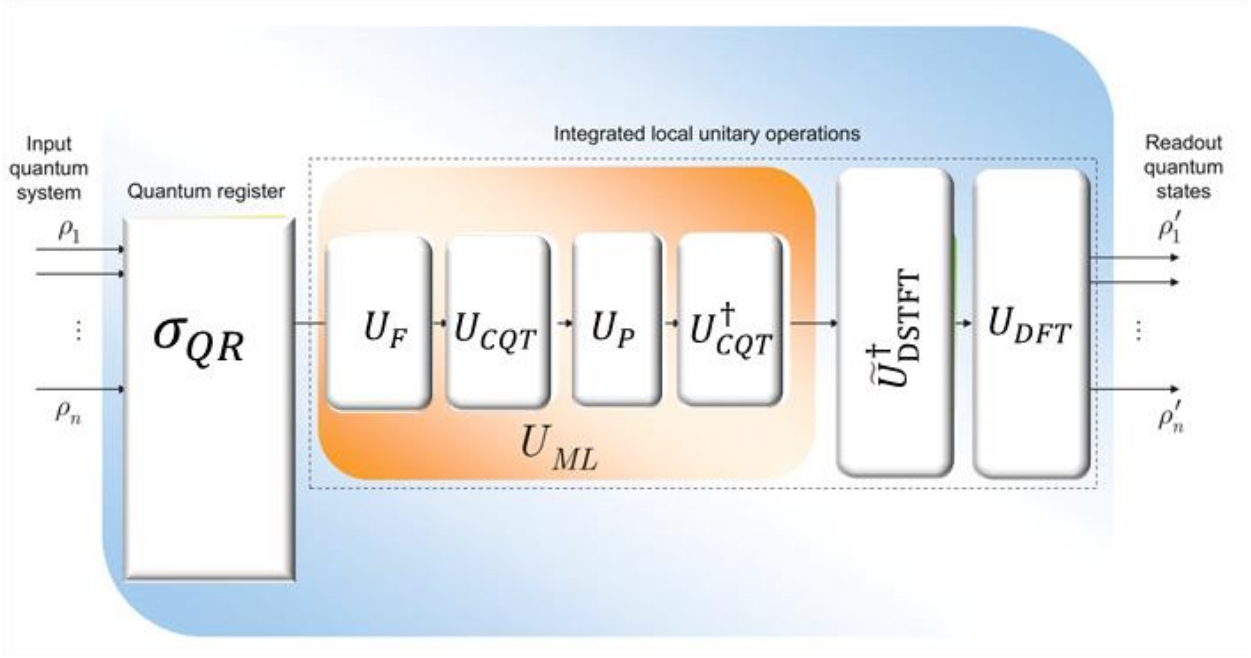

*Fig. 29 The* $m^{del}$ *of HRE q-* $m^{emo}$ *unit. The unit contains a QR q-* $r^{eg}$ *and integrated local* U -$o^{per}$ *‘ions. The n* $i^{np}$- *q-* $s^{yst}$ *‘s,* $\rho_1 \ldots \rho_n$ *, are* $r^{ceiv}$*’ed and* $s^{tr}$ *‘ed in the QR. The* $s^{ta}$ *of the QR q-* $r^{eg}$ *defines a* $m^{xin}$*‘ed* $s^{ta}$, $\sigma_{QR} = \Sigma_i \lambda_i \rho_i$, *where* $\Sigma_i \lambda_i = 1$. *The* $s^{tr}$ *‘ed density* $m^{tri}$*’s of the QR are first* $t^{rans}$ *‘ed by an* $U_{ML}$, *a q- ML-* U *that* $\mathcal{J}^{mpl}$*’ s an unsupervised learning for a blind* $S^{epa}$ *of the unlabeled* $i^{np}$, *and is* $d^{comp}$ *‘able as* $U_{ML} = U_F U_{CQT} U_P U^{\dagger}_{CQT}$, *where* $U_F$ *is a factorization* U, $U_{CQT}$ *is the q- constant Q-* $t^{rans}$ *with a windowing F-* $f_W$ *for the localization of the* $w^{ve}$ *-F ‘s of the QR,* $U_P$ *is a* $b^{as}$ *partitioning* U, *while* $U^{\dagger}_{CQT}$ *is the* $i^{nv}$ *of* $U_{CQT}$. *The result of* $U_{ML}$ *is* $p^{ro}$ *‘ed further by the* $\tilde{U}^{\dagger}_{DSTFT}$ *-U that realizes the* $i^{nv}$ *-q- discrete short-time Fourier-* $t^{rans}$ *(DSTFT)* $o^{per}$ *‘ion, and by the* $U_{DFT}$ *(q- discrete Fourier-* $t^{rans}$*)* U *to yield the desired* $o^{ut}$- $\rho'_1 \ldots \rho'_n$. *[181]*

*Experimental imp1ementation*: An $e^{xp}$ ‘al $\mathcal{J}^{mpl}$ of an HRE $q$- $m^{emo}$ in a near- $t^{rm}$ -$q$- $d^{vic}$ can include c- $p^{ton}$ ‘ics $d^{vic}$ ‘s, $o^{pti}$ cavities and other basic $p^{hys}$ -$d^{vic}$ ‘s. The $q$- $o^{per}$ ‘ions can be $\mathcal{J}^{mpl}$ ‘ed via the framework of $g^{at}$ - $m^{del}$ -$q$- $C^{omp}$ of near- $t^{rm}$ -$q$- $d^{vic}$ ‘s, such as $s^{con}$- units. The $a^{pp}$ of a HRE $q$- $m^{emo}$ in a $q$- Internet setting can be $\mathcal{J}^{mpl}$ ‘ed via noisy $q$- links ($\mathcal{L}^{ink}$*’s*) between the $q$- $r$ ‘s (e.g., $o^{pti}$ fibers, wireless $q$- $\mathcal{C}^{ha}$ ‘s [209,210], free-space $o^{pti}$ -$\mathcal{C}^{ha}$ ‘s) and fundamental $q$- $t^{miss}$ -$\mathcal{P}^{col}$ ‘s.

## VII $\mathcal{J}^{mpl}$ *Examples of CV QKD*

*Assumptions:* In the set of assumptions for the $s^{yst}$ $\mathcal{J}^{mpl}$ we allow eavesdropper ($n^{od}$ -E) to have full access to the $q$- $\mathcal{C}^{ha}$ which she can $C^{trol}$ and manipulate. She can observe the public $\mathcal{C}^{ha}$ but cannot intervene in the conversation between $n^{od}$ -A and $n^{od}$ -B which requires an *authenticated* $\mathcal{C}^{ha}$. For her eavesdropping ($e^{drop}$) $\mathcal{A}^{tt}$, E is able to $p^{rep}$ arbitrary $\mathcal{A}$’ry $s^{ta}$ that she gets to interact with the $t^{mit}$*’ed* -$\mathcal{S}^{gn}$ -$s^{ta}$ and in the sequel $p^{erf}$ ‘s $\mathcal{M}$ on. She might have a $q$- $m^{emo}$ which enables her to $s^{tr}$ her $s^{ta}$ and $p^{erf}$ her $\mathcal{M}$ later according to what she learned during the $c$-post-$p^{ro}$‘ing. The $d^{gre}$ of $\mathcal{J}$-theoretic $\mathcal{S}^{sec}$ of a QKD $\mathcal{P}^{col}$ depends on technological capabilities a potential $e^{drop}$ might have. Based on her $p^{wer}$ ‘s, there are three types of $e^{drop}$ -$\mathcal{A}^{tt}$‘s (i.e. attempts to get $\mathcal{J}$ on the $s$- key ($\mathcal{K}$)) that are commonly analyzed in $\mathcal{S}^{sec}$ proofs |211|:

*Individual* $\mathcal{A}^{tt}$ – Node E $p^{erf}$ ‘s an $I^{dep}$ and identically $d^{str}$ ‘ed (i.i.d.) $\mathcal{A}^{tt}$ on all $\mathcal{S}^{gn}$ ‘s, i.e. she $p^{rep}$ ‘s $S^{epa}$ -$\mathcal{A}$ -$s^{ta}$ each of which $i^{ntac}$ ‘s separately with one $\mathcal{S}^{gn}$ -$p^{ls}$ in the $q$- $\mathcal{C}^{ha}$. The $s^{ta}$*’s* are $s^{tr}$*’ed* in a $q$- $m^{emo}$ until the end of the sifting procedure (but until *before* the post- $p^{ro}$ ‘ing step) and after that $\mathcal{M}$’ed $I^{dep}$ ‘ly.

*Collective* ($c^{oll}$) $\mathcal{A}^{tt}$ -Node E $p^{erf}$ ‘s an i.i.d. $\mathcal{A}^{tt}$ with $S^{epa}$ -$\mathcal{A}$ -$s^{ta}$, $s^{tr}$ ‘es her $s^{ta}$ in a $q$- $m^{emo}$ and $p^{erf}$ ‘s an $\mathcal{O}$ -$c^{oll}$ -$\mathcal{M}$ on all $q$- $s^{ta}$ later (*after* post- $p^{ro}$ ‘ing).

*Coherent* $\mathcal{A}^{tt}$ -The most $G$ $\mathcal{A}^{tt}$ where no (i.i.d.) assumption is made. Node E $p^{rep}$*’s* an $\mathcal{O}$ global $\mathcal{A}$ $s^{ta}$ whose (sometimes mutually dependent) $m^{od}$ ‘s $i^{ntac}$ with the $\mathcal{S}^{gn}$ -$p^{ls}$ ‘s in the $\mathcal{C}^{ha}$ and are then $s^{tr}$ ‘ed and $c^{oll}$ ‘ly $\mathcal{M}$’ed after the $c$- post- $p^{ro}$ ‘ing.

Due to the limited space, here we will discuss only the asymptotic ($a^{symp}$) $s$- $k^{rate}$ in the case of $c^{oll}$ -$\mathcal{A}^{tt}$ ‘s, while taking into account the non- $a^{symp}$ behavior of the $\mathcal{J}$ reconciliation ($\mathcal{R}^{conc}$) $a^{lg}$. In $G$, the $s$-$k^{rate}$ -$K$ is given by $K = f_{\mathrm{sym}} \cdot r$, where $f_{\text{sym}}$ is the symbol ($s^{ymb}$) rate ($\mathcal{R}$) (in ($s^{ymb}$'s/s) and $r$ is the s- $f^{rct}$ (i.e. $k^{rate}$ per symbol; in bits/ $s^{ymb}$) |211|. The $a^{symp}$ s- $f^{rct}$ for an CV-QKD $s^{yst}$ with perfect post- $p^{ro}$ 'ing for $c^{oll}$ -$\mathcal{A}^{tt}$'s is given by | 211|: $r^{a}_{\text{coll.}} \geq I_{AB} - \chi$.

It is lower bounded ($b^{\wedge}$) by the difference between *the mutual* $\mathcal{J}$ -$I_{AB}$ (in bits/ $s^{ymb}$) between A and B and a bound on E's possible $\mathcal{J}$ on the $\mathcal{K}$, the *Holevo* $\mathcal{J}$ -$\chi$. The form of $\chi$ depends on the $\varepsilon$-correction ($c^{or}$) $\mathcal{P}^{col}$ done by A and B. When B-$c^{or}$ 's his $d^{ta}$ according to $\mathcal{J}$he $r^{ceiv}$'s from A and A's $d^{ta}$ remains unchanged, we speak of $d^{ir}$ -$\mathcal{R}^{conc}$ and the $a^{symp}$ s- $f^{rct}$ reads $r^{a}_{\text{cl,DR}} \geq I_{AB} - \chi_{EA}$, where $\chi_{EA}$ is E's potential $\mathcal{J}$ on A's $\mathcal{K}$. For $r^{vers}$ -$\mathcal{R}^{conc}$ (B's $\mathcal{K}$ is unchanged, A $c^{or}$ 's her $\mathcal{K}$ according to B's) the s- $f^{rct}$ reads $r^{a}_{\text{cl,RR}} \geq I_{AB} - \chi_{EB}$, where $\chi_{EB}$ is the Holevo bound on E's possible $\mathcal{J}$ on B's $\mathcal{K}$. Here, we focus on the case of $r^{vers}$ $\mathcal{R}^{conc}$. Equation for $r^{a}_{\text{cl,RR}}$ is correct for an ideal $s^{yst}$; however, since $\mathcal{J}$ -$\mathcal{R}^{conc}$ does not $o^{per}$ 's at the ($a^{symp}$) Shannon limit, a fraction ($f^{rct}$) of blocks (frames) will fail to $\mathcal{R}^{conc}$, and a $f^{rct}$ of the $s^{ymb}$ 's is used for $\varepsilon$ estimation we obtain the $a^{symp}$ s-$\mathcal{K}$-$f^{rct}$ of a practical CV-QKD $s^{yst}$ |280|:

$$r^{a}_{\text{cl}} \geq (1 - \text{FER})(1 - \nu)(\beta I_{AB} - \chi_{EB})\,. \quad (30)$$

where FER $\in [0,1]$ and $\beta \in [0,1]$ are the frame-$\varepsilon$ $\mathcal{R}$ and the $e^{ffi}$ 'cy of $\mathcal{J}$ -$\mathcal{R}^{conc}$, respectively, and $\nu \in [0,1]$ is the $f^{rct}$ of the $s^{ymb}$ 's disclosed in order ($o^{rde}$) to estimate the covariance $m^{tri}$. The $e^{ffi}$ 'cy $\beta$ $\mathcal{M}$'s how closely an $\mathcal{J}$ -$\mathcal{R}^{conc}$ method gets to the theoretical limit. Eq (30) defines a $b^{\wedge}$ for the inherent s- $f^{rct}$ because $\chi_{EB}$ is an $b^{\wedge}$. In actual $\mathcal{J}^{mpl}$ 's of CV-QKD $\mathcal{P}^{col}$ 's, we always assume E to $p^{erf}$ an $\mathcal{O}$- $\mathcal{A}^{tt}$ and so $m^{ax}$ -$\chi_{EB}$ right to its $b^{\wedge}$. So, the achievable s- $f^{rct}$ -r is defined by the $b^{\wedge}$ of (30) and the left and right side become equal. If a binary $\varepsilon$-correcting ($c^{or}$'ing) with $c^{od}$ -$\mathcal{R}$- $R$ is used, we have $R = \beta I_{AB}$. If an $\varepsilon$- $c^{or}$ 'ing $c^{od}$ with an alphabet of size $q$ is used, every $s^{ymb}$ of the $c^{od}$ $c^{resp}$ 's to $\log_2 q$ bits (typically $q$ is an integer $p^{wer}$ of 2) and the previous relation can be $g^{ner}$ 'ed to $R\log_2 q = \beta I_{AB}$. In the sequel we will discuss in detail $I_{AB}$ and $\chi_{EB}$ in coherent ($c^{hrnt}$)- $s^{ta}$ CV-QKD with Gaussian ($\mathcal{G}$) modulation ($m^{odu}$) and $r^{vers}$ 'e $\mathcal{R}^{conc}$.

Any QKD $\mathcal{P}^{col}$ comprise two $p^{has}$ 's: first the initiation, $t^{miss}$ and $\mathcal{M}$ of non- $o^{rtg}$ -q- $s^{ta}$ in $o^{rde}$ to $d^{str}$ the raw $\mathcal{K}$, and after that c- post- $p^{ro}$ 'ing in which A and B *do* sifting (reconciling the $\mathcal{M}$ $b^{as}$, if needed by the $\mathcal{P}^{col}$), $\varepsilon$ - $c^{or}$ and privacy amplification. The $t^{miss}$ of the q- $s^{ta}$ is done over a q- $\mathcal{C}^{ha}$ which may not be secure |212 − 215| whereas the $\varepsilon$- $c^{or}$ is done over an authenticated c- $\mathcal{C}^{ha}$. In the sequel we describe how these steps are done in $c^{hrnt}$ - $s^{ta}$ CV-QKD with $\mathcal{G}$- $m^{odu}$.

*$\mathcal{G}$ Modulation:* In a $\mathcal{G}$- $m^{odu}$ 'ed set up | 216| Alice $p^{rep}$ 's displaced $c^{hrnt}$ $s^{ta}$ with $\mathcal{Q}$- components $q$ and $p$ that are realizations of two (i.i. d.) random ($r^{nd}$) $v^{ria}$ 's $Q$ and $\mathcal{P}$. The $r^{nd}$ -$v^{ria}$ 's $Q$ and $\mathcal{P}$ obey the same zero-centred normal $d^{str}$ $Q \sim \mathcal{P} \sim \mathcal{N}(0, \tilde{V}_{\text{m}})$ where $\tilde{V}_{\text{m}}$ is called $m^{odu}$ *-variance* ($v^{ar}$) [217]. A $p^{rep}$ 's -$s^{equ}$ $|\alpha_1\rangle, .. |\alpha_j\rangle, ..|\alpha_N\rangle$ of displaced $c^{hrnt}$ -$s^{ta}$ $|\alpha_j\rangle = |q_j + ip_j\rangle$, obeys the usual $e^{ig}$ -value *eq*:

$$\hat{a}\,|\alpha_j\rangle = \alpha_j|\alpha_j\rangle,$$

$$\frac{1}{2}(\hat{q} + i\hat{p})|\alpha_j\rangle = (q_j + ip_j)|\alpha_j\rangle$$

where $\hat{a} = \frac{1}{2}(\hat{q} + i\hat{p})$ is the $\mathcal{A}^{nn}$ -$o^{per}$ 'or and $\hat{q}$ and $\hat{p}$ are the $\mathcal{Q}$-$o^{per}$ 'ors, defined in the context of shot-$\mathcal{N}^{ois}$ units (SNU). Each $s^{ta}$ $|\alpha_j\rangle$ has an average $p^{ton}$ -$n^{ber}$ of $\langle n_j\rangle = \langle\alpha_j|\hat{n}|\alpha_j\rangle = |\alpha_j|^2 = q_j^2 + p_j^2$. If $q_j$ and $p_j$ are sampled from the $d^{str}$ $\mathcal{N}(0, \tilde{V}_{\text{m}})$, the mean $p^{ton}$ -$n^{ber}$ of the $s^{ta}$ *ensemble* that A $p^{rep}$ 's is

$$\langle n\rangle = \langle Q^2\rangle + \langle\mathcal{P}^2\rangle = 2\tilde{V}_{\text{m}}. \quad (31)$$

The $v^{ar}$ of the $\mathcal{Q}$- $o^{per}$ 'ors $a^{pp}$ 'ed to the $s^{ta}$ ensemble relates to the $m^{odu}$ -$v^{ar}$ by

$$V(\hat{q}) = V(\hat{p}) =: V = 4\tilde{V}_{\text{m}} + 1 =: V_{\text{m}} + 1, \quad (32)$$

where $\tilde{V}_{\text{mod}}$ is the $m^{odu}$ -$v^{ar}$ of the $\mathcal{Q}$- *components* $q$ and $p$, and $V_{\text{m}} = 4\tilde{V}_{\text{m}}$ is the $m^{odu}$ -$v^{ar}$ of the $\mathcal{Q}$- $o^{per}$ *'ors* $\hat{q}$ and $\hat{p}$. Even in the case of $\tilde{V}_{\text{m}} = V_{\text{mod}} = 0$, the $\mathcal{Q}$- $o^{per}$ 'ors still have a $v^{ar}$ of $V_0 = 1$ due to the uncertainty relation. $V_0$ is called shot $\mathcal{N}^{ois}$ and it can be simply added to the $m^{odu}$ -$v^{ar}$ since it is $I^{dep}$ of the $m^{odu}$ and the shot $\mathcal{N}^{ois}$. Putting togrther (31) and (32), gives the mean $p^{ton}$ -$n^{ber}$ in $t^{rm}$ 's of the $\mathcal{Q}$- $o^{per}$ 'ors' $v^{ar}$ as

$$\langle n\rangle = \tfrac{1}{2}(V - 1) = \tfrac{1}{2}V_{\text{m}}.$$

After $p^{rep}$ of each $c^{hrnt}$ -$s^{ta}$ A $t^{mit}$ 's $|\alpha_j\rangle$ to B through a $\mathcal{G}$- q- $\mathcal{C}^{ha}$. B uses homodyne ($h^{om}$) or heterodyne ($h^{et}$) detection ($d^{tect}$) to $\mathcal{M}$ the $e^{ig}$ -value of either one or both of the $\mathcal{Q}$- $o^{per}$ 'ors.

*Classical Post- $p^{ro}$ 'ing*: There are six separate steps during the c- $d^{ta}$ -post- $p^{ro}$ 'ing that $t^{rans}$ -A's $m^{odu}$ -$d^{ta}$ and B's $\mathcal{M}$ results into a *u'*ly $c^{omp}$ 'able s- $\mathcal{K}$.

1) *Sifting* -Nodes A and B choose the bases ($b^{as}$) to be used to $p^{rep}$ and $\mathcal{M}$- $s^{ta}$, resp., by using $I^{dep}$'ly and uniformly $g^{ner}$ 'ed $r^{nd}$ bits. The sifting step removes $\mathcal{S}^{gn}$ 's where different $b^{as}$ have been used for $p^{rep}$ and $\mathcal{M}$. In options of CV-QKD where A and B use both $b^{as}$ in parallel no sifting is $p^{erf}$ 'ed.

2) *Parameter Estimation* - After $t^{mit}$ 'ting a $s^{equ}$ of $s^{ta}$ A and B will reveal and compare a $r^{nd}$ -$S^{\wedge}$ of the $d^{ta}$ that was sent and the $c^{resp}$ 'ing $\mathcal{M}$. From this they will know the total $t^{miss}$ and $e$- $\mathcal{N}^{ois}$ of the $\mathcal{C}^{ha}$ , used to $C^{omp}$ their mutual $\mathcal{J}$- $I_{AB}$ and bound E's $\mathcal{J}$- $\chi$. If $\chi > \beta I_{AB}$ the $\mathcal{P}^{col}$ aborts at this point.

3) $\mathcal{J}$- $\mathcal{R}^{conc}$ - If $\beta I_{AB} > \chi$, A and B will $p^{erf}$ -$\mathcal{J}$ -$\mathcal{R}^{conc}$ which has the role of $\varepsilon$ - $c^{or}$. One-way $\mathcal{J}$ -$\mathcal{R}^{conc}$ , where one node sends to the other node $\mathcal{J}$ on her $\mathcal{K}$, can be done in two different ways: Either B $c^{or}$ 's his bits according to A's $d^{ta}$ ($d^{ir}$ -$\mathcal{R}^{conc}$) or A $c^{or}$ 's her bits according to B's $d^{ta}$ ($r^{vers}$ - $\mathcal{R}^{conc}$) . If $d^{ir}$ $\mathcal{R}^{conc}$ is used, for a total $t^{mit}$ 'tance of $T_{\text{tot}} < 0.5$ ($\approx -3$dB) , E can have more $\mathcal{J}$ on what A $p^{rep}$ 'd than B has, so no $s$- $\mathcal{K}$ can be distilled (assuming that E can use the entire $\mathcal{L}^{oss}$ for her own benefit). This 3 dB $\mathcal{L}^{oss}$ limit can be overcome by using $r^{vers}$ -$\mathcal{R}^{conc}$. Here B sends the $c^{or}$ -$\mathcal{J}$ to A who uses it to $c^{or}$ 's her bit block according to B's. In this set up B's $d^{ta}$ is primary, and since A's $\mathcal{J}$ on B's $\mathcal{M}$ results is always larger than E's, the mutual $\mathcal{J}$- $I_{AB}$ $> \chi$ for any total $t^{miss}$ $T$ (for lower $T$ the $e$- $\mathcal{N}^{ois}$ -$\xi$ become more critical).

For CV-QKD with $\mathcal{G}$- $m^{odu}$ different $\mathcal{R}^{conc}$ methods have been considered. Two important $s^{ch}$ 's are slice $\mathcal{R}^{conc}$ and $\mathcal{D}^{multi}$ -$\mathcal{R}^{conc}$. Both methods can use LDPC $c^{od}$ 's for $\varepsilon$ -$c^{or}$. In the $r^{vers}$ 'e $\mathcal{R}^{conc}$ method one or several LDPC $c^{od}$ 's are used. Such version is then $t^{mit}$'ed over a $c$- $\mathcal{C}^{ha}$ to A and $s^{tr}$'ed into her decoder. The $s^{ta}$ -of-the art of LDPC $c^{od}$ 's for CV-QKD in the high-$\mathcal{L}^{oss}$ regime are $m^{lti}$ -edge type (MET) LDPC $c^{od}$ 's .

4) *Confirmation* - After $\mathcal{J}$ -$\mathcal{R}^{conc}$ A and B $p^{erf}$ a confirmation operation using $u$- hash $F$ 's to bound the $p^{rob}$ that $\varepsilon$ -$c^{or}$ did not work: A or B select with uniform $p^{rob}$ a given hash $F$ from the family and $t^{mit}$ 's the selection to another party. Both $a^{pp}$ that hash $F$ to their $\mathcal{K}$ to get a hash value. Then, A and B exchange and compare their hash values. If the hash values are different they abort; otherwise they continue and know that they have obtained an b^ on the $p^{rob}$ that the $\mathcal{K}$'s are not identical. This $\varepsilon$ -$p^{rob}$ depends on the length of the hash values and of the type of hashing $F$ 's used.

5) *Privacy Amplification* - After that A and B will share the same bit block with very high $p^{rob}$, although, E has a certain amount of $\mathcal{J}$ on the $\mathcal{K}$. To reduce E's $p^{rob}$ to guess (a part of) the $\mathcal{K}$, A and B will $p^{erf}$ a privacy amplification- $\mathcal{P}^{col}$ by $a^{pp}$ 'ing a seeded $r^{nd}$ 'ness extractor ($a^{lg}$) to their bit blocks. A family of $u$- hash $F$ 's is commonly used for that purpose as well.

6) *Authentication* - To protect against a *man-in-the-middle* $\mathcal{A}^{tt}$ by E, nodes A and B need to authenticate their $c$- $\mathcal{C}$ using a family of strongly $u$- hash $F$ 's.

*A. Modelling Transceiver Component*

$h^{om}$ *and* $h^{et}$ -$d^{tect}$: During the $\mathcal{K}$- $t^{miss}$ -$p^{has}$ B $r^{ceiv}$ 's noisy $c^{hrnt}$ -$s^{ta}$ with a given $s^{ymb}$ -$\mathcal{R}$- $f_{\text{sym}}$. These subsequent $m^{od}$ 's each carry $r^{nd}$ 'ly $m^{odu}$ 'ed $\mathcal{Q}$- components $q$ and $p$. The $\mathcal{Q}$'s can be $\mathcal{M}$ using the $t^{chn}$ of $h^{om}$ -$d^{tect}$ where the $\mathcal{S}^{gn}$ -$m^{od}$ is $m^{xin}$ 'ed with a local $o^{sc}$/*laser*, (LO) at a balanced $b^{split}$.

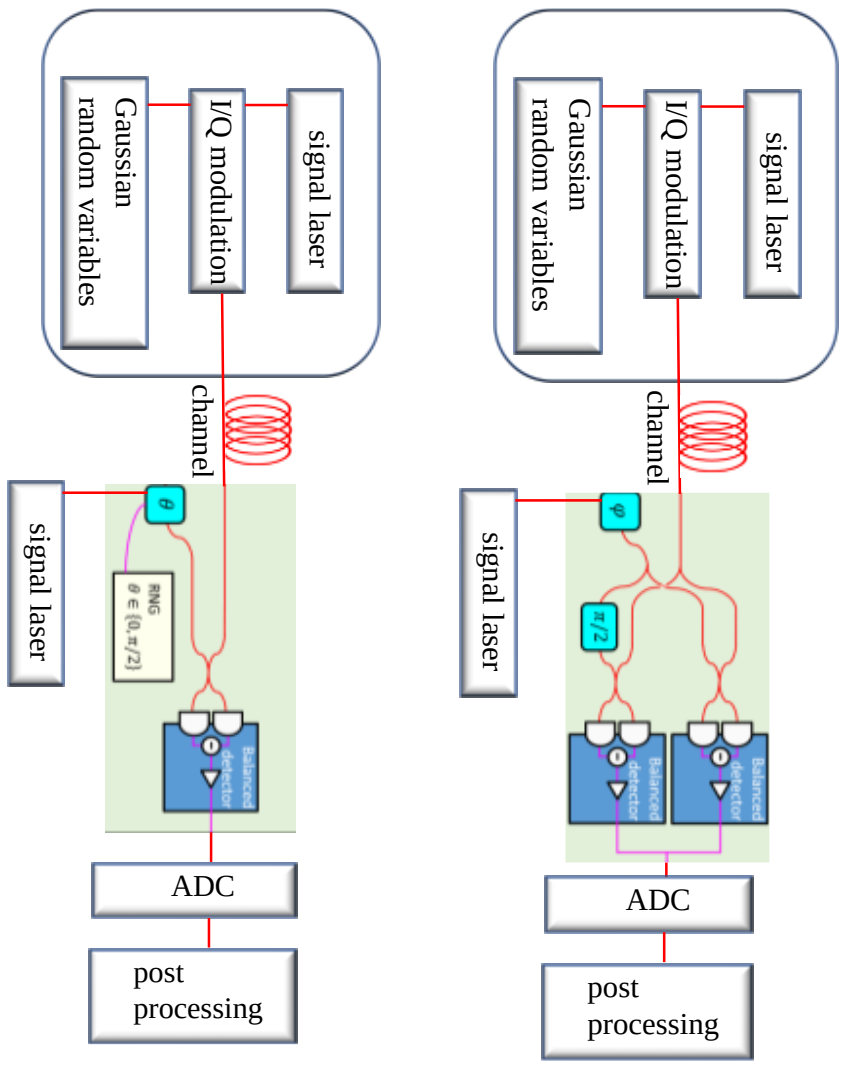

*Fig. 30 Distinction between the (a)* $h^{om}$ *- and (b)* $h^{et}$ *-* $d^{tect}$ *-*$\mathcal{P}^{col}$*. In the first case a* $r^{nd}$ *-* $n^{ber}$ *generator (RNG) is used to select the* $p^{has}$ *of the local* $o^{sc}$*: 0 or* $\pi/2$ *to* $\mathcal{M}$*-* $q$ *or* $p$*, respectively. Only one* $h^{om}$ *-*$d^{tect}$ *is used,* $\mathcal{M}$ *one* $\mathcal{Q}$ *at a time. In the case of* $h^{et}$ *-*$d^{tect}$ *the q-* $\mathcal{S}^{gn}$ *is split using a balanced* $b^{split}$*. One arm is used to* $\mathcal{M}$*-* $q$*, the other one— after a LO-* $p^{has}$ *shift of* $\pi/2 -$ *to* $\mathcal{M}$*-* $p$*.*

Depending on the relative $p^{has}$ -$\theta$ *of* $\mathcal{S}^{gn}$ -$m^{od}$ *and LO*, the $p^{ton}$ -$n^{ber}$ difference at the $o^{ut}$*'s* of the BS is proportional to either the *q- or p* $\mathcal{Q}$ $\triangle \hat{n} = |\alpha_{LO}|(\hat{q}\, cos\, \theta + \hat{p}\, sin\, \theta)$ . Now B can either $\mathcal{M}$ *one* $\mathcal{Q}$- component at a time by $r^{nd}$ *'ly* choosing between $\theta = 0$ *and* $\theta = \pi/2$ for each incoming $m^{od}$, *or he* $\mathcal{M}$*'s both* $\mathcal{Q}$*'s of each* $m^{od}$ in parallel, which is sometimes called "no-switching $\mathcal{P}^{col}$" *[218]* The two options of the $\mathcal{P}^{col}$ are shown in Fig. 30.

For the latter option B will $S^{epa}$ the $i^{np}$ -$s^{ta}$ with a BS and then $\mathcal{M}$ the two $\mathcal{Q}$'s by $h^{om}$ -$d^{tect}$ on each half of the $\mathcal{S}^{gn}$ -one with $\theta = 0$ to $\mathcal{M}$- $q$ and one with $\theta = \pi/2$ to $\mathcal{M}$- $p$. Here, we are speaking of $h^{om}$ -$d^{tect}$ when B -$\mathcal{M}$ only one $\mathcal{Q}$- signal at a time and of $h^{et}$ - $d^{tect}$ when B uses a BS and *two* $h^{om}$ - $d^{tect}$ to $\mathcal{M}$ both $\mathcal{Q}$- signals in parallel. Using $h^{et}$ instead of $h^{om}$ -$d^{tect}$ will double the mutual $\mathcal{J}$ for each $s^{ymb}$ for the price of additional 3 dB $\mathcal{L}^{oss}$ introduced by the $h^{et}$ -BS. Splitting B's $m^{od}$ for $h^{et}$ -$d^{tect}$, with notation $\nu=\sqrt{T(V_{mod}^2+2V_{mod})/2}$ will $t^{rans}$ the covariance $m^{tri}$ to

$$\Sigma_{AB} = \begin{matrix} & A & B_1 & B_2 \\ \begin{matrix} A \\ B_1 \\ B_2 \end{matrix} & \multicolumn{3}{l}{\begin{pmatrix} (V_{\text{mod}}+1)\mathbb{1}_2 & \nu\sigma_z & \nu\sigma_z \\ \nu\sigma_z & (\frac{T}{2}V_{\text{mod}}+1+\frac{\xi}{2})\mathbb{1}_2 & -\frac{1}{2}(TV_{\text{mod}}+\xi)\mathbb{1}_2 \\ \nu\sigma_z & -\frac{1}{2}(TV_{\text{mod}}+\xi)\mathbb{1}_2 & (\frac{T}{2}V_{\text{mod}}+1+\frac{\xi}{2})\mathbb{1}_2 \end{pmatrix}} \end{matrix}.$$

where $V_{\text{mod}} = V_{\text{m}}$ .

---

*B. $\mathcal{J}^{mpl}$ EXAMPLE:* Any $e^{xp}$ 'al $\mathcal{J}^{mpl}$ of CV-QKD (or DV-QKD) must deal with a $n^{ber}$ of issues in $o^{rde}$ to get a required *s-*$k^{\mathcal{R}}$ over a given distance [219]. The *eq* 's and $\mathcal{N}^{ois}$ -$m^{del}$ 's used here [217], enable us to design various $e^{xp}$ 'al setups, test specific $h^{rdw}$ options and see potential problems prior to the $e^{xp}$ [220]. Here we discuss in which way actual $e^{xp}$ 'al $p^{met}$ 's impact the $p^{erf}$ of a CV-QKD setup. Ultimately, the *s-*$k^{\mathcal{R}}$ is proportional to the $s^{ymb}$ -$\mathcal{R}$- $f_{\text{sym}}$ times the *s-* $f^{rct}$ -*r* (i.e. the *s-* $\mathcal{K}$ per $t^{mit}$'ed $s^{ymb}$)

$$K \sim f_{\text{sym}} \cdot r,$$

where $r$ is defined as before:

$$r = \beta I_{AB} - \chi_{EB}. \quad (32a)$$

A first option to enhance the $k^{\mathcal{R}}$ would be to use higher $s^{ymb}$ -$\mathcal{R}$ using the compatibility of CV-QKD with the advanced telecom technology. The latest $w^{ve}$ -form generators, $q/p$-modulators and $c^{hrnt}$ -$r^{ceiv}$ 's enable $s^{ymb}$ -$\mathcal{R}$'s ~ 10 Gbaud. This cannot improve the $t^{miss}$ distance since the amount of $s^{ymb}$ 's per second will have no effect on the *s-* bits *one* $s^{ymb}$ can carry and therefore cannot compensate for $r = 0$. High $\mathcal{R}$'s require high- $b^{nd}$'width $d^{tect}$ 's which, will increase the $d^{tect}$ -$\mathcal{N}^{ois}$ -$l^{ine}$ 'ly with the $b^{nd}$ 'width $B$. As shown in the sequel, the $d^{tect}$ -$\mathcal{N}^{ois}$ already builds up the most of the total $\mathcal{N}^{ois}$ in a CV-QKD $\mathcal{J}^{mpl}$. So, under the assumption that the $\mathcal{N}^{ois}$ -$g^{ner}$ 'ed at the $r^{ceiv}$ is available to E, the setup is prone to $d^{tect}$ - $\mathcal{N}^{ois}$ — and so to the increase of bandwidth and $s^{ymb}$ -$\mathcal{R}$. However, when the $d^{tect}$ -$\mathcal{N}^{ois}$ is classified as trusted $\mathcal{N}^{ois}$, fast $r^{ceiv}$'s are less $p^{blem}$ 'atic. *Exp-* $\mathcal{J}^{mpl}$ of CV-QKD $\mathcal{P}^{col}$ 's demonstrated $s^{ymb}$ $\mathcal{R}$'s in the range of kbaud up to 250 Mbaud [221].

The *s-* $f^{rct}$ (32a) $r$ depends on: 1) the $m^{odu}$ -$v^{ar}$ ($d^{ir}$ 'ly related to the mean $p^{ton}$ -$n^{ber}$ per $s^{ymb}$), 2) the $t^{mit}$ 'tance, 3) the *e-* $\mathcal{N}^{ois}$ and 4) the $\mathcal{R}^{conc}$ -$e^{ffi}$ 'cy. The $m^{odu}$ -$v^{ar}$ -$V_{\text{mod}}$ is $v^{ria}$ that can be adjusted to the $\mathcal{S}^{sec}$ specifications, so the $\mathcal{K}$ $p^{met}$ 's that define a CV-QKD setup are the $t^{mit}$ 'tance $T$, the excess ($e$) -$\mathcal{N}^{ois}$ -$\xi$ and the $\mathcal{R}^{conc}$ -$e^{ffi}$ 'cy $\beta$. Fig. 31 [217] presents the *s-*$k^{\mathcal{R}}$ (i.e. *s-* $\mathcal{K}$ per $s^{ymb}$) versus $t^{mit}$ 'tance and $\mathcal{N}^{ois}$. As indicated by Fig. 31a, if *e-* $\mathcal{N}^{ois}$ is lower higher $\mathcal{L}^{oss}$ can be tolerated and so longer $t^{miss}$ distances can be achieved. As shown in Fig. 31b, the $\mathcal{C}^{ha}$ length sets an b^ on the *e-* $\mathcal{N}^{ois}$ beyond which we cannot get s- $\mathcal{K}$ anymore. As an example, for a $\mathcal{C}^{ha}$ length of $l = 40$ km the threshold *e-* $\mathcal{N}^{ois}$ is lower than 1% of the shot $\mathcal{N}^{ois}$ which is a real $e^{xp}$ 'al challenge. For additional studies of the CV-QKD in the presence of $\mathcal{C}^{ha}$ -$\mathcal{L}^{oss}$ and $\mathcal{N}^{ois}$ see [222,223].

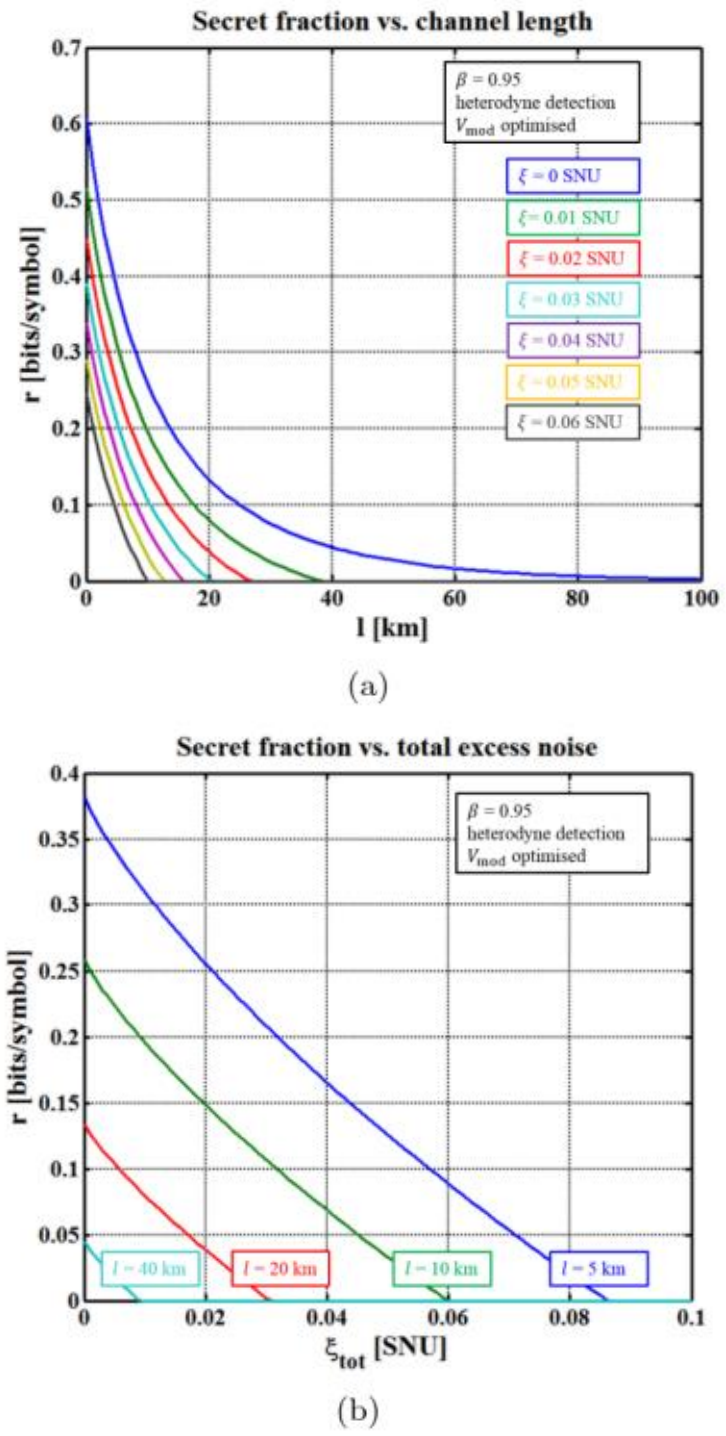

*Fig. 31 s-* $f^{rct}$ *with respect to* $\mathcal{C}^{ha}$ *length and e-* $\mathcal{N}^{ois}$*. The* $m^{odu}$ -$v^{ar}$ -$V_{mod}$ *has been* $d^{yna}$ *'ally* $\mathcal{O}$ *to maximize the simulated* $k^{\mathcal{R}}$ *for each point in the* $g^{ph}$ *'s [ 217].*

Fig. 32a shows how much the $\mathcal{S}^{sec}$ assumptions impact the needed quality of the balanced $d^{tect}$ 's: If the $d^{tect}$-$\mathcal{N}^{ois}$ contributes to an $e^{drop}$'s $\mathcal{J}$, a low $\mathcal{N}^{ois}$ -equivalent $p^{wer}$ (NEP) is required in $o^{rde}$ to establish a non-zero *s-* $\mathcal{K}$ (blue curve, NEP $< 4\text{pW}/\sqrt{\text{HZ}}$ for $T = 0.1$ and $P_{\text{LO}} = 8\text{mW}$, ignoring all other $\mathcal{N}^{ois}$ sources). In the trusted- $d^{tect}$ set up the needed quality of the NEP is more relaxed (green curve) since it only has an impact on the SNR but not the Holevo bound. The $d^{tect}$ -$\mathcal{N}^{ois}$ can be mitigated by increasing the local $o^{sc}$ -$p^{wer}$. Fig. 32a shows the $m^{ini}$ -LO $p^{wer}$ regarding the NEP under strict assumptions. Most low-$\mathcal{N}^{ois}$ balanced $d^{tect}$ 's, have the low $o^{pti}$ saturation limit of the PIN diodes which confines the LO $p^{wer}$ to ~ 10mW. To improve the $p^{erf}$ of CV-QKD in the untrusted- $r^{ceiv}$ set up we must decrease the intrinsic $d^{tect}$ -$\mathcal{N}^{ois}$ (NEP) or increase the saturation limit of the PIN diodes.

The $q^{za}$ -$\mathcal{N}^{ois}$ of ADC converter can, also be classified as trusted and so not be attributed to E. The $q^{za}$ -$\mathcal{N}^{ois}$ is proportional to the $i^{nv}$ of $2^{2n}$ where $n$ is the bit resolution of the ADC. So, adding one bit reduces $\xi_{\mathrm{ADC}}$ by a factor of 4, adding two bits will improve $\xi_{\mathrm{ADC}}$ by factor of 16. In summary, unlike the $d^{tect}$ -$\mathcal{N}^{ois}$, the $q^{za}$ -$\mathcal{N}^{ois}$ does not constitute a serious problem in CV-QKD $\mathcal{J}^{mpl}$, even under strict $\mathcal{S}^{sec}$ assumptions.

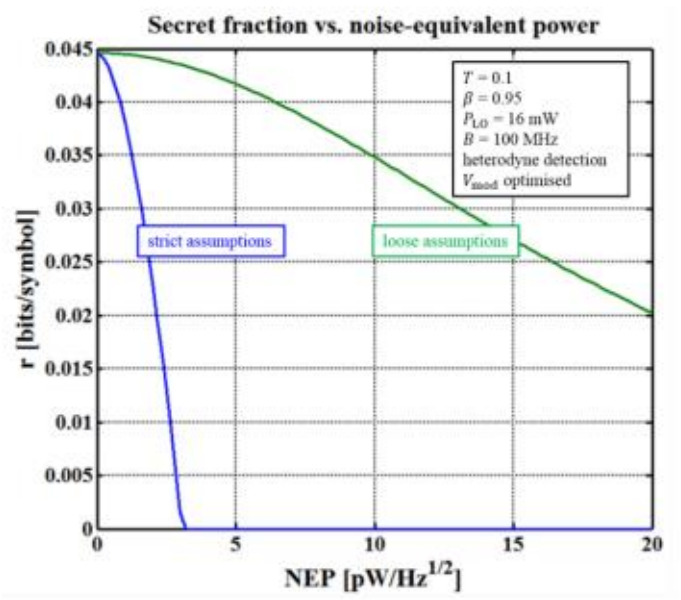

(a)

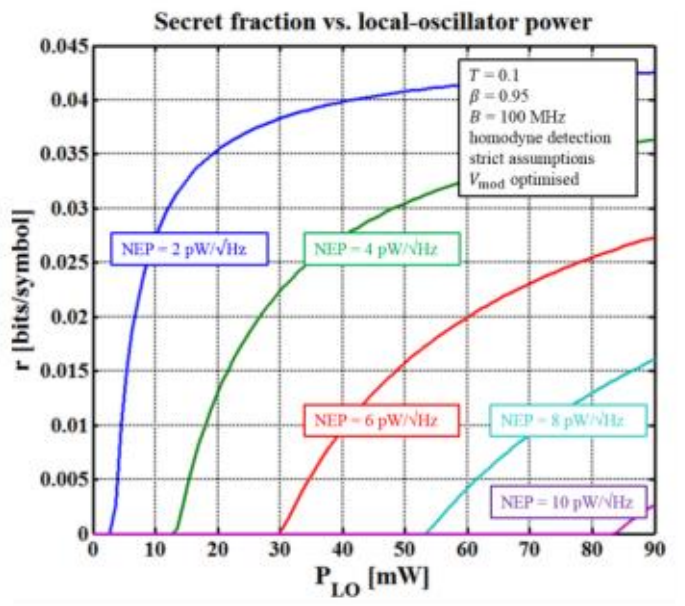

(b)

*Fig. 32(a) s- $f^{rct}$ vs. $\mathcal{N}^{ois}$ -equivalent $p^{wer}$ under strict (untrusted $d^{tect}$) and loose (trusted $d^{tect}$) assumptions. (b) s- $f^{rct}$ vs. local- $o^{sc}$ $p^{wer}$ for different NEP values under strict assumptions. The detector's NEP can be compensated with a higher LO $p^{wer}$, keeping the $e^{lec}$ 'ic $\mathcal{N}^{ois}$ small with respect to the shot $\mathcal{N}^{ois}$. However, in practice this compensation is not always feasible due the saturation limit of conventional PIN diodes (usually in the $o^{rde}$ of $\sim 10mW$) . For both plots all other $\mathcal{N}^{ois}$ sources were neglected.[ 217]*

For $\beta = 1$, Fig.33(a) demonstrates that the $m^{odu}$ -$v^{ar}$ (and therefore the SNR) can be arbitrarily increased since A and B can use the full mutual $\mathcal{J}$ for their $\mathcal{K}$ and do not leave any $\mathcal{J}$ advantage to E. The $m^{odu}$ -$v^{ar}$ -$V_{\mathrm{mod}} = 2\langle n\rangle$ can be adjusted arbitrarily and $\mathcal{O}$ to the other characteristic $p^{met}$ 's $T$, $\xi$ and $\beta$ by using $v^{ria}$ -$o^{pti}$ attenuation,. The dependence of the $\mathcal{O}$ -$m^{odu}$ -$v^{ar}$ and its tolerance on the $\mathcal{R}^{conc}$ -$e^{ffi}$ 'cy and $e$- $\mathcal{N}^{ois}$ are shown in Fig. 33.

### *C. QKD $\mathcal{J}^{mpl}$ at Terahertz $b^{nd}$ 's*

Several QKD $\mathcal{J}^{mpl}$ 's are of specific interest for 6/7G $n^{et}$ 's. As the first step we will look into the $p^{erf}$ of QKD over THz $b^{nd}$ since this $b^{nd}$ is of interest already for 6G and will continue to be of interest for 7G as well. Authors in [224] discuss a CV $q$- $\mathcal{K}$ $d^{str}$ (CVQKD) method at terahertz (THz) $b^{nd}$ 's based on $m^{lti}$ -carrier ($mC^{rri}$) $m^{lti}$ -plexing (MCM) concept.

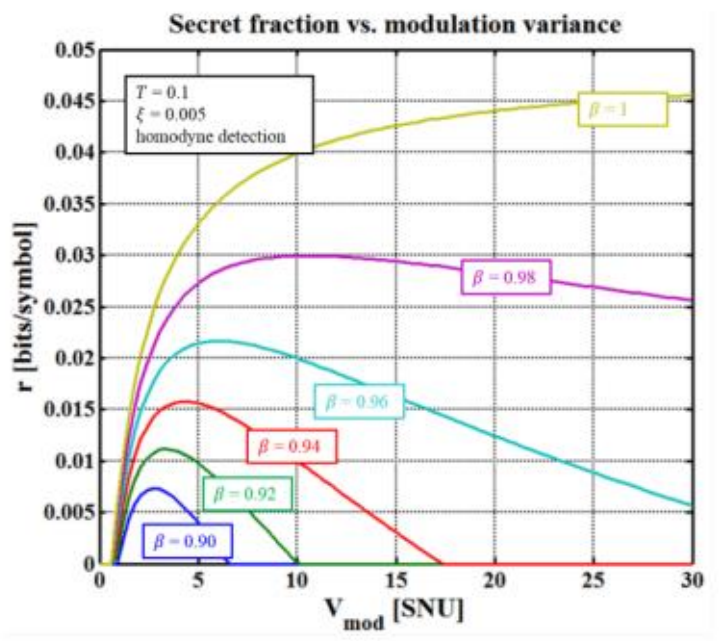

(a)

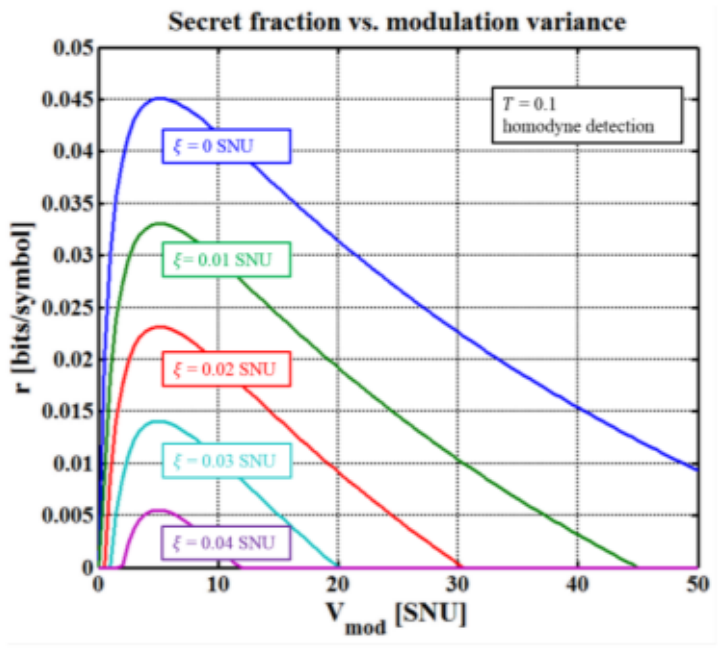

(b)

*Fig. 33 s- $f^{rct}$ vs. $m^{odu}$ -$v^{ar}$. The smaller the $\mathcal{R}^{conc}$ -$e^{ffi}$ 'cy $\beta(a)$ and the higher the e- $\mathcal{N}^{ois}$ $\xi(b)$ , the smaller becomes the interval of the $m^{odu}$ -$v^{ar}$ in which a non-zero s- $\mathcal{K}$ can be established.[217]*

Here, $m^{lti}$ 'ple $\mathcal{G}$- $m^{odu}$ 'ed subcarriers ($sC^{rri}$'s) are $c^{pld}$ to $t^{mit}$ -$m^{lti}$ -$\mathcal{P}^{th}$ superposed thermal ($t^{hrm}$) $\mathcal{G}$- $s^{ta}$. At the $r^{ceiv}$, $o^{pti}$ DFT is used to extract the $r^{ceiv}$ 'ed $sC^{rri}$ 's, and the $\mathcal{K}$'s can be $g^{ner}$ 'ed in parallel by $h^{om}$ -$d^{tect}$ and post $p^{ro}$ 'ing. The $\mathcal{S}^{sec}$ of the $s^{ch}$ against the optimal $c^{oll}$ -$\mathcal{G}$- $\mathcal{A}^{tt}$ is analyzed in indoor $e^{vir}$ and in inter-satellite $\mathcal{L}^{ink}$ 's respectively. It was shown that in MCM schemes, each sub-$\mathcal{C}^{ha}$ may be impacted by crosstalk, causing slight reduction of the $m^{ax}$ -$t^{miss}$ distance while the total s- $k^{\mathcal{R}}$ can be significantly increased. The prospect of higher $k^{\mathcal{R}}$ and longer distance in THz-QKD by using MCM schemes in inter-satellite $\mathcal{L}^{ink}$ ‘s $\mathcal{C}$ is verified as well. This may provide an $e^{ffi}$ -$\mathcal{P}^{th}$ to build a global $q$- $\mathcal{C}$- $n^{et}$.

---

The $\mathcal{C}$’s at THz band (0.1-10THz), combine the features of micro -$w^{ve}$ -$\mathcal{C}$’s and $o^{pti}$ -$\mathcal{C}$’s. With respect to the former, THz $\mathcal{C}$’s provide large capacity ($c^{city}$), $d^{ir}$ ‘ionality and $\mathcal{N}^{ois}$ immunity; and the latter, they have higher $e^{nrgy}$ $e^{ffi}$ ‘cy. The limited coverage of THz $\mathcal{C}$’s is primarily due to the water absorption in the atmosphere [225,226]. The long-distance satellite-to-ground station $\mathcal{C}$’s can be effective at THz bands if the atmosphere is dry enough [227]. THz $\mathcal{C}$’s in inter-satellite $\mathcal{L}^{ink}$ ‘s are possible [228], since in the vacuum the concentration of water is negligible, and the advantages of THz $\mathcal{C}$’s can be fully exploited. The performance of QKD $s^{yst}$ ‘s at various $w^{ve}$ -lengths of the $e^{lm}$ spectrum including the factors affecting $\mathcal{S}^{sec}$ has been studied in [229].

A THz-CVQKD $s^{yst}$ can be used in $\mathcal{S}^{sec}$ proofs. The currently available SKR of the $s^{yst}$ need to be improved, which is consistent with the traditional fiber-based QKD. There are three options to solve this $p^{blem}$: 1) to develop new $\mathcal{P}^{col}$ ‘s to increase the SKR. 2) to improve the $p^{ls}$ repetition $\mathcal{R}$ which better $p^{erf}$ of $d^{tect}$ ‘s and post- $p^{ro}$ ‘ing speed 3) to use $mC^{rri}$ -$m^{lti}$ -plexing (MCM) $t^{chn}$ ‘s to $g^{ner}$ parallel $\mathcal{K}$’s in QKD. QKD based on MCM $t^{chn}$ ‘s has been studied since they provide higher SKR [230-241]. In CVQKD based on MCM $t^{chn}$‘s , MCM divides the high-speed $d^{ta}$ -$s^{equ}$ into several low-speed $s^{equ}$ ‘s and $m^{odu}$ ‘s them on $sC^{rri}$ ‘s to achieve parallel $\mathcal{K}$- $d^{str}$. Two methods to improve the $\mathcal{K}$ -$g^{ner}$ -$\mathcal{R}$ by using $m^{lti}$ -plexing $t^{chn}$ for DVQKD $s^{yst}$ ‘s are discussed in [240]. In one option, $sC^{rri}$ ‘s are $g^{ner}$ ‘ed $d^{ir}$ ‘ly by an $o^{pti}$ comb generator or laser sources [242,243]. After $m^{odu}$ the $sC^{rri}$ ‘s, an $o^{pti}$ coupler merges them to $g^{ner}$ MCM $\mathcal{S}^{gn}$ ‘s. Alternatively, MCM $\mathcal{S}^{gn}$ ‘s are $g^{ner}$ ‘ed by the $o^{pti}$ -$i^{nv}$ discrete Fourier- $t^{rans}$ (OIDFT) $c^{irc}$ [244]. Both $s^{ch}$ ‘s use real-time ODFT at $r^{ceiv}$ ‘s [245].

Motivated by the former $s^{ch}$, authors in [224] present a CVQKD $\mathcal{P}^{col}$ at THz $b^{nd}$ ‘s using MCM concept (THz-MCM- CVQKD). The $S^{po}$ of multi-$\mathcal{P}^{th}$ $\mathcal{G}$- $s^{ta}$ is $\mathcal{J}^{mpl}$ by MCM and $\mathcal{G}$- $m^{odu}$ (GM). In the $r^{ceiv}$, the $sC^{rri}$ ‘s of $\mathcal{G}$- $s^{ta}$ are $S^{epa}$ ‘ed by ODFT, and the $\mathcal{K}$’s are $g^{ner}$ in parallel by $h^{om}$ -$d^{tect}$ and post $p^{ro}$ ‘ing. For discussion on the $\mathcal{S}^{sec}$ of the proposed QKD $\mathcal{P}^{col}$ against the $\mathcal{O}$ -$c^{oll}$ -$\mathcal{G}$- $\mathcal{A}^{tt}$ at THz $b^{nd}$ ‘s in indoor $e^{vir}$ and in inter-satellite $\mathcal{L}^{ink}$ ‘s respectively see [224].

*D QKD Over $o^{pti}$ Backbone $n^{et}$ ‘s*

Q- $\mathcal{K}$- $d^{str}$ (QKD) can provide $\mathcal{S}^{sec}$, for 5G/6G/7G $\mathcal{C}$’s. In practice, $o^{pti}$ -$n^{et}$ ‘s are cost-effective options for QKD over the existing fiber $r^{eso}$ ‘s. $\mathcal{M}$- $d^{vic}$ - $I^{dep}$ QKD can extend the $s$- distance using an *untrusted relay* ($r^{lay}$). Such $r^{lay}$ does not rely on any assumption on $\mathcal{M}$ and even allows to be accessed by an $e^{drop}$. It cannot extend QKD to an arbitrary distance like the trusted $r^{lay}$, unless it is combined with the trusted $r^{lay}$. Such a combination is discussed in [246] with focus on cost $\mathcal{O}$ during the $\mathcal{J}^{mpl}$ -$p^{has}$. A new $n^{et}$ architecture is described along with the positioning of the $n^{od}$ ‘s in such $n^{et}$. The $c^{resp}$ ‘ing $n^{et}$, cost, and $\mathcal{S}^{sec}$ -$m^{del}$ ‘s are presented in detail. To $\mathcal{O}$ the $\mathcal{J}^{mpl}$ cost, an *ILP* $m^{del}$ and a heuristic $a^{lg}$ are developed.

When it comes to the resource ($r^{eso}$) allocation ($a^{loc}$) for different $\mathcal{C}^{ha}$ ‘s (e.g., $q$- and $c$- $\mathcal{C}^{ha}$ ‘s) in a QKD integrated $o^{pti}$ -$n^{et}$, *ML* $t^{chn}$ ‘s are used for $\mathcal{O}$ -$\mathcal{C}^{ha}$ -$a^{loc}$ through real time prediction [247,248], so the $s^{uff}$ -$s$- $k^{\mathcal{R}}$ is provided in real time. In [249-251], TDMA is used in a QKD-over-WDM $n^{et}$ to enhance the $r^{eso}$ utilization for both $q$- and $c$- $\mathcal{C}^{ha}$ ‘s. The offline $r^{eso}$ -$a^{loc}$ -$p^{blem}$ has been studied with the Integer $l^{ine}$ Programming (ILP) $m^{del}$ and heuristics [249,250], and the online version is based on heuristics [251]. In [252], a number of near-$\mathcal{O}$ $w^{ve}$ -length $a^{loc}$ -$s^{ch}$ ‘s were described to enhance the achievable $s$-$k^{\mathcal{R}}$ ‘s, in nonidel $p^{hys}$ -layer. Software defined $n^{et}$ ‘ing (SDN) $t^{chn}$ ‘s can be employed to improve the flexibility of $\mathcal{C}^{ha}$ -$a^{loc}$ and QKD-integrated $o^{pti}$ -$n^{et}$, that was confirmed by $e^{xp}$ ‘s [253] and $f^{iel}$ trials [247,254,255].

Regarding QKD $n^{et}$ -$a^{pp}$ ‘s, $\mathcal{K}$ on demand [256], $m^{lti}$ -tenancy [257-259], QKD as a service [260], and QoS provisioning [261] have been described to offer high $\mathcal{S}^{sec}$. Several examples have used the $a^{pp}$ of QKD $n^{et}$ ‘s, like the $\mathcal{S}^{sec}$ improvement of the 5G service orchestration [262], the virtual optical ($o^{pti}$)- $n^{et}$ [263], and the $m^{lti}$ -cast services [264]. High volume of private $d^{ta}$ and sensitive $\mathcal{J}$ from 5G/6G $a^{pp}$ ‘s are carried on over $o^{pti}$ -$n^{et}$ ‘s, with $\mathcal{S}^{sec}$ improved by QKD [265]. In addition, the $C^{trol}$ plane in SDN and $n^{et}$ -$F$ virtualization $e^{vir}$ ‘s can be $s$’d by QKD $n^{et}$ ‘s [256], [266]. For the $\mathcal{O}$ deployment of QKD $n^{et}$ ‘s, authors in [267] studied the QKD $n^{et}$ topology, where analytical $m^{del}$ ‘s were presented to $\mathcal{O}$ the spatial $d^{str}$ of QKD $n^{od}$ ‘s. In this work, dark fibers were used for QKD instead of using existing $o^{pti}$ -$n^{et}$ ‘s. In [268], a cost- $e^{ffi}$ design was $p^{erf}$ ‘ed for uisng QKD over WDM $n^{et}$ ‘s, in which an ILP $m^{del}$ and a heuristic $a^{lg}$ were developed to achieve cost- $e^{ffi}$ QKD $n^{et}$. In [269], the $\mathcal{O}$ design of QKD for $o^{pti}$ -transport $n^{et}$ ‘s was studied, where two $m^{xin}$‘ed ILP $m^{del}$ ‘s were formulated and evaluated by simulations. The work in [267-269] only focused on the $\mathcal{O}$ of QKD with $p^{ur}$ ‘ly trusted $r^{lay}$ ‘s.

Based on [270-273] using the untrusted $r^{lay}$, the $\mathcal{O}$ deployment of QKD with hybrid trusted/untrusted $r^{lay}$ 's needs to be investigated. The routing $p^{blem}$ in a hybrid trusted/untrusted $r^{lay}$ based QKD- $n^{et}$ has been studied in [274].

*E. Q- $r^{ceiv}$ 's*

In spite of the fact that the $\mathcal{O}$- $e^{cod}$ 'ing, reaching the $\mathcal{E}$ assisted ($a^{sis}$) (EA) $\mathcal{C}^{ha}$ -$c^{city}$, was published earlier [276,277], the design of $\mathcal{O}$- $q$- $r^{ceiv}$ remains an open $p^{blem}$ [275,278]. Ref. [276] shows that in order to achieve the EA $c^{city}$ we need to use two $m^{od}$ -$\mathcal{G}$- $s^{ta}$. Ref. [278] proposes using the $m^{lti}$ 'ple sections of the feedforward sum- $f^{req}$ generation (FF-SFG) $r^{ceiv}$, $i^{nit}$ 'ly used to $d^{tect}$ the $t^{rgt}$ in highly noisy $e^{vir}$ [279]. The $s^{ch}$ is not an EA $\mathcal{C}^{ha}$ -$c^{city}$ achieving $s^{ch}$. To achieve the EA $\mathcal{C}^{ha}$ -$c^{city}$ they have $t^{mit}$'ed the same binary $\mathcal{J}$ over $D = 10^6$ bosonic $m^{od}$ 's thus occupying the whole C and L $b^{nd}$ 's as well as the portion of S- $b^{nd}$.

Authors in [275] discuss the $\mathcal{O}$ -$e^{cod}$ 'ing using $\mathcal{G}$- $m^{odu}$ (GM) of the $\mathcal{S}^{gn}$ -$p^{ton}$ in two- $m^{od}$ -squeezed-vacuum (TMSV) $s^{ta}$ and a generic EA $\mathcal{C}$- $s^{yst}$ of interest. Also they review recent $q$- $d^{tect}$ solutions using the $o^{pti}$ parametric amplifier (OPA) as the building block in $c^{resp}$ 'ing joint $r^{ceiv}$ 's. They also present joint $r^{ceiv}$ -$s^{ch}$ 's having comparable or better $p^{erf}$, with lower $c^{plex}$ 'ity without using the OPA. Then they present the results of evaluation demonstrating significantly better $\mathcal{C}^{ha}$ -$c^{city}$ performance over $c^{resp}$ 'ing Holevo, $h^{om}$, and $h^{et}$ -$\mathcal{C}^{ha}$ -$c^{city}$ 's.

*EXAMPLE:* ______________________________________

*EA c-* $o^{pti}$ -$\mathcal{C}$- $s^{yst}$: All-fiber based $q$- $n^{et}$, illustrated in Fig. 34, is used to $d^{str}$ the $\mathcal{E}$'ed- $s^{ta}$, $s^{tr}$ 'ed in $q$- $m^{emo}$ 'ies . Node A uses her $\mathcal{S}^{gn}$ -$p^{ton}$ of $\mathcal{E}$'ed pair and $t^{mit}$ 's the $c$- $d^{ta}$, imposed on the $\mathcal{S}^{gn}$ -$p^{ton}$ over $\mathcal{L}^{os}$ 'y and noisy $q$- $\mathcal{C}^{ha}$ -$\mathcal{C}$. On $r^{ceiv}$ side, $n^{od}$ -B uses the $\mathcal{E}$'ed idler $p^{ton}$ to find out what was $t^{mit}$'ed on $\mathcal{S}^{gn}$ -$p^{ton}$ in an $\mathcal{O}$- $q$- $r^{ceiv}$ $\mathcal{QR}$.

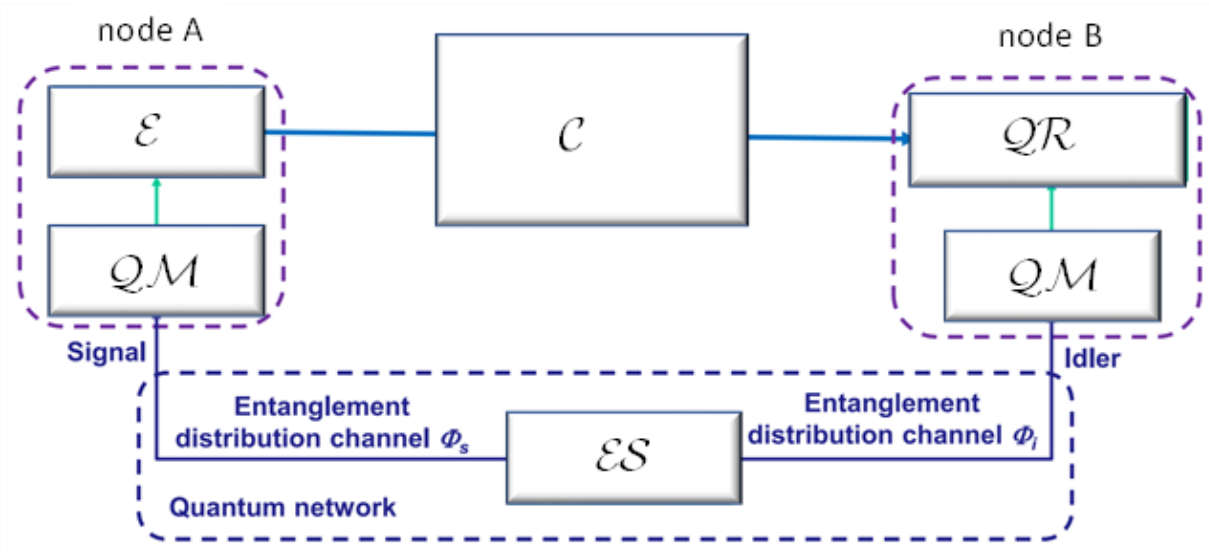

*Fig. 34 The $\mathcal{E}$ $a^{sis}$ -c- $o^{pti}$ -$\mathcal{C}$ where $\mathcal{E}$ $d^{str}$ is done through an all-fiber-based q- $n^{et}$. $\mathcal{E}$-$\mathcal{O}$- $e^{cod}$ 'er, $\mathcal{QR}$-$\mathcal{O}$- q- $r^{ceiv}$, $\mathcal{QM}$- q- $m^{emo}$, $\mathcal{ES}$-$\mathcal{E}$ source, $\mathcal{C}$-$\mathcal{L}^{os}$ 'y and noisy $\mathcal{C}^{ha}$*

The $c^{resp}$ 'ing $m^{del}$ for EA $c$- $\mathcal{C}$ is given in Fig.35, where $\mathcal{E}$ is $d^{str}$ 'ed using two $\mathcal{C}^{ha}$ 's: $\mathcal{S}^{gn}$ -$\mathcal{C}^{ha}$, $\Phi_s$, and the idler $\mathcal{C}^{ha}$, $\Phi_i$. The $q$- $\varepsilon$ -$c^{or}$ is $a^{pp}$ 'ed [275] to protect the $q$- $s^{ta}$ -$s^{tr}$ 'ed in $q$- $m^{emo}$ 'ies imperfections. The A-to-B (main) $\mathcal{C}^{ha}$ is $m^{del}$ 'ed by the 1- $m^{od}$ -$t^{hrm}$ -$\mathcal{L}^{os}$ 'y -$\mathcal{B}$- $\mathcal{C}^{ha}$ -$m^{del}$ $\hat{a}_s = \sqrt{T}\hat{a}_s + \sqrt{1-T}\hat{a}_b$, where $T$ is the transmissivity of the main $\mathcal{C}^{ha}$, while $\hat{a}_b$ is a $t^{hrm}$ (background) $m^{od}$ with the mean $p^{ton}$ -$n^{ber}$ $=N_b/(1-T)$ . This $\mathcal{C}^{ha}$ can also be seen as a AWGN $\mathcal{C}^{ha}$ with $p^{wer}$-spectral density of $N_b$ and attenuation coefficient $T$. Node A $m^{odu}$ 's the $\mathcal{S}^{gn}$ -$m^{od}$ $\hat{a}_s$, by using I/Q modulator (Fig. 35), by $p^{erf}$ 'ing the $\mathcal{G}$- $m^{odu}$ (GM).

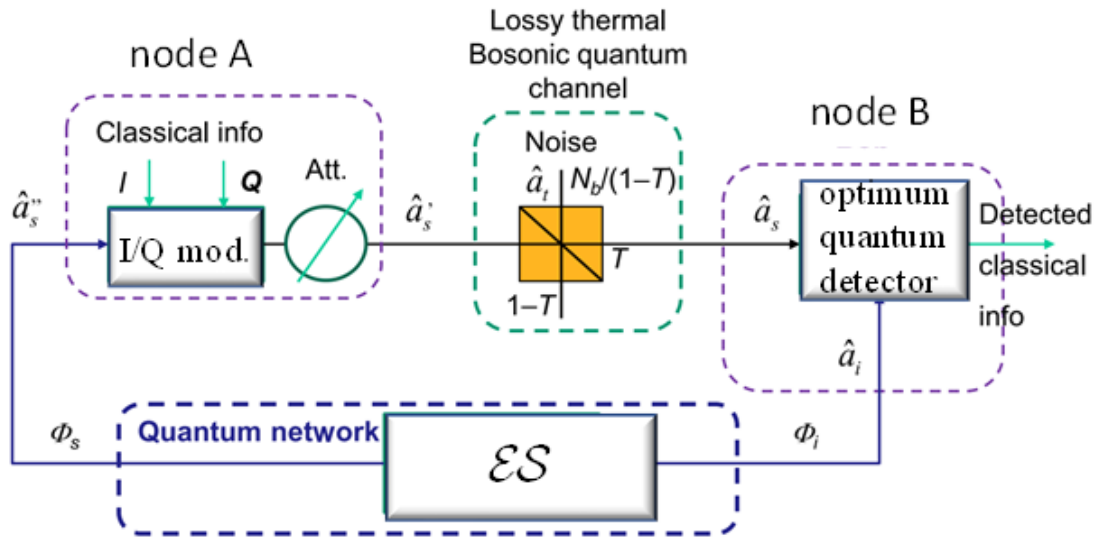

*Fig. 35. Illustrating the $\mathcal{E}$ $a^{sis}$ c- $\mathcal{C}$- $s^{yst}$ $m^{del}$. I/Q mod: I/Q modulator, Att.: attenuator (it is optional).*

---

| Topic | References |
|---|---|
| Eavesdropping attacks | [211] |
| Transmission of the quantum states | [212 – 215] |
| *Gaussian Modulation* | [216,217] |
| CV-QKD *Implementation Example* | [217,219-221] |
| *QKD Implementation at Terahertz Bands* | [224-245] |
| *QKD Over Optical Backbone Networks* | [246-274] |
| *Quantum Receivers* | [275-279] |

*Table 3 Recommended reading on CV QKD $\mathcal{J}^{mpl}$*

For additional details on the $o^{per}$ 'ion principle of OPA based $r^{ceiv}$, non' $l^{ine}$ -$r^{ceiv}$ 's and joint $r^{ceiv}$ 's see [275].

## VIII CONCLUSION

In this paper we provide optimization framework for decision-making on investment partitioning between hardware and network solutions and survey necessary work for an insight into available q-hardware and its performance (imperfections).

As a special contribution of this paper Section II presents new optimization frameworks for energy efficiency optimization in q- satellite networks primarily intended for QKD. The solution is focused on optimizing LEO constellation that enables energy savings *up to factor $10^2$ compared to the existing proposals that also include use of GEO satellites*. The optimization problems have the combinatorial form and leverage the use of quantum computing. In addition to significant speed up in the computing rate (order of $10^8$ announced by Google) they are expected to use q-search algorithms QSA and q-optimization algorithms that enable additional benefits.

1. QSA $\mathcal{A}^{lgrt}$ like Grover's $\mathcal{A}^{lgrt}$ can find the $m^{ax}$ -$v^{lu}$ of the component in the set of N entries in ~$N^{1/2}$ iterations while the c-approach with exhaustive search would require ~N iterations. So, if for example N=$10^6$ , Grover's $\mathcal{A}^{lgrt}$ would find the $m^{ax}$ (optimum $v^{lu}$) $10^3$ time faster than the c- approach.

2. In general, Quantum Approximate Optimization $\mathcal{A}^{lgrt}$ 's can find the $m^{ax}$ of the combinatorial $\mathcal{O}^{ptmz}$ -$p^{rblm}$ in $p^{oly}$ -times so turning the NP hard $p^{rblm}$ 's with exponential times into faster $p^{rblm}$ 's with the price that the optimum $v^{lu}$ is an $a^{prx}$ 'ion. The compromise between the accuracy and speed up in the execution of the $\mathcal{O}^{ptmz}$ -$\mathcal{A}^{lgrt}$ is the design $p^{mtr}$.

The paper also presents the problem of q-network cost optimization considering the specific parameters of the network, type of the used algorithms and network coherency time. As an example, one of the conclusions from that segment is that investment into the network synchronization (synchronous reset of the components' q-states) keeps the probability of network coherency $p_{cn}$ equal to the probability of coherency of an individual q-component $p_{cc}$. Otherwise, the probability of network coherency would decay as $p_{cn}=(p_{cc})^m$ where m is the number of processing q-hardware components used in the q-state processing. *As illustration, for m=10 and* $p_{cc}=0.9$ *we have* $p_{cn}=0.348$ *and for* $p_{cc}=0.99$ *the probability of network coherency reduces to 0.904.*

Finally, having in mind the channel capacities and secret key rates, the paper presents the optimization framework for optimum allocation of these resources in the network.

We believe that the insight into the available q-technology (presented in the way close to the mindset and terminology of the network designers) and optimization frame work for its use will help 7G network designers to decide if, when and how to deploy it.